\documentclass[iop,apj]{emulateapj}
\usepackage{amsmath,amssymb}
\usepackage{graphicx}
\usepackage{natbib}
\usepackage{color}
\usepackage[colorlinks,citecolor=blue,linkcolor=red]{hyperref}
\graphicspath{{./}{figs/}}

\newcommand{\angstrom}{{\rm \mathring A}}
\newcommand{\actaa}{{\rm AcA}}

\begin{document}
\title{
Simulating the timescale dependent color variation in quasars \\
with a revised inhomogeneous disk model 
}
\author{Zhen-Yi Cai\altaffilmark{1}, Jun-Xian Wang\altaffilmark{1}, Wei-Min Gu\altaffilmark{2}, Yu-Han Sun\altaffilmark{1}, Mao-Chun Wu\altaffilmark{1}, Xing-Xing Huang\altaffilmark{1} \& Xiao-Yang Chen\altaffilmark{1} }
\altaffiltext{1}{CAS Key Laboratory for Researches in Galaxies and Cosmology, University of Science and Technology of China, Chinese Academy of Sciences, Hefei, Anhui 230026, China; zcai@ustc.edu.cn, jxw@ustc.edu.cn}
\altaffiltext{2}{Department of Astronomy and Institute of Theoretical Physics and Astrophysics, Xiamen University, Xiamen, Fujian 361005, China}

\begin{abstract}
The UV/optical variability of active galactic nuclei and quasars is useful for understanding the physics of the accretion disk and is gradually attributed to the stochastic fluctuations over the accretion disk. Quasars generally appear bluer when they brighten in the UV/optical, the nature of which remains controversial. Recently \citeauthor{Sun2014} discovered that the color variation of quasars is timescale dependent, in the way that faster variations are even bluer than longer term ones. While this discovery can directly rule out models that simply attribute the color variation to contamination from the host galaxies, or to changes in the global accretion rates, it favors the stochastic disk fluctuation model as fluctuations in the innermost hotter disk could dominate the short-term variations. In this work, we show that a revised inhomogeneous disk model, where the characteristic timescales of thermal fluctuations in the disk are radius-dependent (i.e., $\tau \sim r$; based on the one originally proposed by \citeauthor{DexterAgol2011}), can well reproduce a timescale dependent color variation pattern, similar to the observed one and unaffected by the un-even sampling and photometric error. This demonstrates that one may statistically use variation emission at different timescales to spatially resolve the accretion disk in quasars, thus opens a new window to probe and test the accretion disk physics in the era of time domain astronomy. Caveats of the current model,  which ought to be addressed in future simulations, are discussed.
\end{abstract}


\section{Introduction}

Since they were discovered \citep{MatthewsSandage1963}, quasars have been found to be aperiodically variable from radio to X-ray and gamma-ray \citep{Ulrich1997}, whose UV/optical emissions vary in flux on an order of $\sim 10$-20\,\% and on characteristic timescales ranging from days to years \citep{Hook1994,VandenBerk2004,Sesar2007}. Currently, the studies of quasar variability are generally based on long-term monitoring of individual sources \citep{Sesar2007,Kelly2009,Kozlowski2010,MacLeod2010,Sun2014}, such as the Optical Gravitational Lensing Experiment \citep[OGLE,][]{Udalski1997} and the Sloan Digital Sky Survey Stripe 82 \citep[SDSS,][]{York2000,Sesar2007}, and/or on the ensemble properties of quasars \citep{VandenBerk2004,deVries2005,Wilhite2005,Wilhite2008,Sesar2006,Bauer2009,MacLeod2011}. 
On one hand, the former approach mainly employing spectral techniques, e.g., power density spectra, structure functions, etc., has revealed that the power density spectra of quasar optical light curves are of form of $P(\nu) \propto \nu^{-2}$, where $\nu$ is the frequency, on timescales of 100 to 1000 days \citep{Giveon1999,CollierPeterson2001}
and could become flattened on longer timescales \citep{Czerny1999}. Motivated by these results and the similar power density spectrum implied by the damped random walk (DRW) process, \citeauthor{Kelly2009} (\citeyear{Kelly2009}, see also \citealt{Kozlowski2010,MacLeod2010,Zu2013}) have demonstrated that the optical variations in AGNs could be well modeled with the DRW process on timescales of weeks to years, though the power density spectra of some Kepler AGNs appear much steeper, with slope of -2.6 to -3.3, on shorter timescales \citep{Mushotzky2011,ChenWang2015}.
On the other hand, the ensemble studies have uncovered several relationships between the optical variability amplitude and the physical properties of AGNs, including the anti-correlations with the observed wavelength, the bolometric luminosity, and the Eddington ratio, the positive correlation with the BH mass, as well as no significant correlation with the redshift, etc. \citep{Hook1994,Cristiani1996,Garcia1999,VandenBerk2004,Wold2007,Wilhite2008,Bauer2009,MacLeod2010,Zuo2012,MeusingerWeiss2013}.

In addition, the amplitudes of UV/optical variations are larger in bluer bands, which is also coined as the so-called bluer-when-brighter trend, that is, AGNs normally appear bluer when they get brighter \citep[][]{Cutri1985,Wamsteker1990,Clavel1991,Giveon1999,WebbMalkan2000,Trevese2001,TreveseVagnetti2002,VandenBerk2004,Wilhite2005,Meusinger2011,Sakata2011,Schmidt2012,Zuo2012,Bian2012,Ruan2014,Sun2014,Guo2016}. More strikingly, the variations across the UV/optical continuum almost simultaneously occur in phase, with a time lag less than 1-2 days inferred from local Seyfert galaxies \citep{Krolik1991,Clavel1991,Korista1995,Edelson1996,Edelson2000,Crenshaw1996,Wanders1997,Collier1998,Collier2001,Peterson1998,Kriss2000,Doroshenko2005,Sergeev2005,Breedt2009} and more luminous quasars \citep{Giveon1999,Hawkins2003}.

Although the physical origin for quasar variability and then the aforementioned observations are still unclear, it is widely accepted that the variations are generally intrinsic, thanks to the results from reverberation mapping showing that the variation of broad emission lines closely responds to that of the continuum just after some time lag \citep[e.g.,][]{Peterson2004,Peterson2005}. 
Despite the changes of global accretion rate or the {\it global temperature fluctuations} could account for many aforementioned correlations and the bluer-when-brighter trend \citep[e.g.,][]{Pereyra2006,LiCao2008,Sakata2011,Zuo2012,GuLi2013}, 
the hint that the characteristic timescale of quasar optical light curves could be identified with the thermal timescales of accretion disk is found by \citet{Kelly2009}, when modeling the light curves as DRW process. 
This stimulates \citet{DexterAgol2011} to consider a simple inhomogeneous disk model with {\it localized temperature fluctuations} undergoing DRW process to simultaneously explain the observed larger size of accretion disk than that predicted by the standard thin disk, and the amplitudes of the optical variability.
Later on, \citet{Ruan2014} demonstrated that, rather than changing the global accretion rate in the thin disk, the inhomogeneous disk model can better account for the bluer-when-brighter trend constructed from a sample of 604 variable quasars. 

Recently, an interesting timescale dependence of the color variability in quasars, that is, 
short-term variations are even bluer than longer term ones at all redshifts up to $z \sim 3.5$, was discovered by \citet{Sun2014}, using the SDSS $g$- and $r$-band photometric monitoring data for quasars in Stripe 82. Such timescale dependency of the color variation can rule out the models that simply attribute the color variations to mixture of a variable disk emission with blue but constant color and a redder stable emission such as from the host galaxy \citep[e.g.,][]{Hawkins2003}. It can not either simply be explained by changes in the global accretion rate.

A natural mechanism to the observed timescale dependent color variations, as \cite{Sun2014} already pointed out, is that short-term variations could be dominated by thermal fluctuations in the inner most region of the accretions disk where the disk is hotter and the disk emission is bluer, while longer term variations are produced over larger scales with lower effective disk temperatures.
This yields an interesting consequence that quasar variations at different timescales correspond to different disk regions, and one could possibly utilize this effect to spatially resolve the otherwise un-resolvable accretion disk. 

It is therefore essential to numerically testify whether localized disk temperature fluctuations can reproduce a timescale dependent color variation pattern in quasars as observed. 
In the present work, starting from the inhomogeneous disk model of \citet{DexterAgol2011} we simulate localized disk temperature fluctuations and investigate the yielded color variations and the timescale dependency.
The inhomogeneous disk models (and necessary revisions) and their implications are completely demonstrated in Sections~\ref{sect:model} and \ref{sect:results}, respectively, and discussed in Section~\ref{sect:discussion}. Finally, a brief summary is presented in Section~\ref{sect:conclusion}.

\section{Inhomogeneous accretion disk model}\label{sect:model}

The disk fluctuation has been proposed to be responsible for the UV/optical quasar variability, which has been shown to be sufficiently described by the DRW process \citep{Kelly2009,Zu2013}. Besides directly describing the optical light curves, the DRW model has also been assumed for the temperature fluctuation of independent zones on the accretion disk to depict the optical quasar variability \citep{DexterAgol2011,Ruan2014}. Following \cite{DexterAgol2011}, we construct similar inhomogeneous disk models to test whether they can present the timescale dependent color variability of quasars recently discovered by \cite{Sun2014}.

As reference, we consider a standard thin disk surrounding a Schwarzschild black hole (BH) with mass $M_\bullet$, whose innermost radius, $r_{\rm in}$, is assumed to be the same as the innermost stable circular orbit, $r_{\rm ms} = 6 r_{\rm g}$, where $r_{\rm g} \equiv G M_\bullet / c^2$ is the gravitational radius.
The temperature profile of the standard thin disk as a function of radius, $r$, is given by
{\small
\begin{align}
	T_{\rm d}(r) &= \left\{ \frac{3 G M_\bullet \dot M}{8\pi r^3 \sigma_{\rm SB}} \left[1 - \left(\frac{r_{\rm in}}{r} \right)^{1/2}\right] \right\}^{1/4}~{\rm K},
\end{align}}\noindent
where $\sigma_{\rm SB}$ is the Stefan-Boltzmann constant and $\dot M$ the accretion rate. Unless otherwise stated, we will consider our reference models with typical BH mass $M_\bullet = 5 \times 10^8\,M_\odot$ and accretion rate $\dot M = 1\,M_\odot\,{\rm yr}^{-1}$, corresponding to the Eddington ratio $\lambda_{\rm Edd} = 0.075$ for radiation efficiency $\eta = 1/12$. Note that our main results {on the timescale dependent color variations} do not depend on these selections (see Section~\ref{sect:parameters} for discussion).

\begin{figure}[!t]
\centering
\includegraphics[width=\columnwidth]{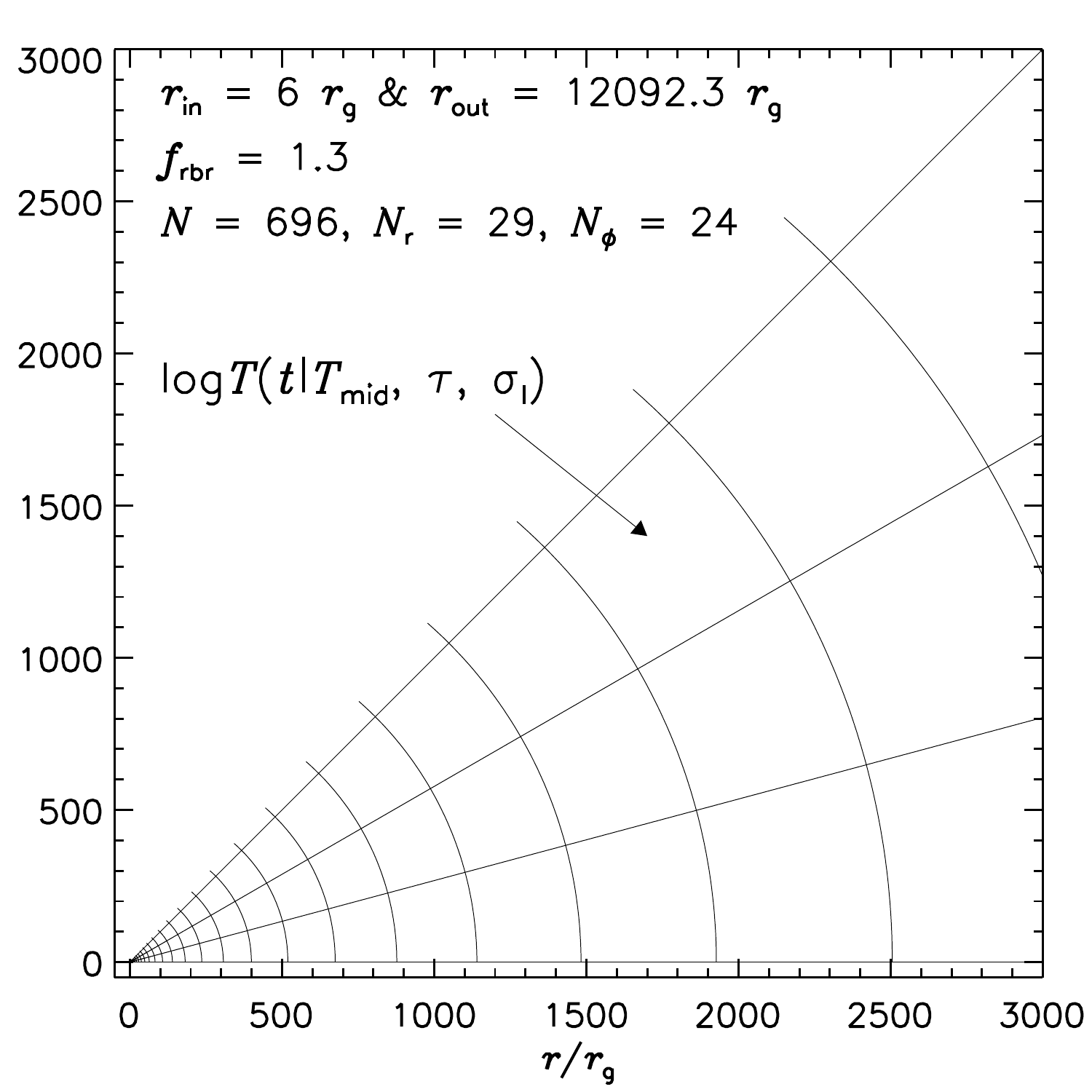}
\caption{Splitting accretion disk into totally $N = 696$ square-like zones in $r$ and $\phi$ with $N_{\rm r} = 29$ layers and $N_{\phi} = 24$ zones per each layer, according to the given inner boundary $r_{\rm in} = 6\,r_{\rm g}$, the outer boundary $r_{\rm out} \sim 10^4\,r_{\rm g}$, and the radial boundary ratio of each layer $f_{\rm rbr} = 1.3$. Damped random walk process is adopted to generate the fluctuations of $\log T$ of each zone once the mean logarithmic temperature, $\log T_{\rm mid}$, the characteristic timescale for the fluctuations to return to $\log T_{\rm mid}$, $\tau$, and the characteristic amplitude of variations on long timescales ($\gg \tau$), $\sigma_{\rm l}$, are provided.}\label{fig:disk}
\end{figure}

To avoid the fluctuating regions being stretched along the radial direction, the accretion disk is split into totally $N$ square-like zones in $r$ and $\phi$ 
with $N_{\rm r}$ layers and $N_\phi$ zones per each layer from the inner boundary $r_{\rm in}$ to a given outer boundary $r_{\rm out} \simeq 10^4 r_{\rm g}$. The number of layer $N_{\rm r}$ equals the integral part of $1+\log_{f_{\rm rbr}}(r_{\rm out}/r_{\rm in})$, where $f_{\rm rbr}$ is the radial boundary ratio of each layer. The number of zones per each layer $N_\phi$ equals the nearest integer of $\pi(f_{\rm rbr}+1)/(f_{\rm rbr}-1)$. The final adopted $r_{\rm out}$ is then updated with $N_{\rm r}$ and $f_{\rm rbr}$, but the exact value is not so relevant once it is set to be large enough. Note that given the inner and outer radial boundaries the only free parameter $f_{\rm rbr}$ characterizes the extent of disk inhomogeneity. An example of splitting disk is illustrated in Figure~\ref{fig:disk}.
\citet{DexterAgol2011} divided the disk into $n$ evenly spaced zones in $\log r$ and $\phi$ per factor of two in radius and found that $n \simeq 10^2-10^3$ is preferred by the data.
Equivalently, $n \simeq N_\phi \log_{f_{\rm rbr}}(2) \simeq 64$ if $f_{\rm rbr} = 1.3$. Although our reference equivalent number of zones is marginally smaller than that suggested by \citet{DexterAgol2011}, we do not expect that our conclusions would be altered (see Section~\ref{sect:parameters} for discussion).

Applying the DRW process to the temperature fluctuation of any zone on the accretion disk, the expected value and variance of logarithmic temperature, $\log T(t)$, of the zone at an epoch $t$ given $\log T(s)$ at a previous epoch $s$ ($s < t$) are, respectively \citep{Kelly2009},
{\footnotesize
\begin{align}\label{eq:DRW}
	E[\log T(t) | \log T(s)] &= \log T_{\rm mid} + \mbox{e}^{-\Delta t/\tau} [ \log T(s) - \log T_{\rm mid} ] \nonumber\\
	\mbox{Var}[\log T(t) | \log T(s)] &= \frac{\tau \sigma^2_{\rm s}}{2} (1 - \mbox{e}^{-2\Delta t/\tau}),
\end{align}}\noindent
where $\Delta t = t -s $, $\log T_{\rm mid}$ is the mean logarithmic temperature to which the fluctuation of $\log T$ would return at a characteristic timescale $\tau$, and $\sigma_{\rm s}$ the characteristic amplitude of variations per day$^{1/2}$ (the variances on short timescales $\approx \sigma^2_{\rm s} \Delta t$ for $\Delta t \ll \tau$ and on long timescales $\sigma^2_{\rm l} \simeq \tau \sigma^2_{\rm s} /2 $ for $\Delta t \gg \tau$). 
Since the temperature fluctuations have been simulated under the DRW process, which is only one of the vast array of autoregressive models, one may worry about that the results discussed in this work may depend on the assumed specific autoregressive model. However, this simple DRW process has been shown to be able to provide significant good description of quasar light curves on timescales of days to years at the level of data quality of the OGLE and Stripe 82 surveys (\citealp{Kozlowski2010,MacLeod2010,Zu2013,Andrae2013}; however, see \citealp{Mushotzky2011,Graham2014}). Especially, \citet{Andrae2013} have shown that the simple DRW process is by far the best model for quasar light curves from Stripe 82 and is favored over many other deterministic and stochastic models, including multiple DRW process, higher order continuous autoregressive processes, composite models, etc. Complementing that the structure functions of the simulated light curves in all considered models are generally consistent with that of a singel DRW process (see Section~\ref{sect:SF}), we do not expect that the following results would be significantly sensitive to the assumed proper stochastic process.

To set up the relationship between $T_{\rm mid}$ and $T_{\rm d}$, we assume that the energy of each zone undergoing temperature fluctuations is separately conserved over a sufficient period. Therefore, the optimized mean temperature $\log T_{\rm mid}$ of a zone spanning from $r_{\rm min}$ to $r_{\rm max}$, is chosen such that $\langle T^4 \rangle_t = T^4_{\rm eff} $, where $T_{\rm eff} = T_{\rm d}(r_{\rm eff})$ is the effective temperature of that zone at the effective radius $r_{\rm eff}$ given by $T^4_{\rm d}(r_{\rm eff}) \equiv \int^{r_{\rm max}}_{r_{\rm min}} T^4_{\rm d}(r) r \mathrm dr / \int^{r_{\rm max}}_{r_{\rm min}} r \mathrm dr$. Furthermore, according to the DRW model, the temperature distribution of a long enough series of $\log T(t)$ satisfies a normal distribution with mean temperature, $\log T_{\rm mid}$, and variance on long timescales, $\sigma^2_{\rm l} = \tau \sigma^2_{\rm s}/2$, that is\footnote{This distribution has been justified using our simulated fluctuations of $\log T$, considering that the standard deviation $\sigma_{\rm sd}$ of the distribution of temperature fluctuations in logarithm approaches $\sigma_{\rm l}$ with increasing simulated time.},
{\footnotesize
\begin{equation}
	P[\log T | \log T_{\rm mid}]\,\mathrm d \log T = \frac{\exp[-\log^2(T/ T_{\rm mid})/2 \sigma^2_{\rm l}]}{\sqrt{2 \pi \sigma^2_{\rm l}}}\,\mathrm d \log T.
\end{equation}}\noindent
Consequently, we have
{\footnotesize
\begin{align}
	\langle T^4 \rangle_t &= \int^\infty_{\log T=-\infty} T^4 P[\log T | \log T_{\rm mid}]\,\mathrm d \log T \nonumber\\
	&= T^4_{\rm mid} \int^\infty_{\ln w=-\infty} \frac{\mathrm d\ln w}{\sqrt{2\pi (\sigma_{\rm l} \ln 10)^2}} \times \nonumber\\
	&\,\,\,\,\,\,\,\,\mbox{e}^{ - \{ [\ln w - 4 (\sigma_{\rm l} \ln 10)^2]^2 - [4(\sigma_{\rm l} \ln 10)^2]^2 \} / 2 (\sigma_{\rm l} \ln 10)^2 } \nonumber\\
	&= T^4_{\rm mid} \mbox{e}^{8 (\sigma_{\rm l} \ln 10)^2},
\end{align}}\noindent
where $w \equiv T / T_{\rm mid}$, or $\log T_{\rm mid} = \log T_{\rm eff} - 2 \sigma^2_{\rm l} \ln 10$. This energy conservation is applied to every zone of the disk over time under the assumption that the temperature of each zone fluctuates independently. 

To obtain the spectrum energy distribution (SED) and the monochromatic luminosity of the fluctuating disk, we simply assume the disk emission is locally blackbody and viewed face-on.
At this stage we neglect relativistic effects and disk atmosphere radiative transfer (see Section \ref{sect:caveats} for discussion). The specific monochromatic luminosity with temperature fluctuations, emitted by one side of a disk zone in an annulus at $r \sim r + \Delta r$ with the effective radius $r_{\rm eff}(r, \Delta r) \in (r, r + \Delta r)$, 
is
\begin{align}
 \Delta \tilde L_\nu (\nu, t|r, N_\phi) = \sum^{N_\phi}_{i=1} \pi B_\nu[\nu, T_i(t, r)] \frac{2\pi}{N_\phi} r \Delta r,
\end{align}
where $B_\nu(\nu, T)$ is the blackbody radiation intensity.
When $N_\phi \rightarrow \infty$, the distribution of temperature fluctuations of this annulus at an epoch is equivalent to that of a single zone of the same annulus over a long enough period. Therefore, the specific monochromatic luminosity emitted from this annulus with temperature fluctuations is independent of time but only of the amplitude of variations, e.g., $\sigma_{\rm l}$, as
{\footnotesize
\begin{align}\label{eq:Lnu_Nphi_infty}
& \frac{\Delta \tilde L_\nu(\nu|r)}{2\pi r \Delta r} = \frac{1}{2\pi r \Delta r} \lim_{N_\phi \rightarrow \infty} \Delta \tilde L_\nu (\nu, t|r, N_\phi) \nonumber\\
	&=  \int^\infty_{\log T = -\infty} \pi B_\nu(\nu, T) P[\log T|\log T_{\rm mid}]\,\mathrm d\log T \nonumber\\
	&= \int^\infty_{\log T=-\infty} \frac{2\pi h_{\rm P} \nu^3}{c^2} \frac{\mbox{e}^{-\log^2(T/T_{\rm mid})/2\sigma^2_{\rm l}}}{\mbox{e}^{h_{\rm P} \nu / k_{\rm B} T}-1} \frac{\mathrm d\log T}{\sqrt{2\pi\sigma^2_{\rm l}}} \nonumber\\
	&= \frac{2\pi h_{\rm P} \nu^3}{c^2} \int^\infty_{w=0} \frac{\mathrm dw}{\sqrt{2\pi} \sigma_{\rm l} \ln 10\,w} \frac{ \mbox{e}^{-\frac{[\ln w + 2(\sigma_{\rm l} \ln 10)^2]^2}{2(\sigma_{\rm l} \ln 10)^2} }}{ \mbox{e}^{x/w}-1 }
\end{align}}\noindent
where $x \equiv h_{\rm P} \nu / k_{\rm B} T_{\rm eff}$, $w \equiv T / T_{\rm eff}$, frequency $\nu$, Planck constant $h_{\rm P}$, and Boltzmann constant $k_{\rm B}$. 
Comparing the above equations to that of Equation~(2) of \citet{DexterAgol2011}, we would get $\sigma_{\rm T} = \sqrt{2} \sigma_{\rm l}$, where the quantity\footnote{Such quantity was also adopted as a parameter of DRW process by some authors, e.g., \citet{MacLeod2010}.} $\sigma_{\rm T}$ quoted by them means the value of structure function of $\log T$ when the time separating two observations is much larger than the damping timescale $\tau$ of DRW process. For the characteristic $\sigma_{\rm T} = 0.35$ and $\tau = 200$ days suggested by \citet{DexterAgol2011} and \citet{Kelly2009}, respectively, we have $\sigma_{\rm l} \simeq 0.25$ dex and $\sigma_{\rm s} = \sigma_{\rm T}/\sqrt{\tau} \simeq 0.025$ dex day$^{1/2}$.

At any epoch $t$, summing up all the assumed blackbody emissions of each zone under the fluctuations of $\log T(t)$, we will obtain the corresponding fluctuated SED as well as the fluctuated specific monochromatic luminosity, $\tilde L_\nu(\nu, t|N)$, emitted by one side of the disk.

\begin{figure*}[!t]
\centering
\includegraphics[width=\columnwidth]{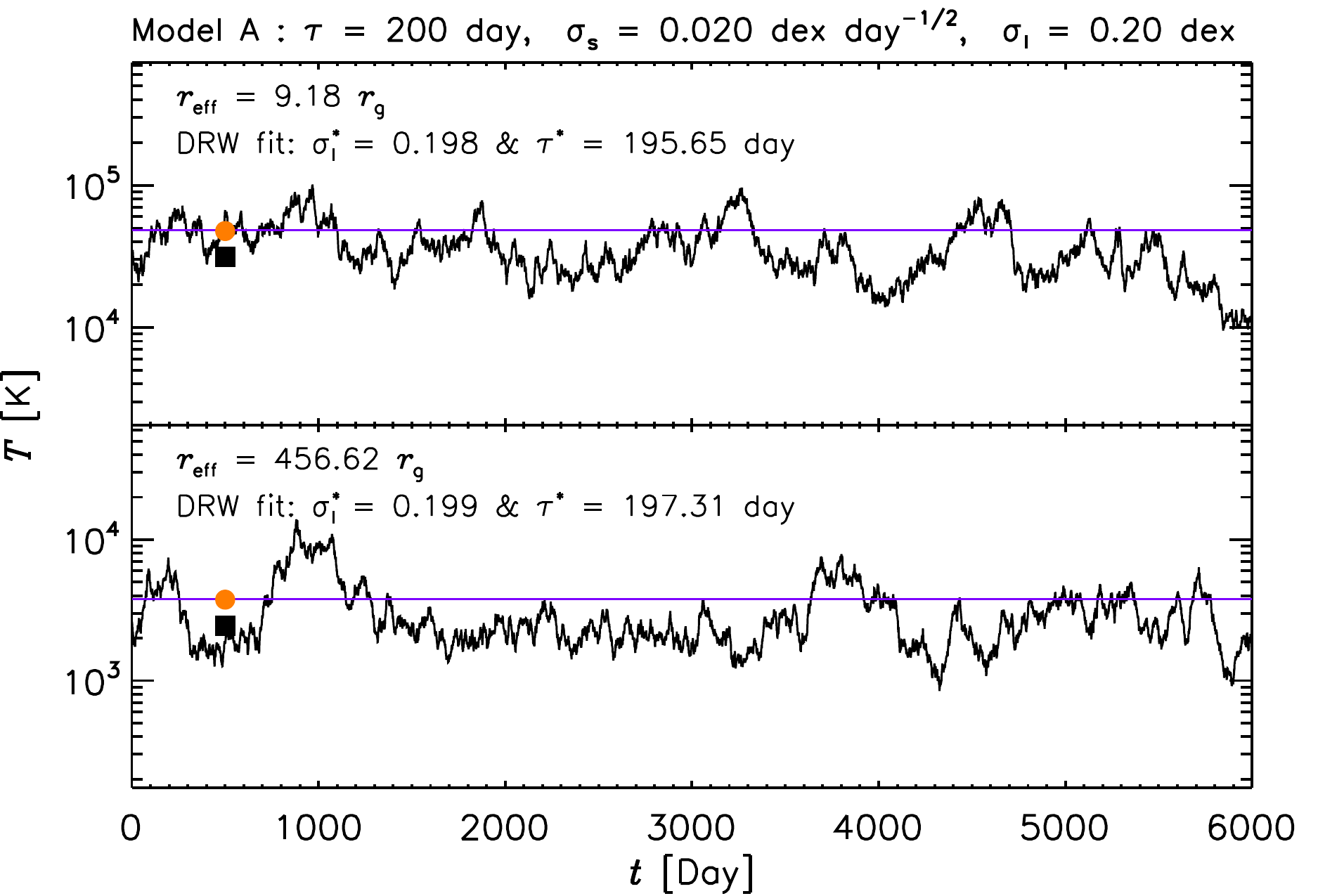}
\includegraphics[width=\columnwidth]{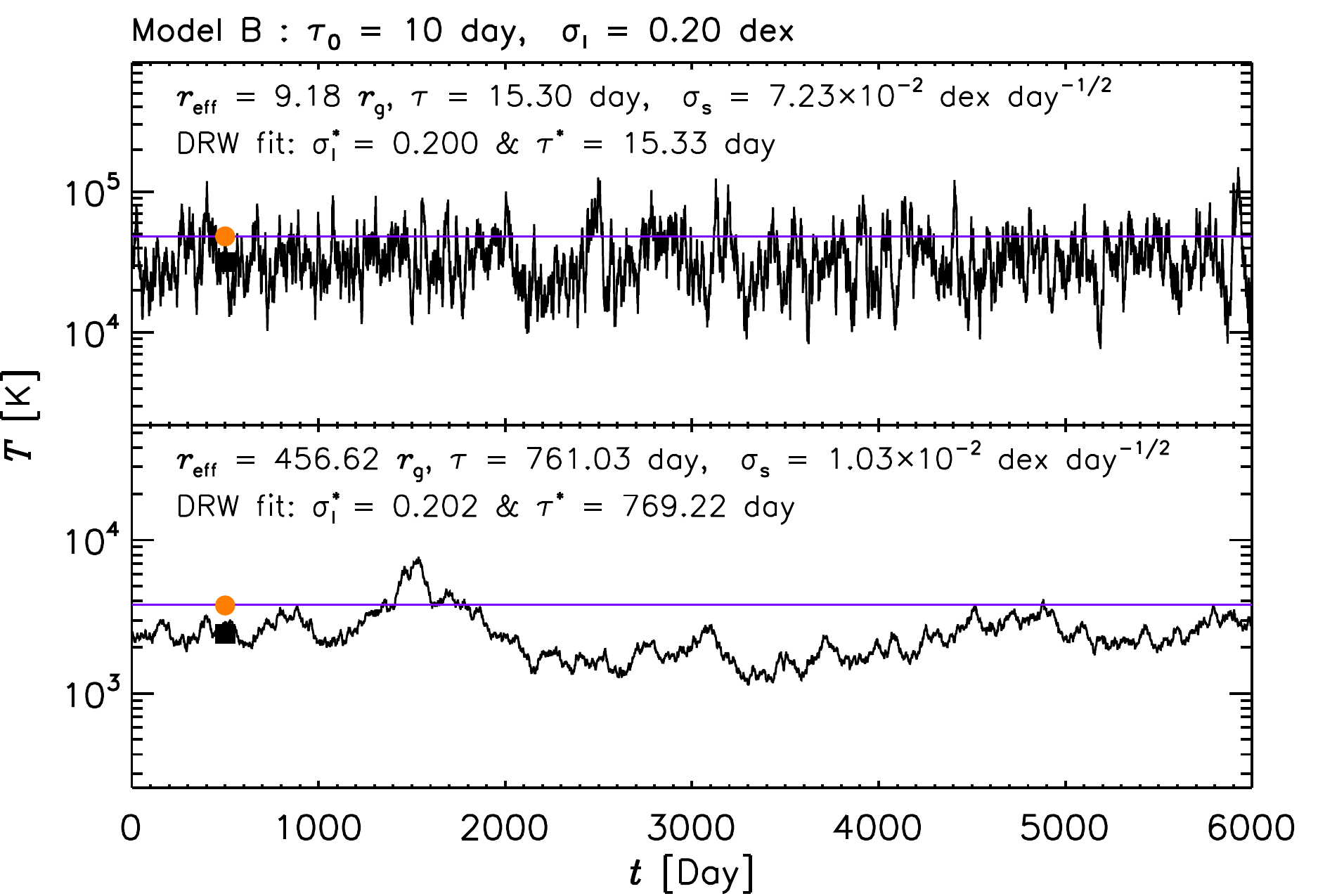}
\caption{An illustration of the fluctuations of $\log T$, simulated with Equation~(\ref{eq:DRW}) in steps of $\Delta t = 1$ day up to $t_{\rm max} = 10^5$ days, for two zones at different effective radii, $r_{\rm eff}$. The left panels show the fluctuations of the reference {\bf model A} with radius-independent $\tau = 200\, {\rm days}$ and long-term fluctuation amplitude $\sigma_{\rm l} = 0.2\,{\rm dex}$, while the right ones for the reference {\bf model B} with radius-dependent $\tau = \tau_0 (r / r_{\rm in})^\alpha$, where $\tau_0 = 10$ days, $\alpha = 1$, and constant $\sigma_{\rm l} = 0.2\,\rm dex$. The horizontal lines represent the effective temperatures, $T_{\rm eff}$, given by the standard thin disk at the corresponding effective radii, $r_{\rm eff}$, while the black filled squares and orange filled circles represent the average, over time and all zones per layer, of the whole fluctuations of $ \log T(t, \phi|r_{\rm eff}) $ and $T^4(t, \phi|r_{\rm eff})$, i.e., ${\langle \log T \rangle_{t,\phi}}$ and $[\langle T^4 \rangle_{t, \phi}]^{1/4}$, approaching $\log T_{\rm mid}$ and $T_{\rm eff}$, respectively. It is clear that {\bf model A} yields disk temperature fluctuations (relative to the mean value) independent to radius, while {\bf model B} yields much faster variations at smaller radius and slower variations at larger radius, which is more natural as the physical size of the individual disk zone (see Figure~\ref{fig:disk}) is proportional to the radius. Consistence between the simulated temperature fluctuation and the assumed DRW process is checked by fitting the structure function of DRW process with two parameters, i.e., the long-term fluctuation amplitude, $\sigma^*_{\rm l}$, and the characteristic timescale, $\tau^*$, to that of $\log T$ averaged over $\phi$ at given effective radius.
}\label{fig:qv_logT}
\end{figure*}

\section{Results}\label{sect:results}

In the following shown are the results, consisting of the simulated temperature fluctuations (Figure~\ref{fig:qv_logT}), the luminosity fluctuations (Figures~\ref{fig:qv_SED_Lbol_Llambda}-\ref{fig:qv_sed_comp_f_rbr_B}), the simulated light curves and structure functions (Figures~\ref{fig:qv_drw_fit}-\ref{fig:qv_sed_comp_sf}), the blue-when-brighter trend (Figures~\ref{fig:qv_BWB_Ruan14}-\ref{fig:qv_BWB_Ruan14_wrange}), and the color variation versus timescale (Figure~\ref{fig:qv_BWB_tau}), implied by our two reference models with typical BH mass $M_\bullet = 5 \times 10^8\,M_\odot$, accretion rate $\dot M = 1\,M_\odot\,{\rm yr}^{-1}$ (or Eddington ratio $\lambda_{\rm Edd} = 0.075$), and $f_{\rm rbr} = 1.3$.
To be more specific, the timescale dependent color variability of quasars is considered within two distinguishable inhomogeneous disk models: one (thereafter, {\bf model A}) analogous to that of \citet{DexterAgol2011} is parameterized with radius-independent $\tau = 200$ days and $\sigma_{\rm l} = 0.2$ dex, the other revised one (thereafter, {\bf model B}) is of radius-dependent $\tau = \tau_0 (r/r_{\rm in})^\alpha$, where $\tau_0 = 10$ days, $\alpha = 1$, and constant $\sigma_{\rm l} = 0.2$ dex. 
Adjusting these parametric values would not cast down our conclusion, although they are selected to be, as much as possible, consistent with those suggested by \citet[][]{DexterAgol2011} when comparing with observational data. 
We raise {\bf model B} as the physical size of the individual disk zone is not constant but proportion to the radius, and simply set the characteristic timescale of the variation in each zone correlates linearly with its physical size (see Section~\ref{sect:caveats} for more discussion).
Thereafter, we will focus on comparing these two models as reference in this section since they exhibit distinguishable difference on timescale dependent UV/optical color variabilities and the reader is referred to Section~\ref{sect:parameters} for more discussions on these parameters.

Firstly, Figure~\ref{fig:qv_logT} illustrates several fluctuations of $\log T$, simulated with Equation~(\ref{eq:DRW}) in steps of $\Delta t = 1$ day up to $t_{\rm max} = 10^5$ days, at inner and outer radii for the two reference models to demonstrate the simulation conditions considered in Section~\ref{sect:model}. The temperature fluctuations of the zones are considered independently and all the following properties are examined from the fluctuations after an initial ``burn-in'' time of $t_{\rm skip} = 500$ days (larger than the selected characteristic $\tau$) to allow the disk to become inhomogeneous \citep[cf.][]{Ruan2014}.

\subsection{Fluctuations of luminosities}\label{sect:fluctuation}

\begin{figure*}[!t]
\centering
\includegraphics[width=\columnwidth]{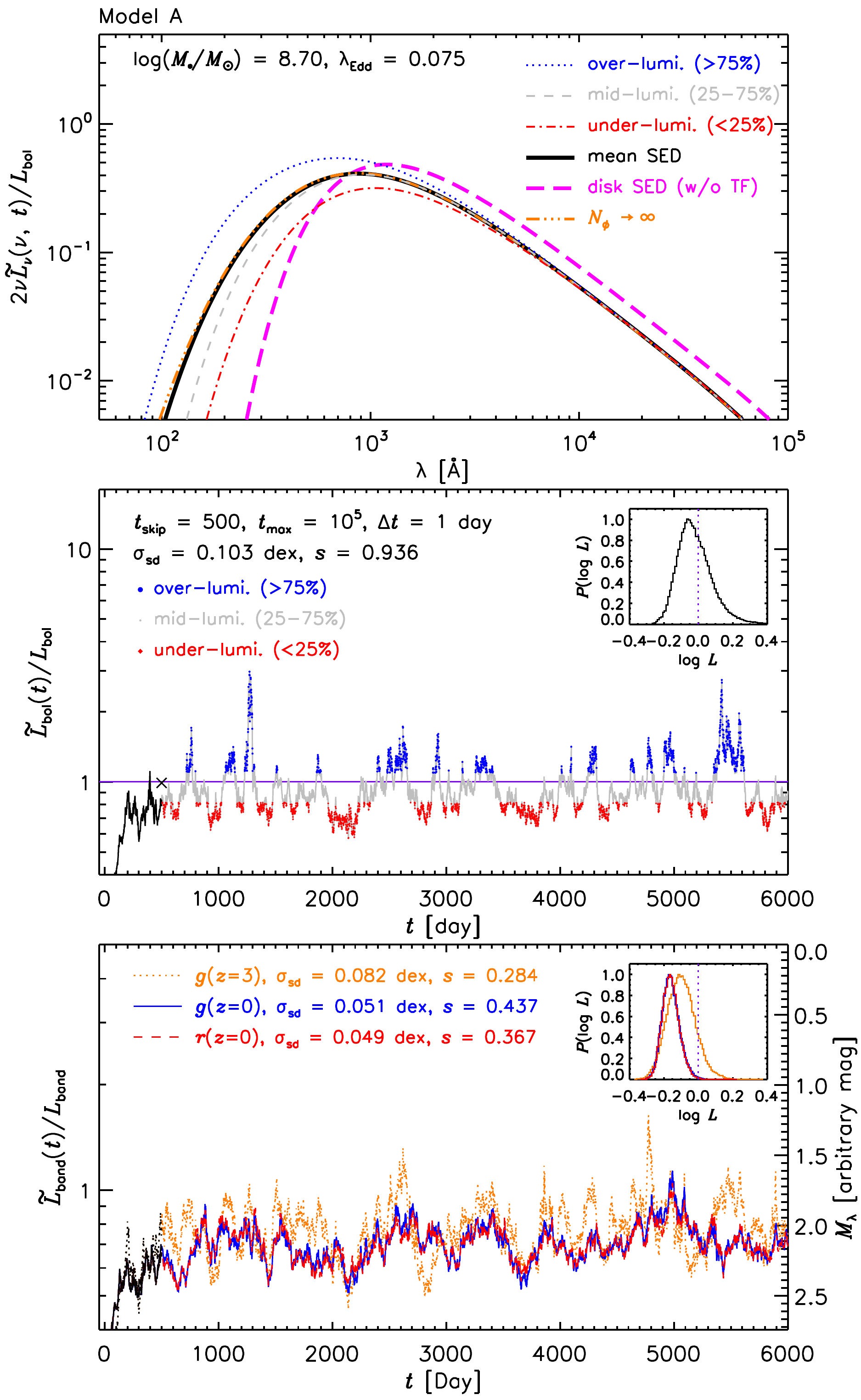}
\includegraphics[width=\columnwidth]{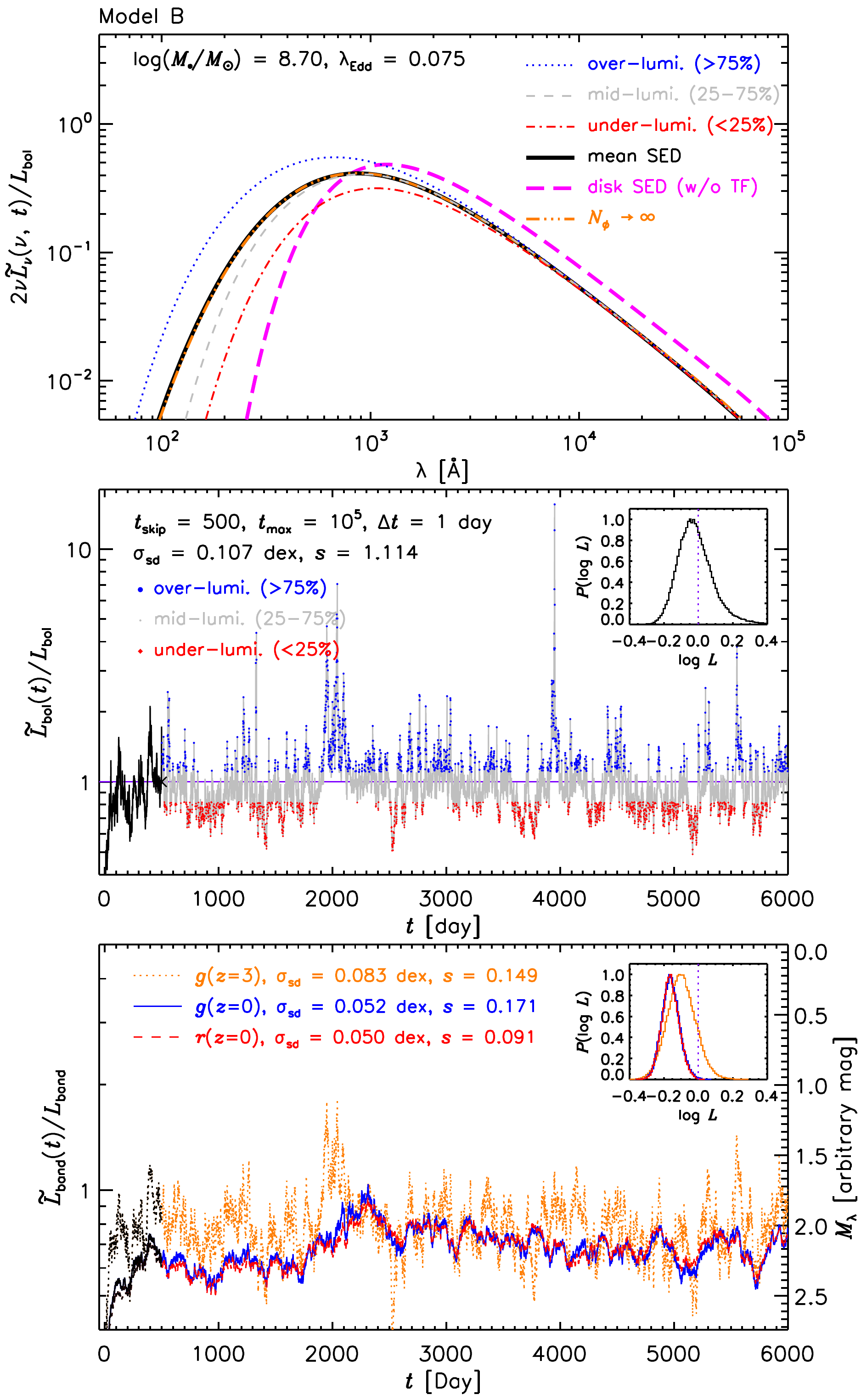}
\caption{An {\it ideal} example of the simulated SEDs, $ \nu \tilde L_\nu(\nu, t)$, the fluctuation of the bolometric luminosity, $\tilde L_{\rm bol}(t)$, and the fluctuations of the UV/optical-band luminosities, $\tilde L_{\rm band}(t)$, from upper to bottom panel, respectively, for {\bf model A} (left panels) and {\bf model B} (right panels) with BH mass of $5 \times 10^{8}\,M_\odot$ and Eddington ratio of $\lambda_{\rm Edd} = 0.075$.  The luminosities are normalized by the corresponding ones given by the thin disk without temperature fluctuations. 
In the top panels, the fluctuated SEDs averaged within three luminosity ranges, corresponding to the over-, mid-, and under-luminous states (blue dotted ($> 75\%$), light-gray dashed ($25-75\%$), and red dot-dashed ($<25\%$) lines, respectively) as indicated by the same color dots in the middle panels, are illustrated to demonstrate that the mean fluctuated SED becomes bluer with increasing luminosity. The arithmetic mean SED (thick black solid line) over all luminosities is compared to the one predicted by the standard thin disk without temperature fluctuations (thick magenta long-dashed line). The orange triple-dot-dashed line indicates the SED for the case of $N_\phi \rightarrow \infty$ directly calculated from the Equation~(\ref{eq:Lnu_Nphi_infty}) with $\sigma_{\rm l} = 0.2$ dex and almost overlaps the mean one. 
In the middle panels, the black cross symbol represents the mean bolometric luminosity over time, compared to the bolometric luminosity given by the standard thin disk (horizontal blue line). 
In the bottom panels, three band luminosities of SDSS $r(z=0)$, $g(z=0)$, and $g(z=3)$, with effective rest-frame wavelengths of $\sim 6122\,\angstrom$, $\sim 4640\,\angstrom$, and $\sim 1160\,\angstrom$, are illustrated as red dashed, blue solid, and orange dotted lines, respectively. The normalization factors, $L_{\rm band}$, are estimated in the corresponding bands from the standard thin disk. The normalized luminosities, $\tilde L_{\rm band}/L_{\rm band}$, are systematically lower than 1 at these considered bands because the temperature fluctuation results in more emission at shorter wavelengths, and, consequently, $\tilde L_{\rm band}/L_{\rm band}$ would be larger than 1 at short enough wavelengths (cf. the wavelength range of $< 1000\,\angstrom$ in the top panels of this figure).
The right scale gives the magnitude scale: $M_\lambda = -2.5 \log L_{\rm band} + \,\rm const.$ 
The distributions of these fluctuating $\log L$ are shown in the inserted figures within the lower panels, together with their standard deviations, $\sigma_{\rm sd}$, and the skewness, $s$.
}\label{fig:qv_SED_Lbol_Llambda}
\end{figure*}

\begin{figure}[!t]
\centering
\includegraphics[width=\columnwidth]{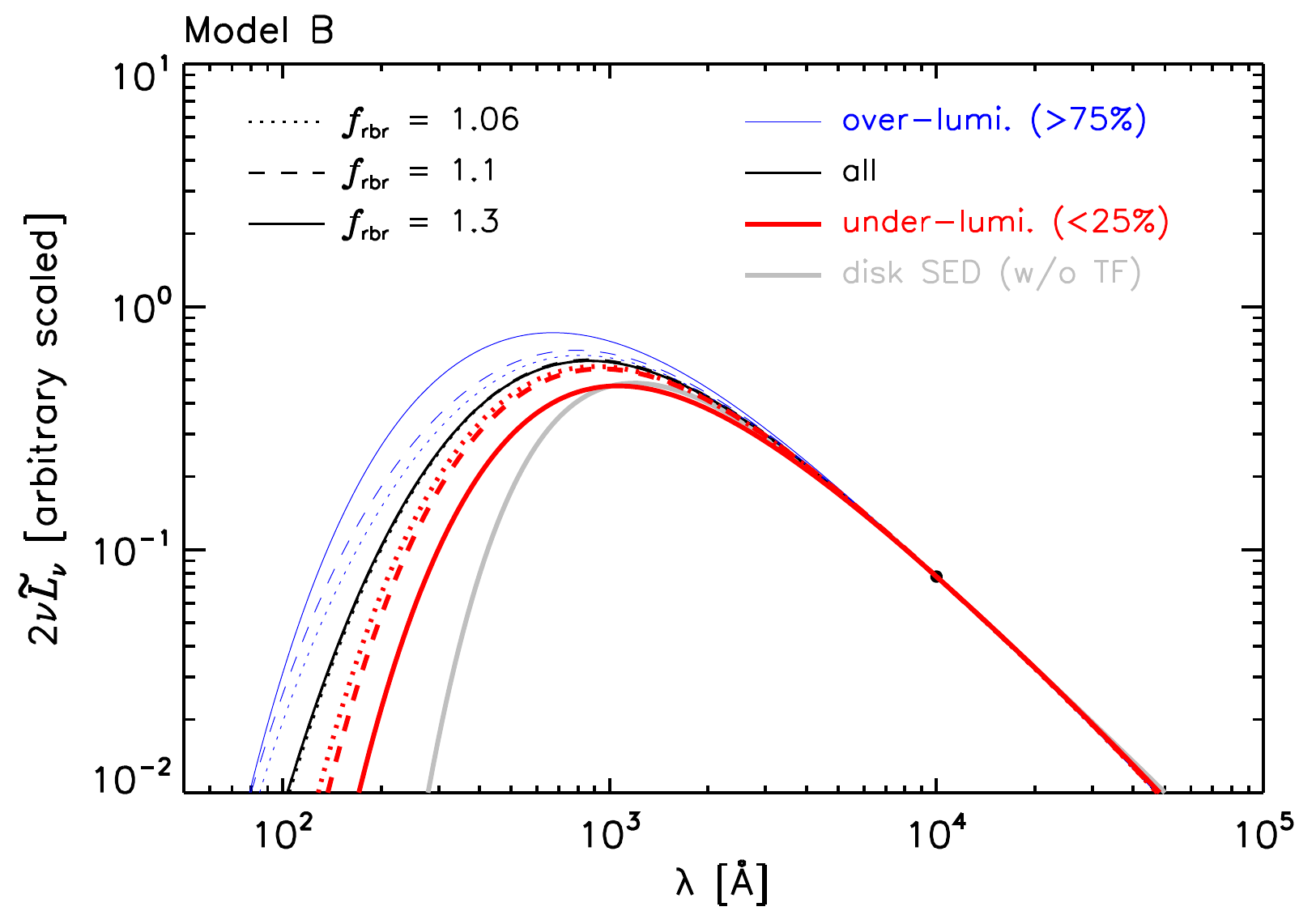}
\includegraphics[width=\columnwidth]{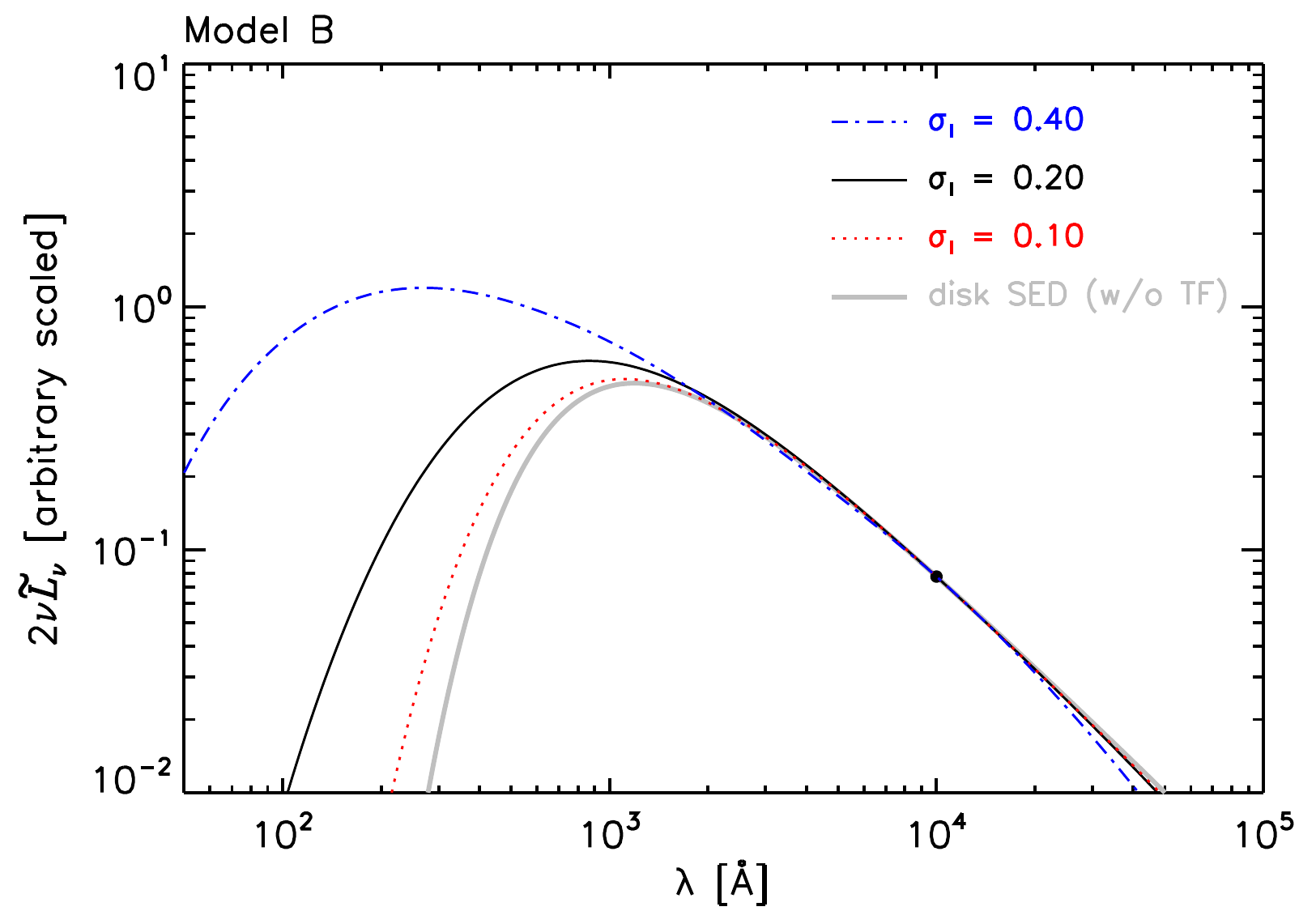}
\caption{Here shows the dependences of mean simulated SEDs implied by {\bf model B} on two model parameters, i.e., $f_{\rm rbr}$ (top panel) and $\sigma_{\rm l}$ (bottom panel). All the simulated SEDs are normalized to that of the thin disk model without thermal fluctuation at $10^4\,\angstrom$ (black filled circle), to highlight the difference in the slopes.
The mean simulated SEDs (black lines in the top panel) appear insensitive to $f_{\rm rbr}$, meanwhile larger $f_{\rm rbr}$ yields slightly stronger difference between the mean SEDs from over-luminous and under-luminous states.
}\label{fig:qv_sed_comp_f_rbr_B}
\end{figure}

From upper to bottom panels, Figure~\ref{fig:qv_SED_Lbol_Llambda} shows, respectively, an {\it ideal} example of the simulated SEDs, the light curves of the bolometric luminosity, and the light curves of the UV/optical luminosities at three different bands, for the two reference models. These light curves are {\it ideal} owing to their properties of long-enough (up to $\simeq 10^5$ days), no-error, and even-sampling ($\Delta t = 1$ day).

In the top panels of Figure~\ref{fig:qv_SED_Lbol_Llambda}, the time-averaged SEDs for both models, which are solely 
determined by $\sigma_{\rm l}$, appear identical to that of $N_\phi \rightarrow \infty$. The averaged SEDs from simulations
are blued at short wavelength compared to the standard thin disk model without temperature fluctuations, and get bluer with increasing 
$\sigma_{\rm l}$ (cf. the bottom panel of Figure~\ref{fig:qv_sed_comp_f_rbr_B} and also the Figure 4 of \citealt{DexterAgol2011}).
In fact the mean SEDs averaged over different luminosity ranges, i.e., the over-luminous ($>75\%$ of luminosity distribution), mid-luminous ($25-75\%$), and the under-luminous states ($<25\%$), are almost always blued at shorter wavelength, comparing with the standard thin disk model. Meanwhile the under-luminous states ($<25\%$)  could be slightly redder at $\sim$ 2000\AA, particularly with larger $f_{\rm rbr}$ (thus smaller $N_\phi$; cf. the top panel of Figure~\ref{fig:qv_sed_comp_f_rbr_B}). At longer wavelengths, all the simulated SEDs show little difference in the slope comparing with the thin disk model. 
Therefore, firstly determining the time-averaged SED would help constraining $\sigma_{\rm l}$ and then the scatter around the time-averaged SED would be able to provide further constraint on $f_{\rm rbr}$ and/or $N_\phi$.

In the middle panels of Figure~\ref{fig:qv_SED_Lbol_Llambda}, the requirement of energy conservation over time is justified by the agreement between the mean fluctuated bolometric luminosities averaged from $t_{\rm skip}$ up to $t_{\rm max}$ (black cross symbols) and those given by the standard thin disk (blue horizontal lines). The other color symbols represent the three typical epochs considered previously. 
Although there are rare extremely over-luminous epochs in the bolometric luminosity light curve, the radiation at these epochs exclusively concentrates at the very short wavelengths, leaving mild variations of luminosity on the other side (cf. the top panels of the same figure). 
The distributions of the fluctuating bolometric luminosity, $\log \tilde L_{\rm bol}$, are asymmetric and possesses a higher tail toward over-luminous states with skewness\footnote{The skewness of a distribution, $P(x)$, is defined as $s \equiv \int^{+\infty}_{-\infty} \left[ ({x - \mu})/{\sigma_{\rm sd}} \right]^3 P(x) dx$, where $\mu$ and $\sigma_{\rm sd}$ are the mean and standard deviation of the distribution, respectively. Skewness determines whether a distribution is symmetric about its maximum and positive skewness indicates the distribution is skewed to the right with a longer tail to the right of the distribution's maximum.} $s \sim 1 $ for both models. Here, this skewness is estimated using the distribution constructed from 100 {\it ideal} simulations to take the scatter of skewness among simulations into account, while those nominated values in Figure~\ref{fig:qv_SED_Lbol_Llambda} are solely based on that single {\it ideal} simulation.
The result that the distribution of $\log \tilde L_{\rm bol} \sim \log \sum^{N}_{i=1} T^4_i (r, \phi)$ is not gaussian is because the distributions of $\log T_i$ are assumed to be gaussian but $T_i(r)$ spans a wide range over the whole disk (see Section~\ref{sect:SF} for further discussion).

The bottom panels of Figure~\ref{fig:qv_SED_Lbol_Llambda} show the light curves of three UV/optical bands, $\tilde L_{\rm band}$, corresponding to the SDSS $r(z=0)$
\footnote{Throughout the paper, we treat SDSS $r(z)$ acting as a SDSS $r$-filter with wavelengths divided by $(1+z)$ or as a source at redshift $z$ observed with SDSS $r$-filter, et cetera.} 
(red dashed lines), $g(z=0)$ (blue solid lines), and $g(z=3)$ (orange dotted lines) bands with rest-frame effective wavelengths of $\sim 6122\,\angstrom$, $\sim 4640\,\angstrom$, and $\sim 1160\,\angstrom$, respectively. The larger fluctuations of band luminosities at shorter effective wavelengths are intuitively illustrated and quantified by the corresponding standard deviations $\sigma_{\rm sd}$ of the fluctuating luminosities in logarithm. This result is qualitatively consistent with the observed anti-correlation between variation amplitude and wavelength. 
Furthermore, the other observed correlations between the UV/optical variation amplitude and the physical properties of AGNs, such as bolometric luminosity, accretion rate, BH mass, etc., can also be qualitatively argued. 
Particularly, for sources with smaller SMBHs and/or higher Eddington ratios, the peak of the disk SED should move toward shorter wavelength, and the resulted variation amplitude at given wavelengths longer than the peak would decrease. This would yield clear mass and/or Eddington ratio dependence of the variation amplitude at given wavelengths.
However, quantitative comparison with observational data is out of the scope of this paper and will be exhaustively discussed in another upcoming paper (see Section~\ref{sect:caveats} for further discussion).

By the way, owing to that the band luminosity at a given effective wavelength, $\lambda_{\rm eff}$, is approximately contributed by the emission from a specific narrow ring at $r_{\lambda_{\rm eff}}$ of the disk with similar $T(r_{\lambda_{\rm eff}})$, the amount of the skewness of the distributions of $\log \tilde L_{\rm band}$ reduces to $\sim 0.3$ for both models at those bands illustrated in the bottom panels of Figure~\ref{fig:qv_SED_Lbol_Llambda}, when averaging over 100 {\it ideal} simulations. We note that in Figure~\ref{fig:qv_SED_Lbol_Llambda} the nominated skewness of light curves implied by {\bf model B} superficially tends to be smaller by coincidence (see Section~\ref{sect:SF} for further discussion).

Comparing the two reference models, when their temperature fluctuation amplitudes on long timescale are identical, their overall mean SEDs are almost identical, but there would be few extraordinarily luminous epoch in the {\bf model B} with radius-dependent $\tau$, due to the relatively large temperature fluctuation amplitudes on short timescales resulting from the inner regions.

\subsection{Comparison with DRW}\label{sect:SF}

\begin{figure}[!t]
\centering
\includegraphics[width=\columnwidth]{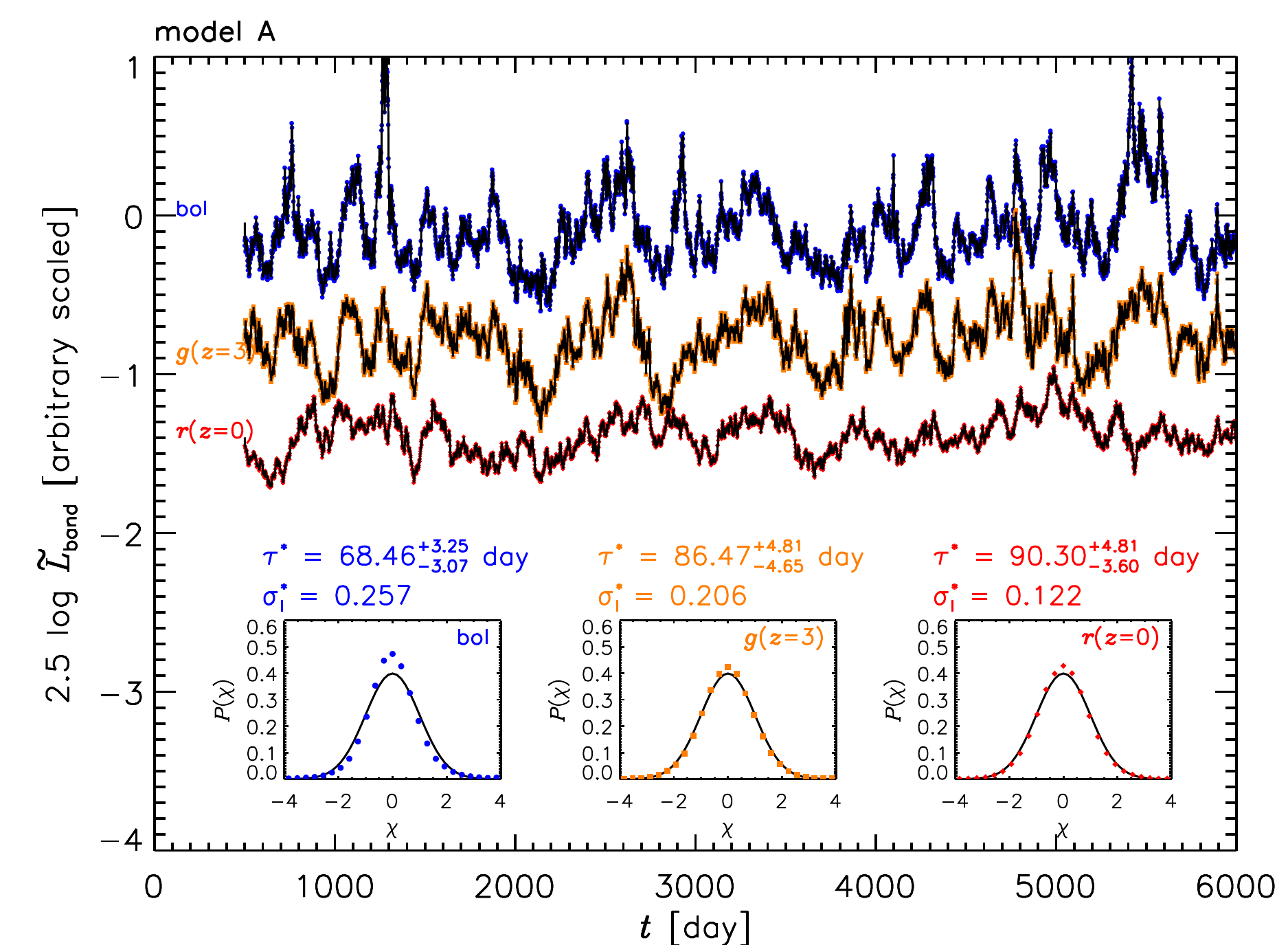}
\includegraphics[width=\columnwidth]{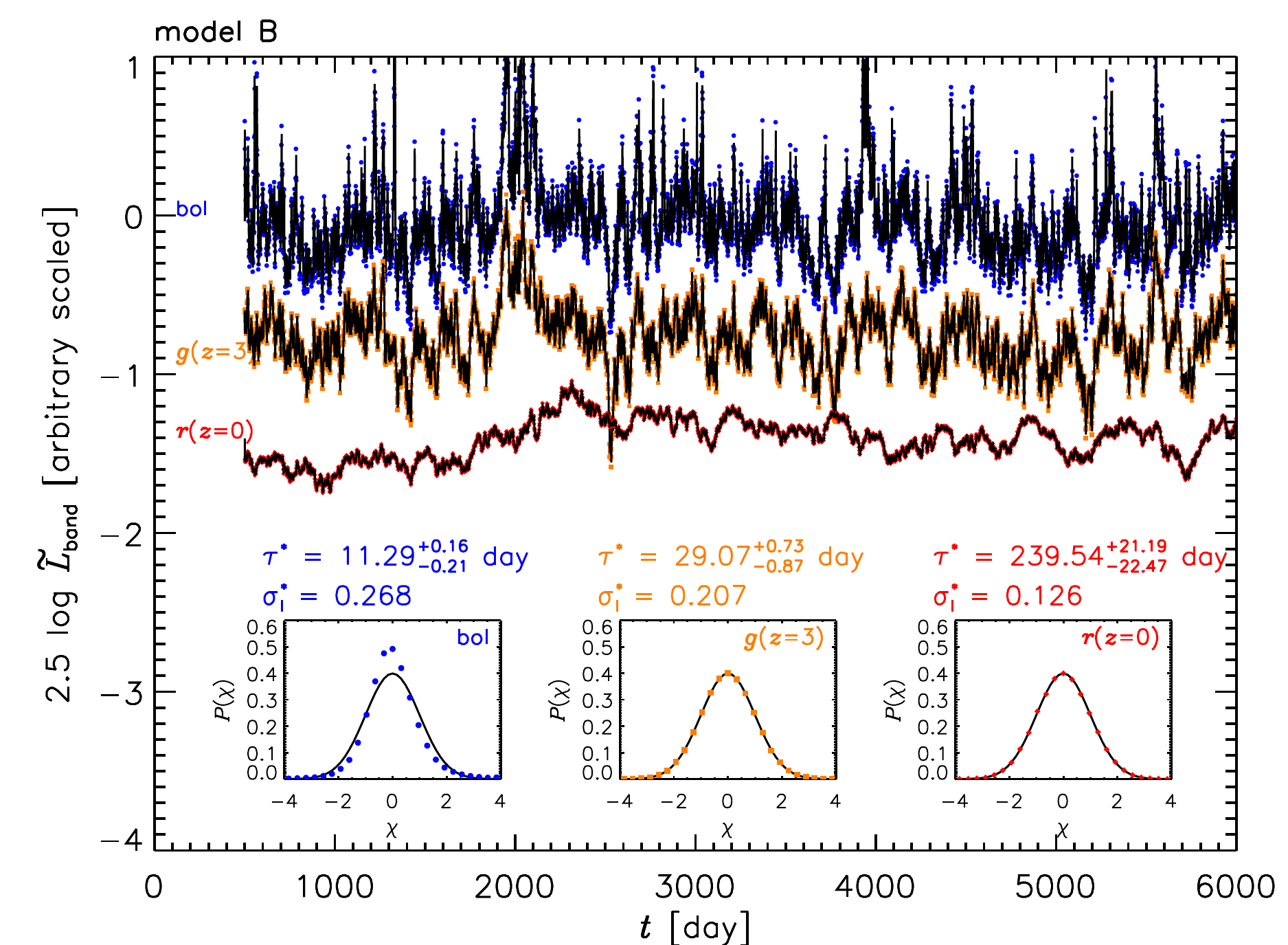}
\caption{The simulated bolometric, $g(z=3)$, and $r(z=0)$ band light curves (colored symbol lines from up to down; see the middle and bottom panels of Figure~\ref{fig:qv_SED_Lbol_Llambda}), implied by {\bf model A} (top panel) and {\bf model B} (bottom panel), are superimposed by the corresponding fitted ones (thin black lines), using the recipe of \citet{Kelly2009}. The retrieved $\tau^*$ (with 1 $\sigma$ error), $\sigma^*_{\rm l}$, and the distributions of the standardized residuals (color symbols) of these light curves are presented with insets from left to right in the same colors. The accuracy of the fit is assessed through comparison between the distributions of the residuals (color symbols in insets) and a standard normal distribution \citep[black curves in insets;][]{Kelly2009}.
}\label{fig:qv_drw_fit}
\end{figure}

\begin{figure}[!t]
\centering
\includegraphics[width=\columnwidth]{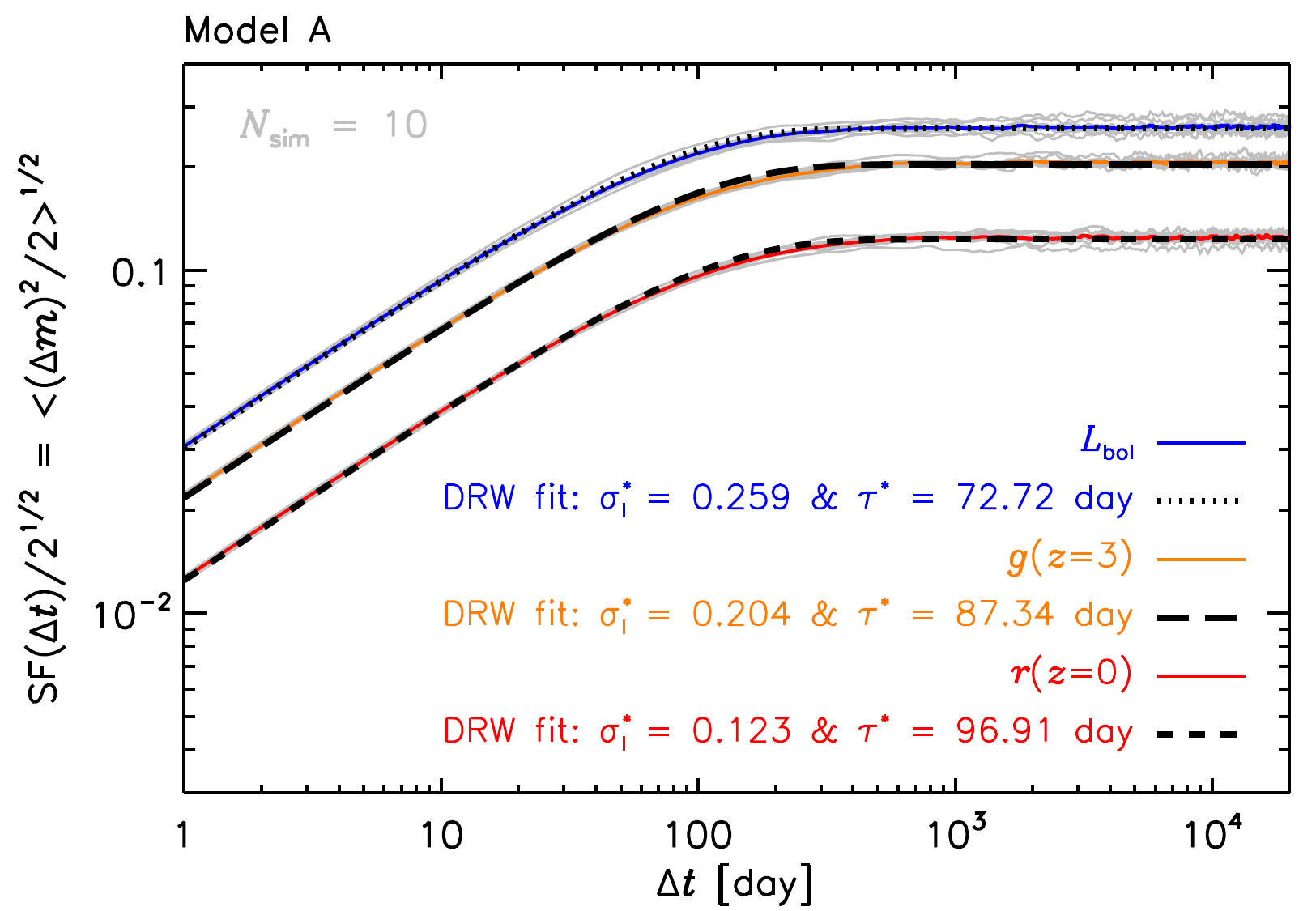}
\includegraphics[width=\columnwidth]{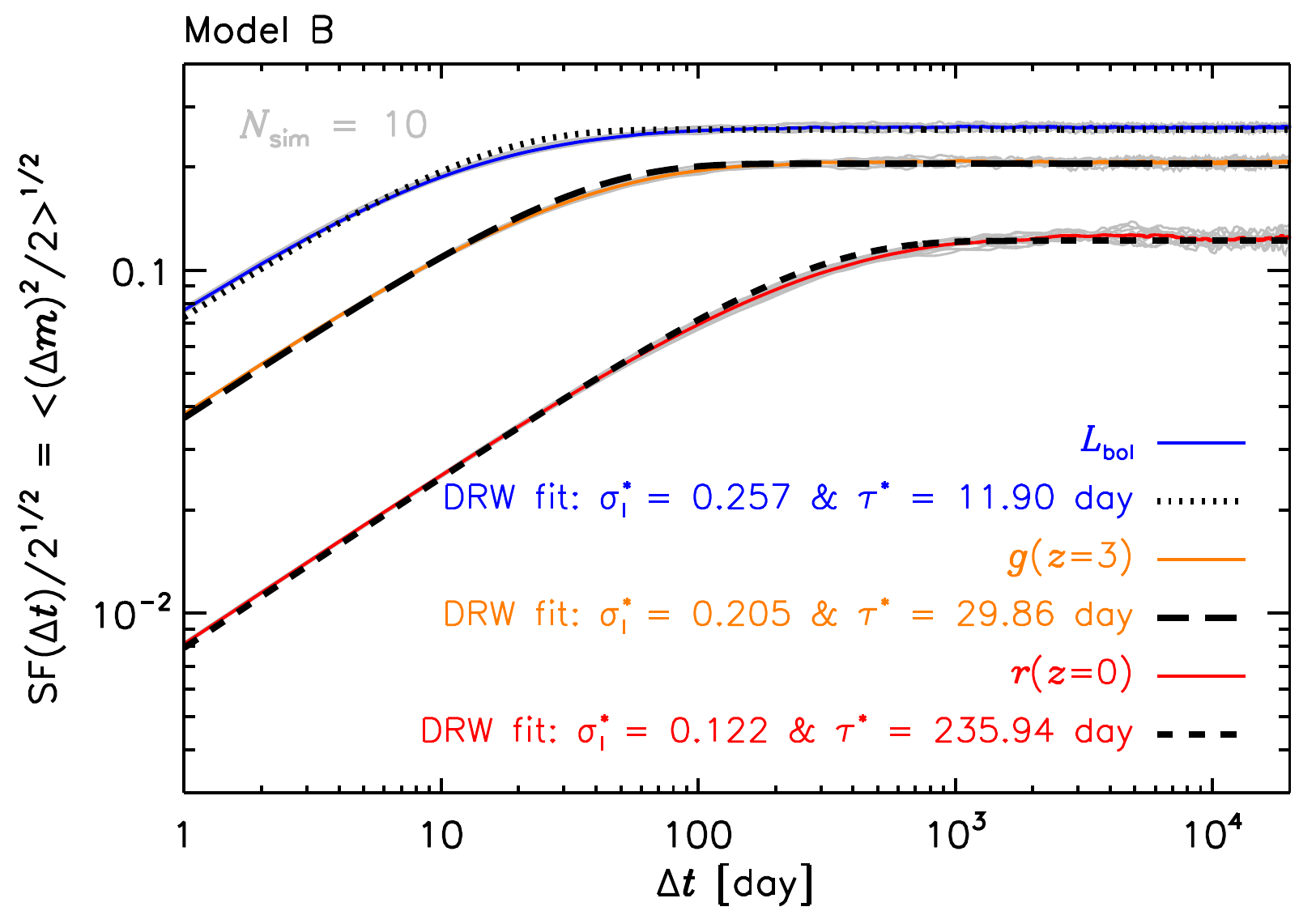}
\caption{The mean simulated structure functions of bolometric, $g(z=3)$, and $r(z=0)$ band light curves (colored solid lines; see the middle and bottom panels of Figure~{\ref{fig:qv_SED_Lbol_Llambda}} for the corresponding light curve whose structure function is one of the thin light-gray lines), implied by {\bf model A} (top panel) and {\bf model B} (bottom panel) through averaging over $N_{\rm sim} = 10$ times of {\it ideal} simulation (thin light-gray lines), are fitted to that of a single DRW process, with $\sigma^*_{\rm l}$ and $\tau^*$ nominated in the corresponding panel (colored broken lines).
}\label{fig:qv_sed_comp_sf}
\end{figure}

\begin{figure}[!t]
\centering
\includegraphics[width=\columnwidth]{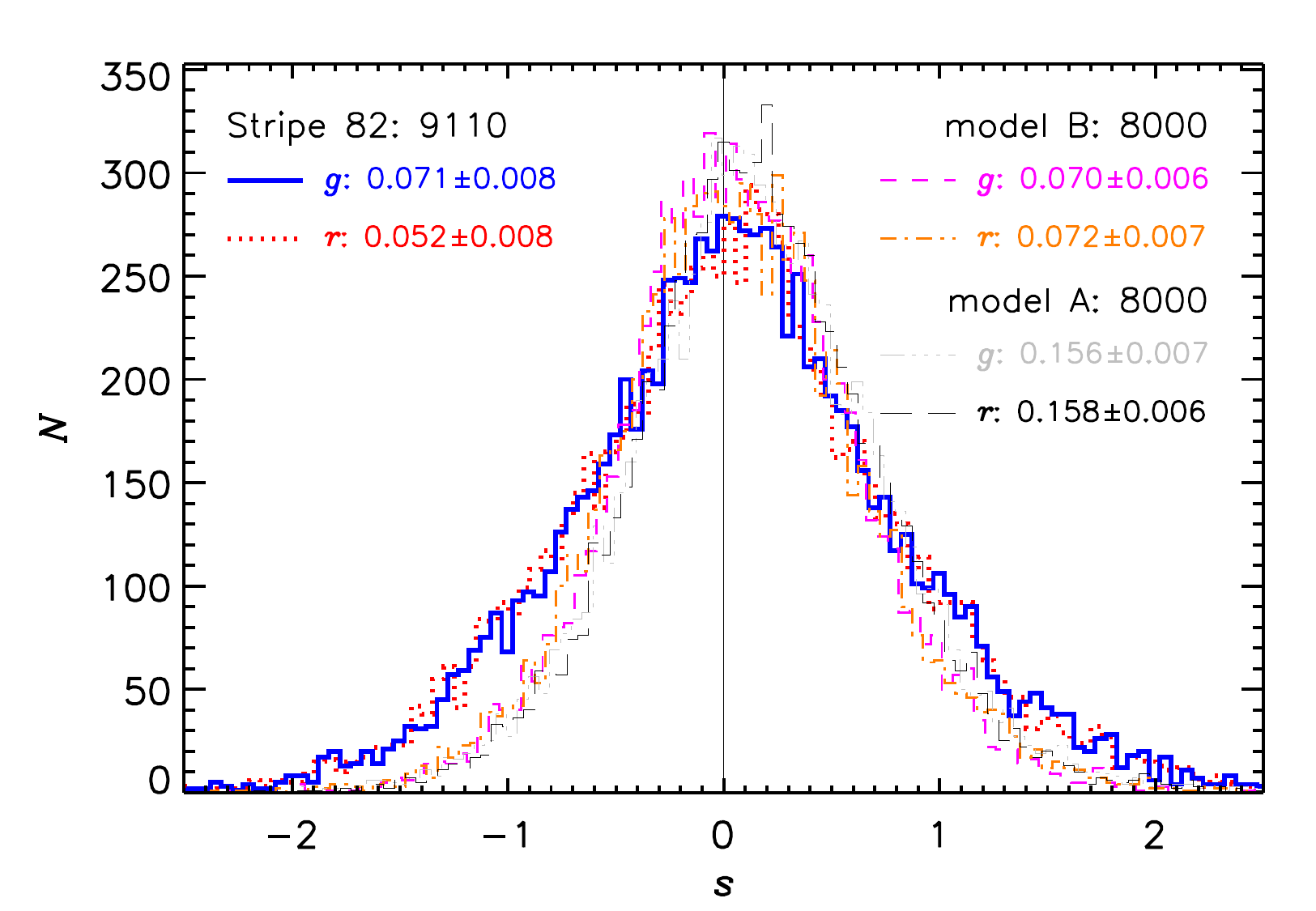}
\caption{The skewness distribution of 9110 Stripe 82 quasars possessing 10 or more good epochs with photometric errors in $g$- and $r$-bands $\leqslant 0.1$ mag, selected from the original sample of 9258 quasars presented by \citet{MacLeod2012}. 
Those for 8000 simulated ``{\it real}" light curves are over plotted for comparison (see Section~\ref{sect:simulatedsample} for their construction).
}\label{fig:qv_s82_skewness}
\end{figure}

Following \citet{DexterAgol2011}, we have assumed $\log T$ of spatially independent zones undergoing DRW process. The thermal emission from each single disk zone therefore also follows DRW process in logarithm form ($\log L \sim 4 \log T$).
Although the bolometric or monochromatic emission integrated throughout the entire disk is no longer strict DRW mathematically,
we show below they can be approximately modeled with DRW, by analyzing the simulated light curves (Figure~\ref{fig:qv_drw_fit}), and the corresponding structure functions (Figure~\ref{fig:qv_sed_comp_sf}).

Considering the recipe of \citet{Kelly2009}, we fit the simulated light curves implied by both models to retrieve the corresponding characteristic timescales, $\tau^*$, and the long-term fluctuation amplitudes, $\sigma^*_{\rm l}$ (Figure~\ref{fig:qv_drw_fit}). For band light curves, good agreement with DRW model is reached according to the consistence between the distribution of the standardized residuals and a standard normal distribution, while for the bolometric light curves minor departures are found for both models.

Figure~\ref{fig:qv_sed_comp_sf} plots the structure functions of the simulated light curves of bolometric and band (at different effective wavelength) luminosities, together with the best-fitted DRW models,  i.e., $\mbox{SF}(\Delta t) = \sqrt{2} \sigma^*_{\rm l} \sqrt{1-\exp(-|\Delta t|/\tau^*)}$, with long-term fluctuation amplitude, $\sigma^*_{\rm l}$, and the characteristic timescale, $\tau^*$.
The structure functions are generally consistent with that of a single DRW process in shape, with only a minor departure of $\lesssim 2-4\%$ for {\bf model A} and $\lesssim 4-6\%$ for {\bf model B}. The slightly larger departure in shape implied in {\bf model B} is attributed to the addition of fluctuations of different characteristic timescales.

We note for {\bf model A}, the best DRW fits yield $\tau^*$ significantly smaller than the input $\tau$ for $\log T$ (200 days), and emission at shorter wavelength tends to have lower $\tau^*$. For the structure function of bolometric light curve, $\tau^*$ is even smaller owing to fact that the major contributions are dominated by the most inner regions.
For {\bf model B}, the difference in $\tau^*$ at different wavelength is more significant, due to different input $\tau$ at different disk radii, and the retrieved $\tau^*$ are also smaller than that of the temperature fluctuation around those radii from which the radiations dominate. 
For example, the characteristic timescale of the temperature fluctuation, $\sim 260$ days at $\sim 150\,r_{\rm g}$ which dominantly contribute to $r(z=0)$ band emission, is larger than the retrieved $\tau^* \simeq 230$ days. 
The difference in the characteristic timescale between the temperature fluctuations and the simulated light curves can be attributed
to the complicated transformation from $\log T$ in individual disk zones to integrated luminosities.
This suggests that the timescale of temperature fluctuation should be re-calibrated, instead of being simply assumed to be the same as that inferred from the observed light curves of quasars. 

Weak deviation from DRW can also be seen through the skewness in the luminosity distributions of the simulated light curves (see Figure~\ref{fig:qv_SED_Lbol_Llambda} and Section~\ref{sect:fluctuation}), as the luminosity distributions of perfect DRW light curves are symmetric\footnote{Note the asymmetry in the luminosity distribution discussed here is different from the variation asymmetry in AGN light curves studied by \cite{ChenWang2015}, where the asymmetry is to describe whether a light curve favors rapid rise and gradual decay, or vice versa.}. While the simulated {\it ideal} bolometric luminosity distributions show obvious skewness ($s$ $\sim$ 1), the skewness in the $\log L_{\rm band}$ distribution is much weaker ($s$ $\sim 0.3$ for both models). 

We calculate the skewness for real $g$- and $r$-band light curves of quasars in Stripe 82, possessing 10 or more epochs with photometric errors in both bands $\leqslant 0.1$ mag (Figure~\ref{fig:qv_s82_skewness}). The observed values span a wide range for individual sources, with mean values of 0.071 and 0.052 for $g$- and $r$-bands respectively, showing weak but statistically significant positive skewness. 
Note the observed skewness is affected by the photometric errors and un-even sampling of SDSS Stripe 82 data. Applying the same sampling and photometric uncertainties to the simulated light curves (see Section~\ref{sect:simulatedsample} for details), we build a large sample of ``{\it real}" $g$- and $r$-band light curves for both models and obtain the corresponding skewness (Figure~\ref{fig:qv_s82_skewness}).  
The simulated ``{\it real}" light curves produce weaker skewness in $\log L$ distribution comparing with the {\it ideal} light curves,
and the decrease of skewness is primarily due to the introduction of un-even sampling.
The mean value from {\bf model B} appears to be well consistent with observations (although with slightly narrow distribution), while {\bf model A} yields a slightly larger value.

\subsection{Bluer-when-brighter trend}\label{sect:BWB_trend}

\begin{figure*}[!t]
\centering
\includegraphics[width=\columnwidth]{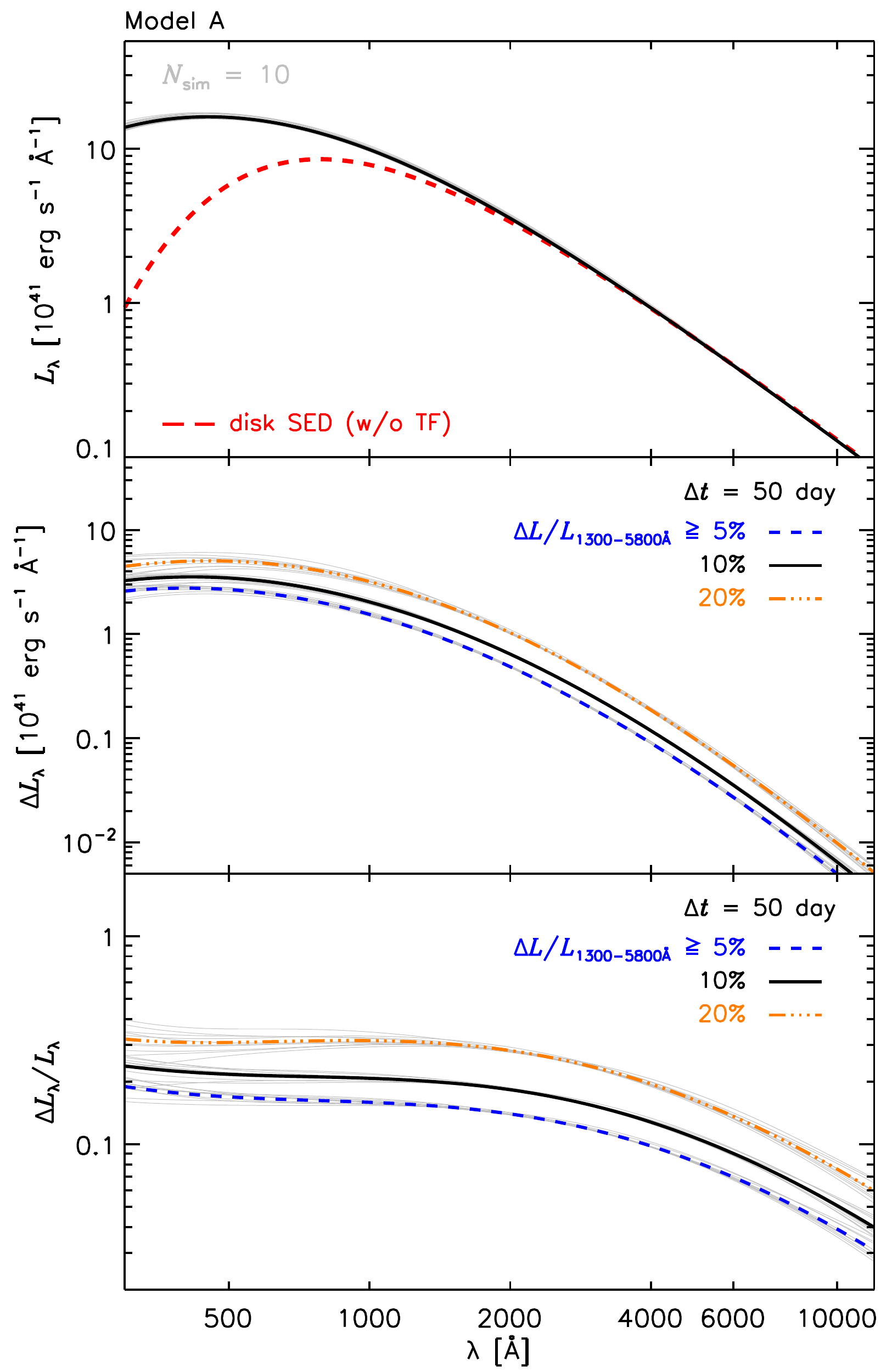}
\includegraphics[width=\columnwidth]{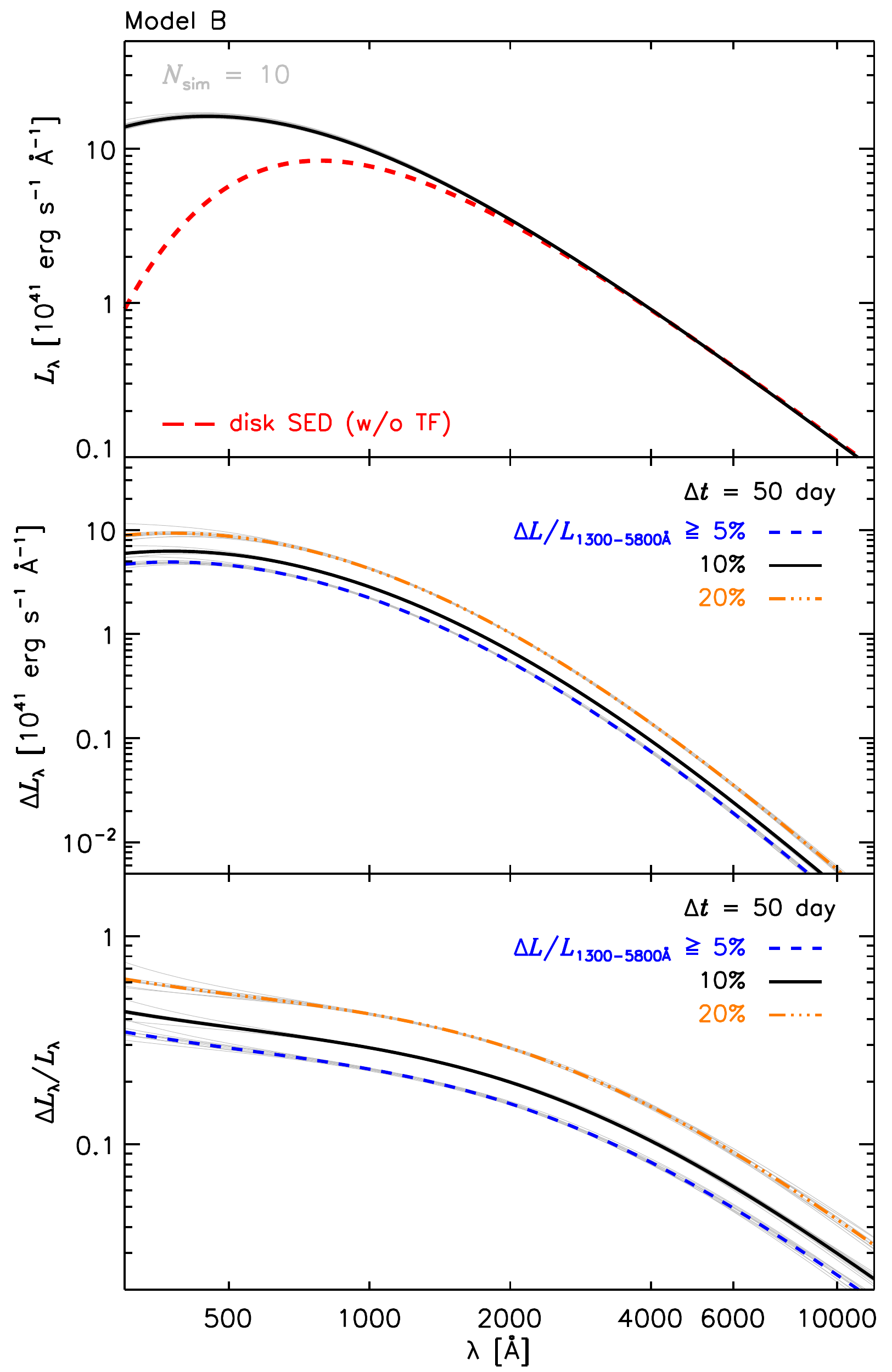}
\caption{From top to bottom panels, shown are the arithmetic mean composite spectra, the arithmetic mean composite difference spectra, and the composite relative variability spectra, for {\bf model A} (left panels) and {\bf model B} (right panels), respectively. 
In the top panels, the composite spectra are compared to the one predicted by the standard thin disk without temperature fluctuation (red 
dashed line), which has been normalized to the composite one at 5000 $\angstrom$ to highlight the difference at shorter wavelengths.
The middle and bottom panels show the composite difference spectra and the composite relative variability spectra, constructed from pair spectra selected from the simulated ones (see Figure~\ref{fig:qv_SED_Lbol_Llambda}) by an interval of $\Delta t = 50$ days and the total luminosity within 1300-5800 $\angstrom$ changed by more than $\Delta L/L_{1300-5800\angstrom} = 5$\% (blue dashed line), 10\% (black solid line), and 20\% (orange triple-dot-dashed line), respectively.
The thick lines are results of averaging over $N_{\rm sim} = 10$ times of {\it ideal} simulations, while results of each {\it ideal} simulation are also presented as the bundles of light-gray lines around the corresponding averaged ones.
For two given wavelengths, say $\lambda_1 < \lambda_2$, the bluer-when-brighter trend is indicated by the ratio of $\Delta L_\lambda/L_\lambda (\lambda_1)$ to $\Delta L_\lambda/L_\lambda (\lambda_2)$ larger than 1.
}\label{fig:qv_BWB_Ruan14}
\end{figure*}

\begin{figure*}[!ht]
\centering
\includegraphics[width=\columnwidth]{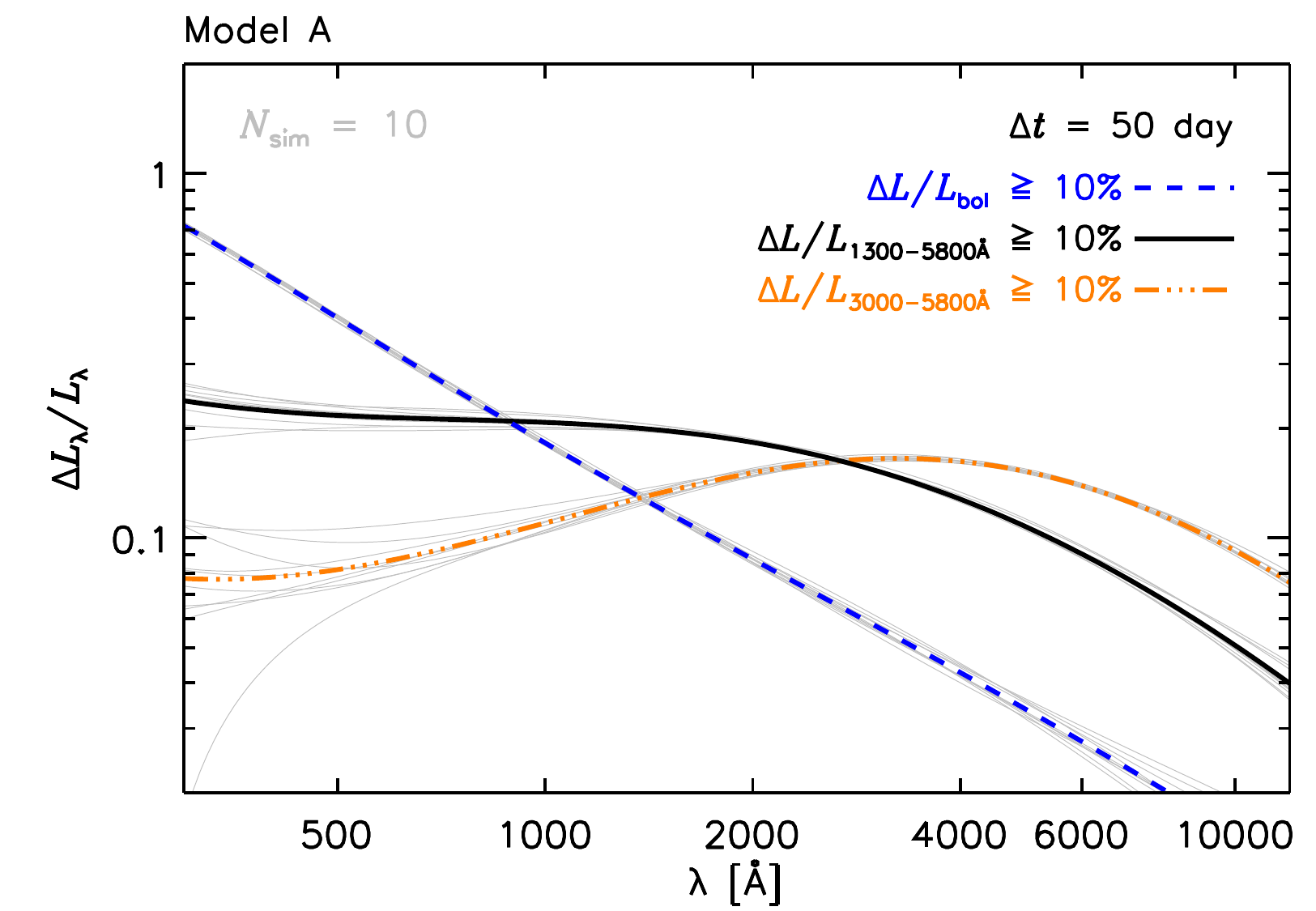}
\includegraphics[width=\columnwidth]{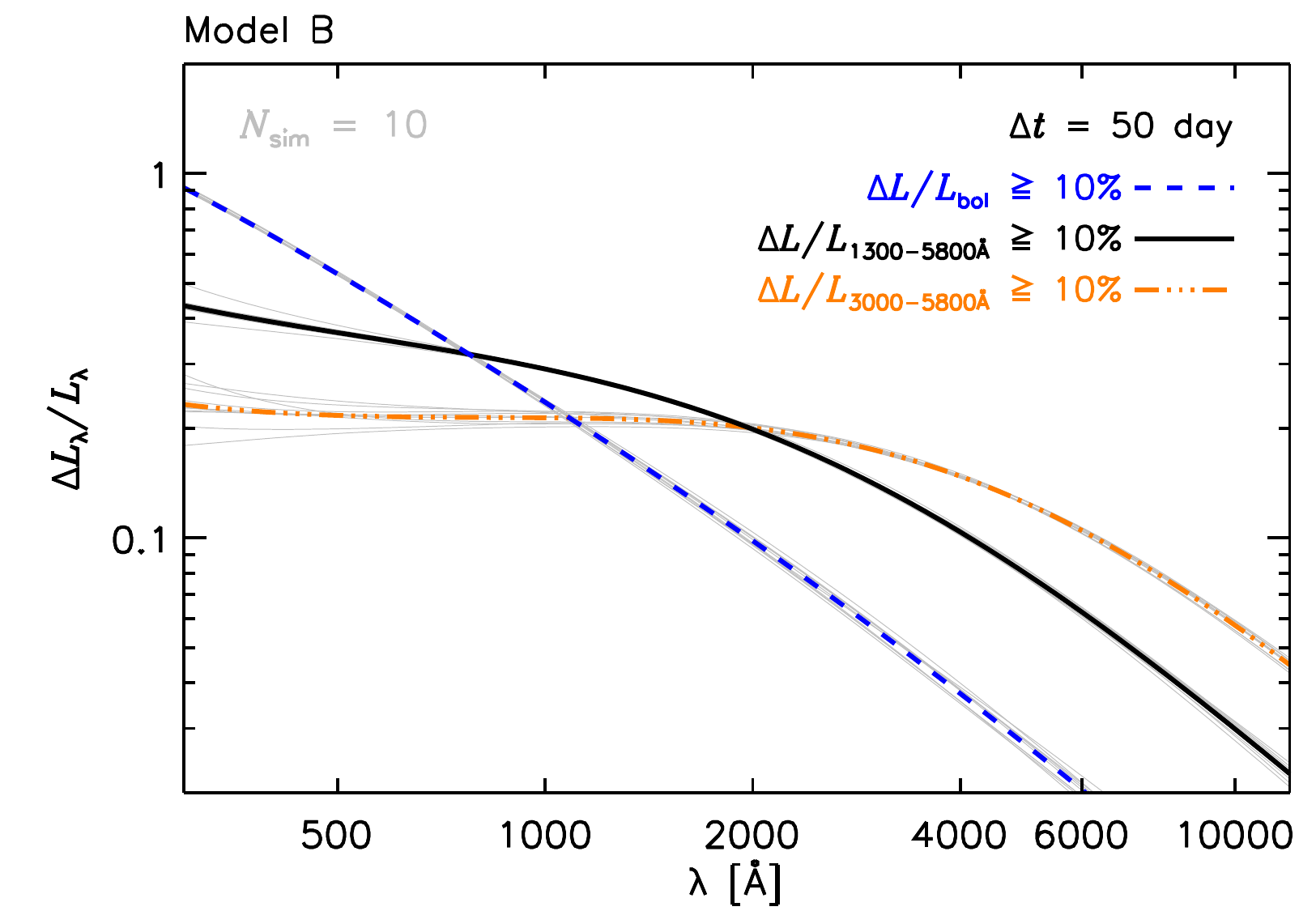}
\caption{Same as the bottom panels of Figure~\ref{fig:qv_BWB_Ruan14} for {\bf model A} (left panel) and {\bf model B} (right panel) with $\Delta t = 50$ days, but for different luminosity criteria larger than 10\% with broader (panchromatism; blue dashed lines) and narrower (3000-5800 $\angstrom$; orange triple-dot-dashed lines) wavelength ranges to illustrate that whether or not there is a broken or bump in the composite relative variability spectrum depends on the adopted wavelength range of luminosity criterion.
}\label{fig:qv_BWB_Ruan14_wrange}
\end{figure*}

To demonstrate the well-known bluer-when-brighter trend observed in quasar variability using the {\it ideal} simulations, Figure~\ref{fig:qv_BWB_Ruan14} shows, from top to bottom panels, the arithmetic mean composite spectra, the arithmetic mean composite difference spectra, and the arithmetic relative variability spectra, for the two reference models, respectively. We select spectral pairs by an interval of 50 days and the total luminosity in the wavelength range of 1300-5800 $\angstrom$ changed by more than 5\%, 10\%, and 20\% from our simulated spectra \citep[cf.][for the selections of time interval and wavelength range]{Ruan2014}. The relative change in luminosity between $t_1$ and $t_2$ is defined as $\Delta L/L_{1300-5800\angstrom} (\Delta t = |t_1 - t_2|) \equiv 2|L_1 - L_2|/(L_1+L_2)$, where $L_1$ and $L_2$ are the integrated luminosities in the wavelength range of 1300-5800 $\angstrom$ at $t_1$ and $t_2$, respectively. If $L_2 > L_1$, the $t_2$ epoch is defined as the brighter one and the difference spectrum of the two epochs is $\Delta L_\lambda \equiv L_\lambda(t_2) - L_\lambda(t_1)$. The illustrated results are further arithmetically averaged over $N_{\rm sim}$ times of {\it ideal} simulations, while results of each {\it ideal} simulation are also presented as the bundles of light-gray lines around the corresponding averaged ones.

In the top panels of Figure~\ref{fig:qv_BWB_Ruan14}, the composite spectra (black lines) are bluer than the spectrum predicted by the standard thin disk (red dashed line), owing to the introduction of temperature fluctuations. In the reference {\bf model B} with radius-dependent $\tau$, the composite spectrum is almost identical to that implied by the reference {\bf model A} with radius-independent $\tau$, because the average has been done over all epochs and the same $\sigma_{\rm l}$ has been assumed for both models.

For each model, requiring larger $\Delta L/L_{1300-5800\angstrom}$ yields difference spectra with larger amplitudes, but the shapes of the difference spectra as well as the relative variability spectra, in the middle and bottom panels of Figure~\ref{fig:qv_BWB_Ruan14}, respectively, are all analogous. The bluer-when-brighter trend is represented by the result that the relative variability continuum monotonically increases with decreasing wavelength. At short time interval, e.g., the shown $\Delta t = 50$ days, {\bf model B} implies bluer relative variability spectra than {\bf model A}, because in {\bf model B} the larger temperature fluctuations happen at inner regions than outer ones for given short time interval (e.g., 50 days).
For sufficiently long time interval or when the temperature fluctuations at all radii saturate, the two models would present the same difference and relative variability spectra as long as $\sigma_{\rm l}$ is identical (see further discussion at the end of Section~\ref{sect:stripe82}).

The relative variability spectra exhibit a shape of broken power-laws with characteristic broken wavelengths of $\sim$ 2000 \AA.
This in fact is a selection effect since these spectra are constructed requiring the luminosity in the wavelength range of 1300-5800~$\angstrom$ changed by more than some percentages. Figure~\ref{fig:qv_BWB_Ruan14_wrange} better demonstrates such bias, in which we see that the characteristic broken wavelength
in the relative variability spectrum changes after we adopt different wavelength ranges to calculate luminosity variations.
The prominence of the broken, or even a bump, becomes more significant with narrower wavelength range (i.e., 3000-5800 \AA),
and disappears if we adopt bolometric luminosity to determine  the brighter epoch. 
Therefore, the shapes of both the difference spectrum and the relative variability spectrum depend on the wavelength ranges adopted to determine the relative change in luminosity. One has to be cautious of such bias while comparing observed difference spectra or relative variability spectra with theoretical calculations. Note that a redder-when-brighter trend could even be observed at the bluer part of the bump if the luminosity change is determined in a narrow enough wavelength range, such as 3000-5800 \AA\ (cf. Figure~\ref{fig:qv_BWB_Ruan14_wrange}).

For both models, the shapes of the relative variability spectrum are broadly consistent with that presented in the explored wavelength range by \citet[][cf. their Figure 6]{Ruan2014}, in which a narrow wavelength range 1300-5800 \AA\ was explored using a sample of 602 variable quasars.

\subsection{Timescale-dependent color variation}\label{sect:color_variation}

\begin{figure*}[!t]
\centering
\includegraphics[width=\columnwidth]{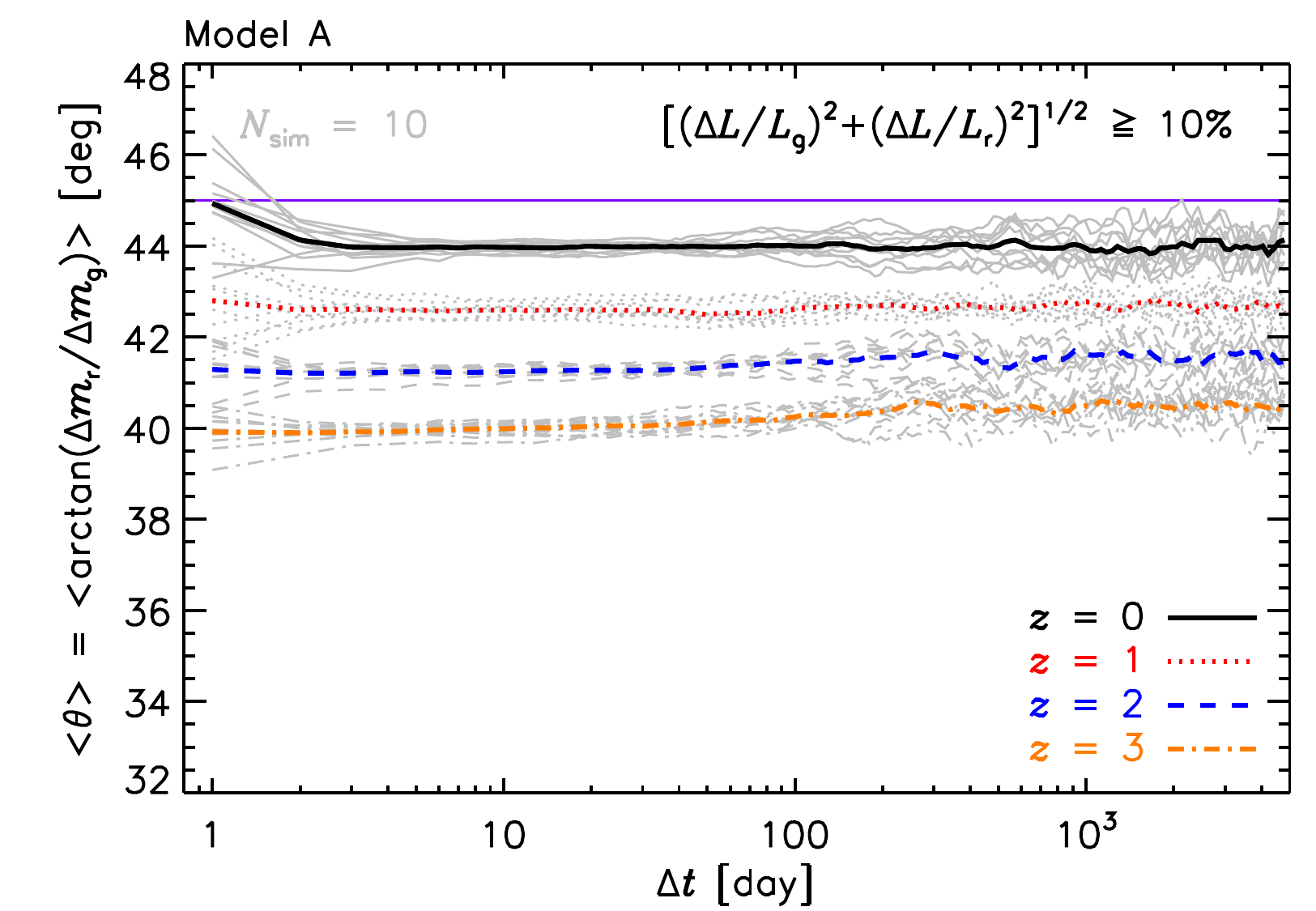}
\includegraphics[width=\columnwidth]{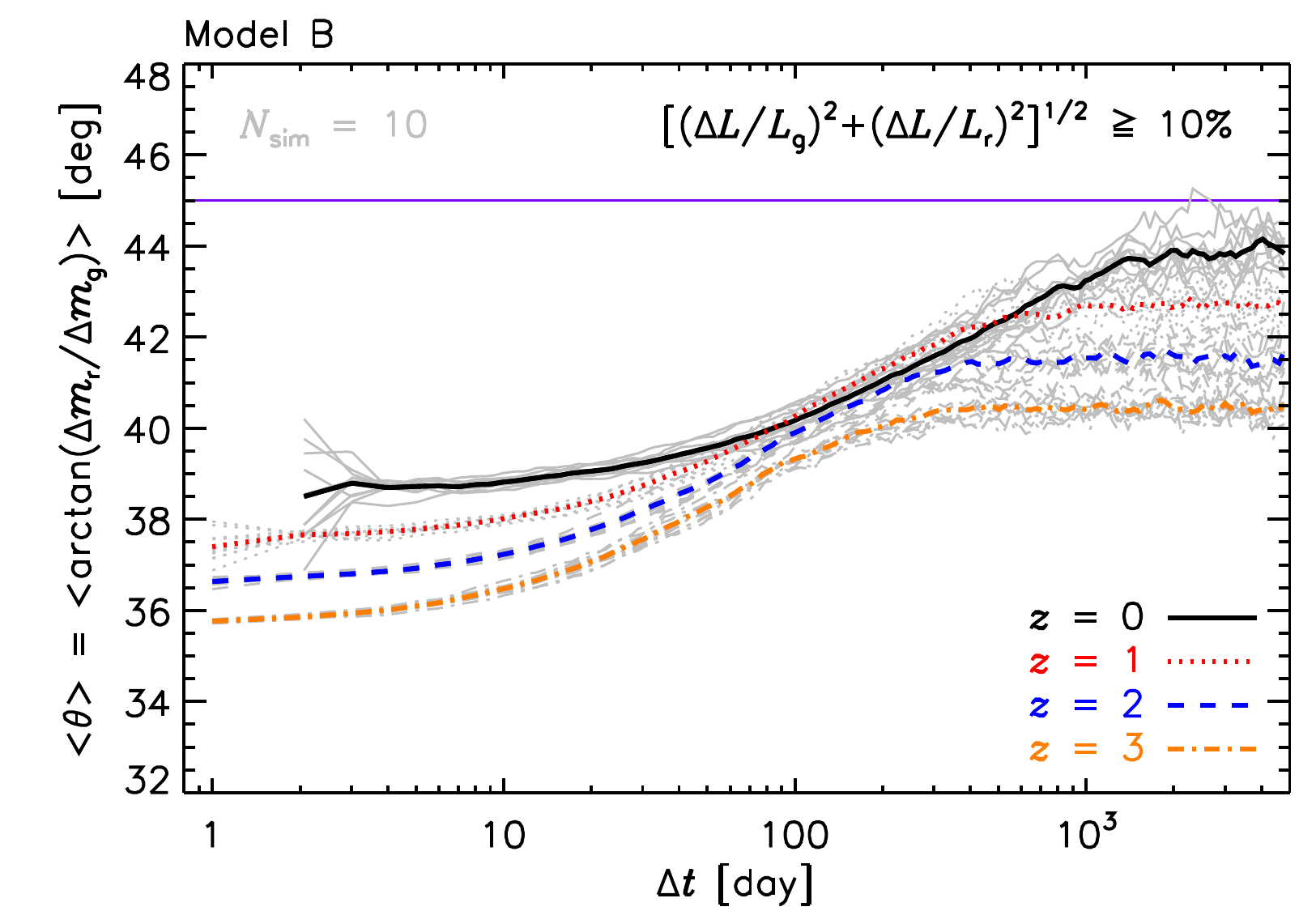}
\includegraphics[width=\columnwidth]{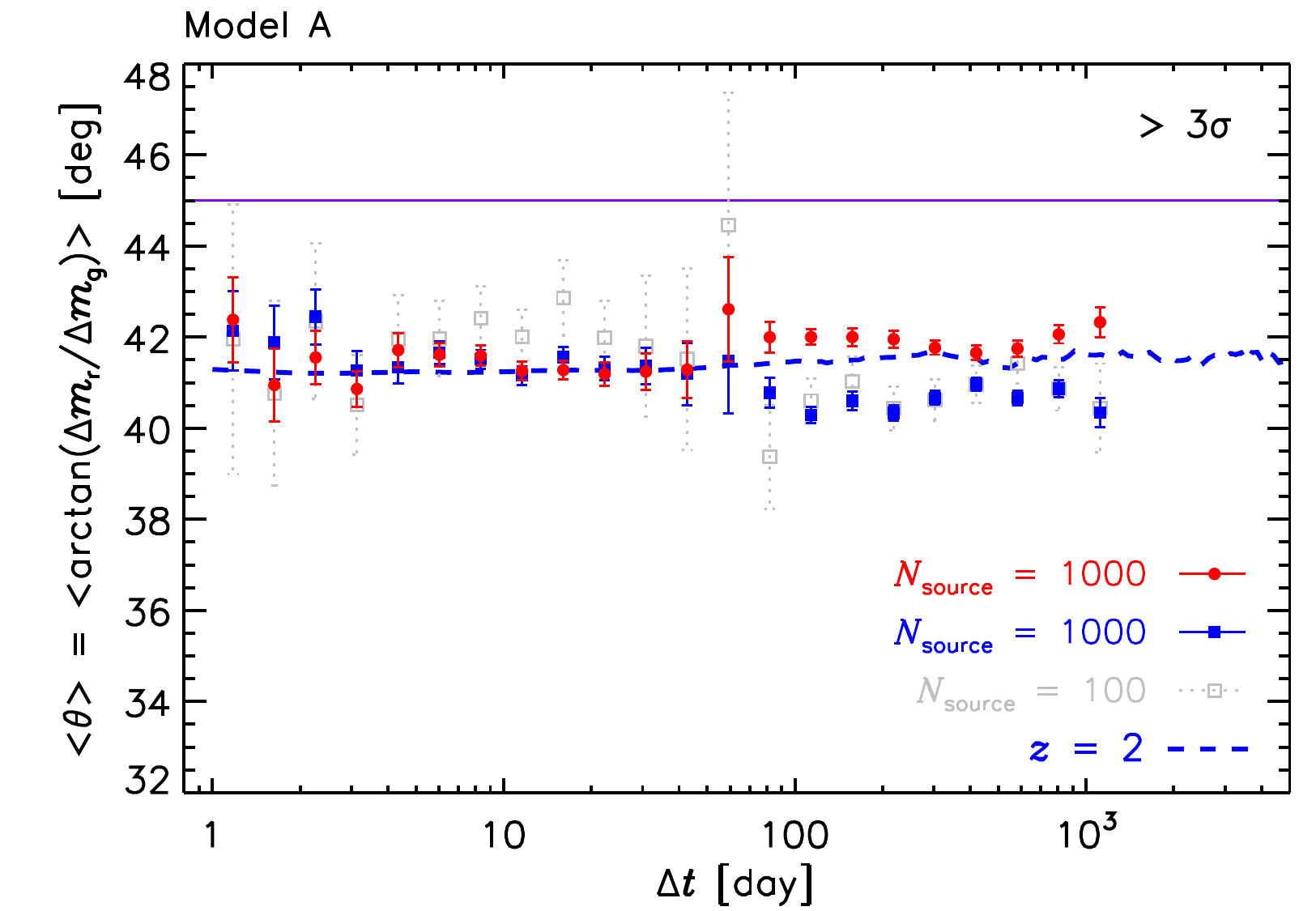}
\includegraphics[width=\columnwidth]{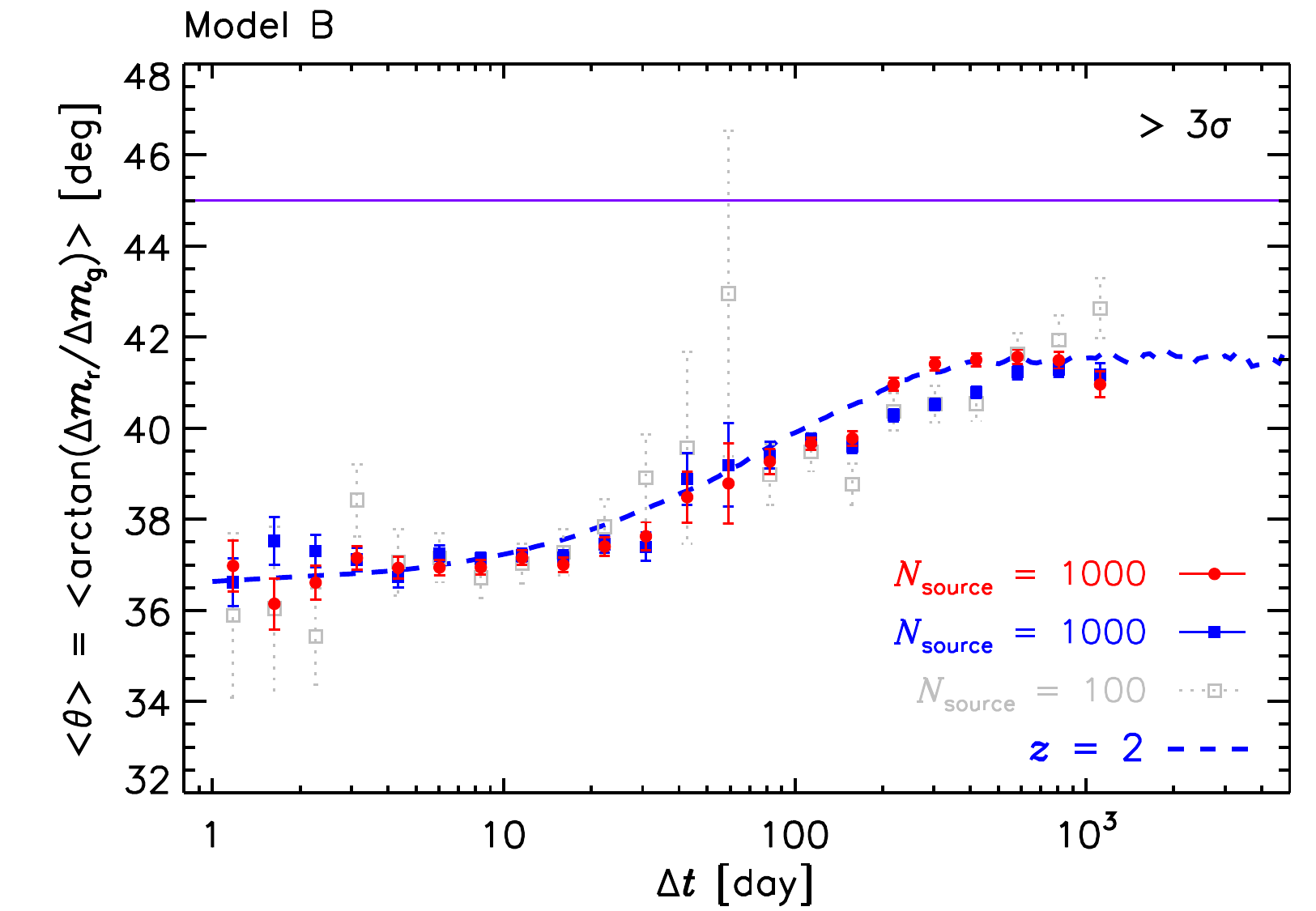}
\caption{Top panels: Amplitudes of color variation, estimated at SDSS $g$- and $r$-bands for {\it ideal} quasars at $z = 0$ (black solid line), 1 (red dotted line), 2 (blue dashed line), and 3 (orange dot-dashed line), as a function of the rest-frame timescale, $\Delta t$, for the reference {\bf model A} with radius-independent $\tau$ (left panel) and {\bf model B} with radius-dependent $\tau$ (right panel).
The shown color variations require $[(\Delta L/L_{\rm g})^2 + (\Delta L/L_{\rm r})^2]^{1/2} \geqslant 10\%$, {which only induces tiny timescale dependence on $\theta$ such as shown in the left panel}. $\theta < 45^\circ$ stands for a bluer-when-brighter trend and $\langle \theta \rangle$ results from averaging over $N_{\rm sim} = 10$ times of {\it ideal} simulations (light-gray lines). 
Bottom panels: Amplitudes of color variation, estimated at SDSS $g$- and $r$-bands for three samples of simulated ``{\it real}" quasars at $z = 2$ with un-even sampling interval and photometric errors borrowed from the SDSS Stripe 82 data, as a function of the rest-frame timescale, $\Delta t$, for the reference {\bf model A} with radius-independent $\tau$ (left panel) and {\bf model B} with radius-dependent $\tau$ (right panel). One sample contains 100 sources (light-gray open squares), while the other two containing completely different 1000 sources (blue filled squares and red filled circles) are considered in order to demonstrate the possible variation between samples with the same source size. Within each $\Delta t$-bin, the averaged $\theta$ is obtained after applying the ``$3\sigma$'' criterion on the variation and its ``$1\sigma$'' error is estimated through bootstrap. 
For both models, regardless of real un-even sampling interval and photometric errors, the global dependence of $\langle \theta \rangle$ to timescale $\Delta t$ is guaranteed after sample averaging and is consistent with that estimated using the {\it ideal} quasars. For smaller sample size, the similar dependence still exists but with larger scatter.
}\label{fig:qv_BWB_tau}
\end{figure*}

The top panels of Figure~\ref{fig:qv_BWB_tau} shows the color variations, estimated at SDSS $g$- and $r$-bands for quasars at $z = 0$ (black solid line), 1 (red dotted line), 2 (blue dashed line), and 3 (orange dot-dashed line), as a function of the rest-frame timescale for the two reference models, requiring the square root of the quadratic sum of the relative changes of successive luminosities in the two bands, i.e., $[(\Delta L/L_{\rm g})^2 + (\Delta L/L_{\rm r})^2]^{1/2}$
\footnote{Here, $\Delta L/L_{\rm band} \equiv 2|L_{\rm band}(t+\Delta t)-L_{\rm band}(t)|/[L_{\rm band}(t+\Delta t)+L_{\rm band}(t)]$ is the change of  luminosities, $L_{\rm band}$, in a given band between epochs $t$ and $t+\Delta t$. Therefore, the corresponding difference in magnitude is $\Delta M_{\rm band} =| M_{\rm band}(t+\Delta t) - M_{\rm band}(t)| = 2.5 \log[(2+f)/(2-f)]$, where $f \equiv \Delta L/L_{\rm band} \in [0, 2)$. For our interested $f \lesssim 0.2$, $\Delta M_{\rm band} \simeq f$.}
, larger than 10\%, and averaging over $N_{\rm sim}$ times of {\it ideal} simulations to reduce the fluctuations of color variation at given timescale.
Following \citet{Sun2014}, the amplitude of the color variation between two epochs separated by timescale $\Delta t$ is represented by a parameter $\theta$ defined as $\theta(\Delta t) \equiv \arctan (\Delta m_{\rm r}/\Delta m_{\rm g})$, where $\Delta m_{\rm g}$ and $\Delta m_{\rm r}$ are the differences of apparent magnitudes between successive epochs separated by $\Delta t$ in $g$- and $r$-bands, respectively. Accordingly, $\theta < 45^\circ$ and $\theta > 45^\circ$ stand for bluer-when-brighter and redder-when-brighter, respectively. 

For the reference {\bf model A} with radius-independent $\tau$, the averaged $\theta$ over all epoch pairs separated by the timescale $\Delta t$ does not exhibit statistically significant dependence on $\Delta t$ as expected (see the top-left panel of Figure~\ref{fig:qv_BWB_tau}). The tiny timescale dependence results from the requirement of $[(\Delta L/L_{\rm g})^2 + (\Delta L/L_{\rm r})^2]^{1/2} \geqslant 10\%$, because the relative change of luminosity at shorter wavelength, e.g., $\Delta L/L_{\rm g}$, is generally larger than that at longer one, e.g., $\Delta L/L_{\rm r}$, and therefore the smaller $\theta$ is preferred under the requirement, especially at shorter timescales.
In short, considering this kind reference models with radius-independent $\tau$, there is no significant timescale dependent color variations and they are hard to explain the timescale dependent color variations of quasars recently discovered by \citet{Sun2014}.

For given observed wavelength pair, {\bf model A}, as well as {\bf model B}, imply smaller $\theta$ with increasing the source redshift (see the upper panels of Figure~\ref{fig:qv_BWB_tau}). 
Interesting, Figure~4 of \citet{Schmidt2012} may have presented a similar trend, where Schmidt et al. demonstrated that the observed color variance of Stripe 82 quasars 
is strongly redshift-dependent. They show such redshift dependence can be largely contributed to contaminations from the broad emission lines which do not respond to the continuum variations. Nevertheless, in their Figure~4 we can still see the observed color variation more prominent than their model at $z > 1$, while at $z < 1$ the model and the observed values are generally consistent.
This may hint that the observed intrinsic color variation of continuum is stronger at higher redshift (i.e., smaller \citeauthor{Schmidt2012}'s $s_{\rm gr}$ or our $\theta$), similar to the pattern shown in the upper panels of our Figure~\ref{fig:qv_BWB_tau}.
However, cautions should be kept in mind that both the neglected reverberation time delays between the continuum and the lines in the simple spectral variability model of \citet{Schmidt2012} and the likely significant contributions of the variability of Balmer continuum from the broad line region in the quasar near-UV spectra \citep[e.g.,][]{Edelson2015} may all be responsible for the discrepancy between the observed redshift dependence of color variability and the simulated one by \citet{Schmidt2012}.

Obviously, the reference {\bf model B} with radius-dependent $\tau$ (the top-right panel of Figure~\ref{fig:qv_BWB_tau}) possesses qualitatively similar timescale dependent color variations as found by \citet{Sun2014}. Moreover, we see that the expected dependence of $\langle \theta \rangle$  to timescale $\Delta t$ is more prominent at intermediate timescales, and is obviously weaker toward much shorter or longer timescales.

\section{Discussion}\label{sect:discussion}

\subsection{Constructing simulated ``real'' sample}\label{sect:simulatedsample}

\begin{figure}[!t]
\centering
\includegraphics[width=\columnwidth]{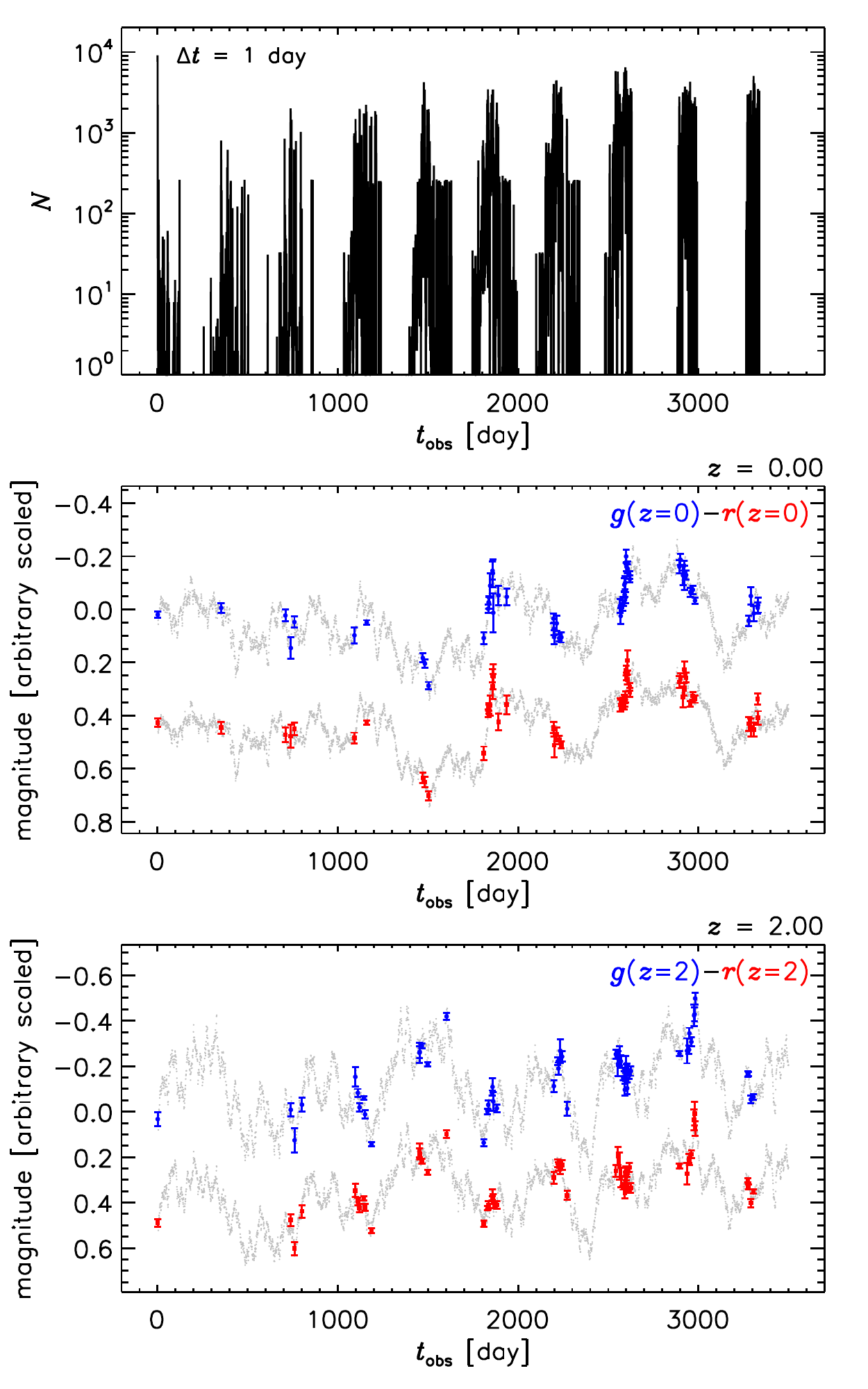}
\caption{An illustration for constructing the simulated ``{\it real}" light curves. The top panel shows the probability distribution of sampling in the observed frame with a time bin of one day, constructed using all the Stripe 82 sources considered by \citet{Sun2014} and the first observing epoch of each source is shifted to be the same. The middle and bottom panels show examples of the simulated ``{\it real}" $g$- (blue points) and $r$-band (red points) light curves,
by convolving the probability distribution of sampling with the corresponding {\it ideal} light curves (light-gray points, in steps of one day in the observed frame). 
}\label{fig:qv_lc_sampling}
\end{figure}

Although a qualitatively similar timescale dependent color variation as found by \citet{Sun2014} has been discussed using the {\it ideal} light curves implied by {\bf model B} in the previous section, some other observational facts, such as photometric errors and un-even sampling \citep[e.g.,][for SDSS photometries]{PalanqueDelabrouille2011}, should be properly taken into account before the delicate comparison with real data can be made and to assess their effects on the measured $\langle \theta \rangle$ -- $\Delta t$ relation. 

To be compared with the data by \citet{Sun2014}, we use all the sources considered by \citet{Sun2014} to get the $\langle \theta \rangle$ -- $\Delta t$ relation averaged over all redshifts, to derive a probability distribution of sampling in the observed frame with a time bin of one day (cf. the top panel of Figure~\ref{fig:qv_lc_sampling}). For a source at redshift $z$ observed at SDSS $g$- and $r$-bands, we first simulate the {\it ideal} $g(z)$ and $r(z)$ light curves in steps of one day in the observed frame, which are randomly convolved with the probability distribution of sampling.
Random photometric uncertainties are added to every pair of both light curves of each simulated source, with a pair of photometric errors of $g$- and $r$-bands taken randomly from real quasars in the same redshift bin, and furthermore, Gaussian deviates with zero mean and standard deviation equal to the randomly selected photometric uncertainty of each point are added to mimic the measurement errors (cf. the middle and bottom panels of Figure~\ref{fig:qv_lc_sampling} for examples of $g$- and $r$-band light curves of sources at $z = 0$ and 2, respectively).
We consider the same redshift bins as explored by \citet{Sun2014} in order to compare the $\langle \theta \rangle$ -- $\Delta t$ relation averaged over quasars within different redshift bins. This sample of $N$ sources at redshift $z$ is then used to calculate the $\langle \theta \rangle$ -- $\Delta t$ relation at $z$ after applying the ``$3\sigma$'' criterion on the variation as done by \citet{Sun2014}, and bootstrapped to estimate its ``$1\sigma$'' errors.

In the bottom panels of Figure~\ref{fig:qv_BWB_tau} , we show that the simulated ``{\it real}" light curves, after applying photometric errors and un-even sampling borrowed from the SDSS Stripe 82 data and the ``$3\sigma$'' criterion on the variation, yield $\langle \theta \rangle$ -- $\Delta t$ relations well consistent with those from the {\it ideal} light curves. This also demonstrates that the observed timescale dependent color variation by \citet{Sun2014} can not be attributed to such observational effects.

\subsection{Dependence on model parameters}\label{sect:parameters}

\begin{figure*}[!t]
\centering
\includegraphics[width=0.92\columnwidth]{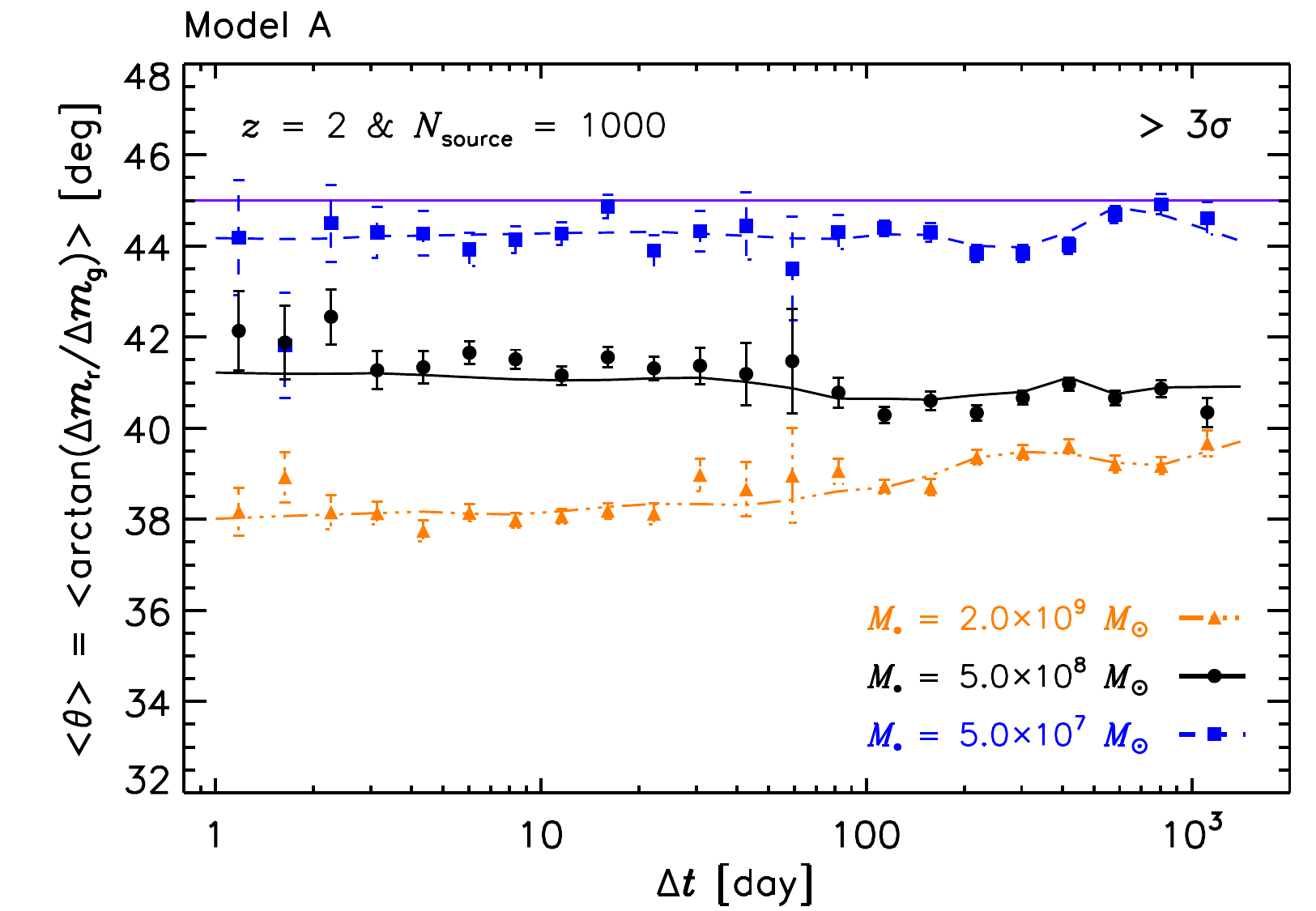}
\includegraphics[width=0.92\columnwidth]{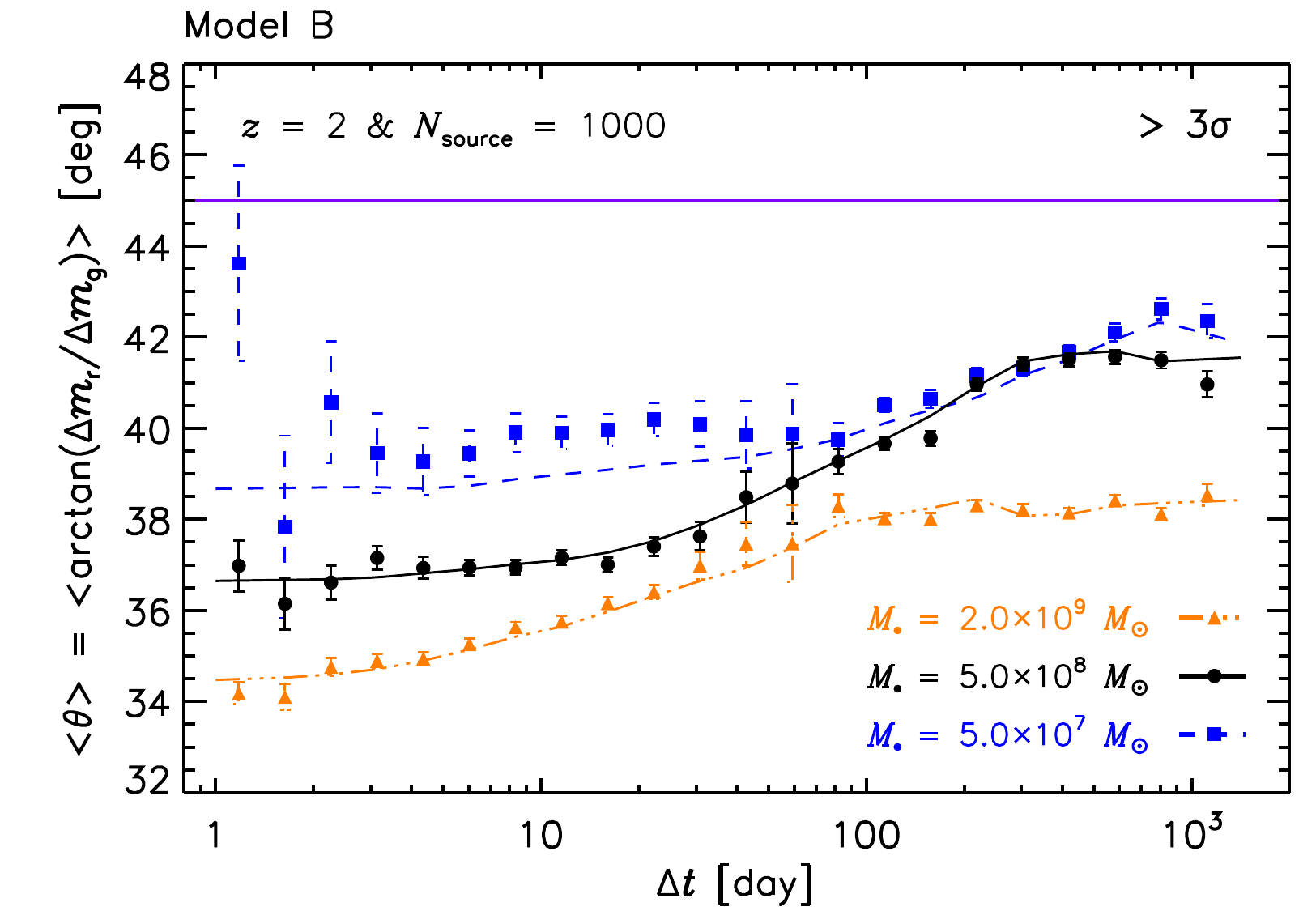}
\includegraphics[width=0.92\columnwidth]{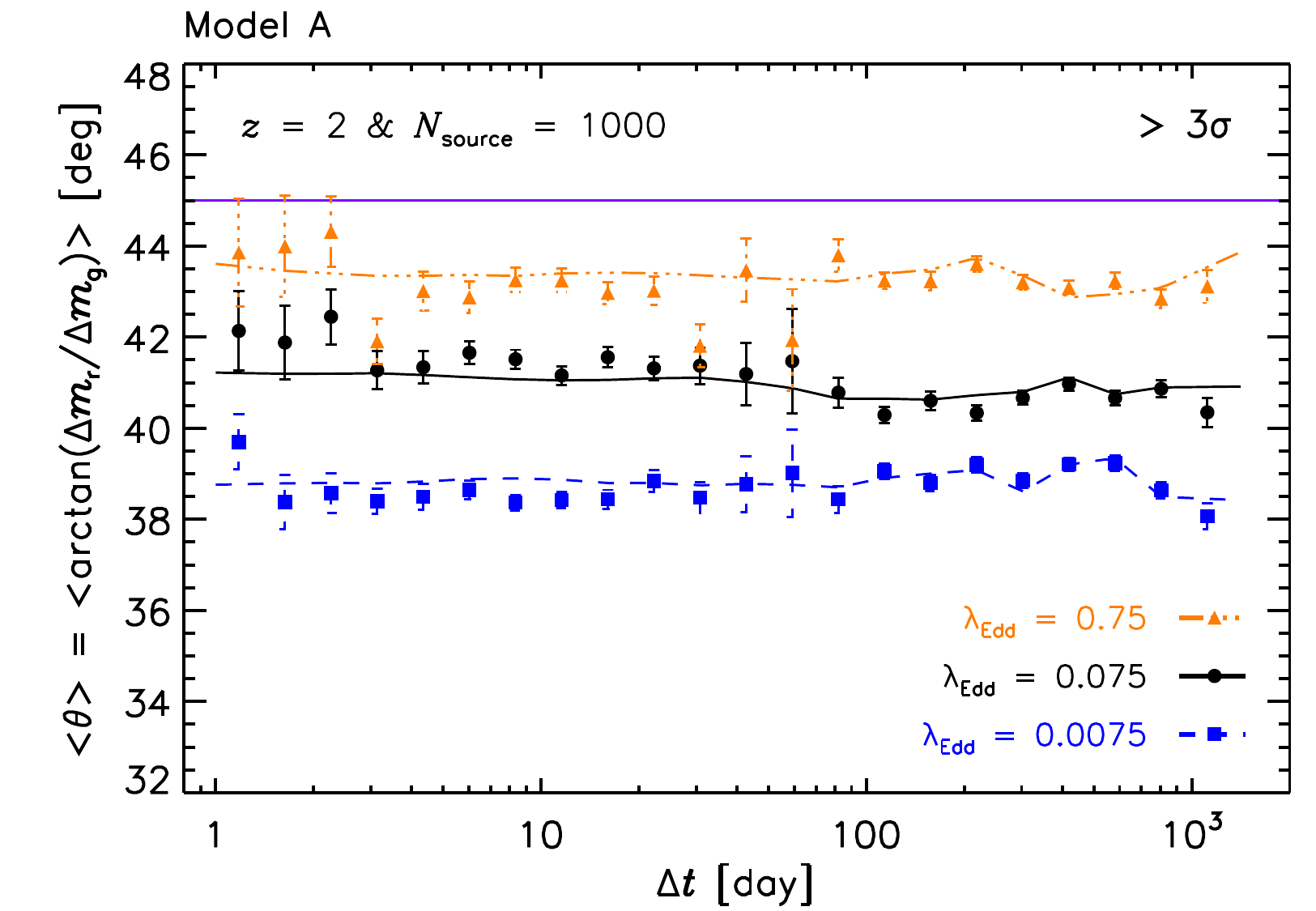}
\includegraphics[width=0.92\columnwidth]{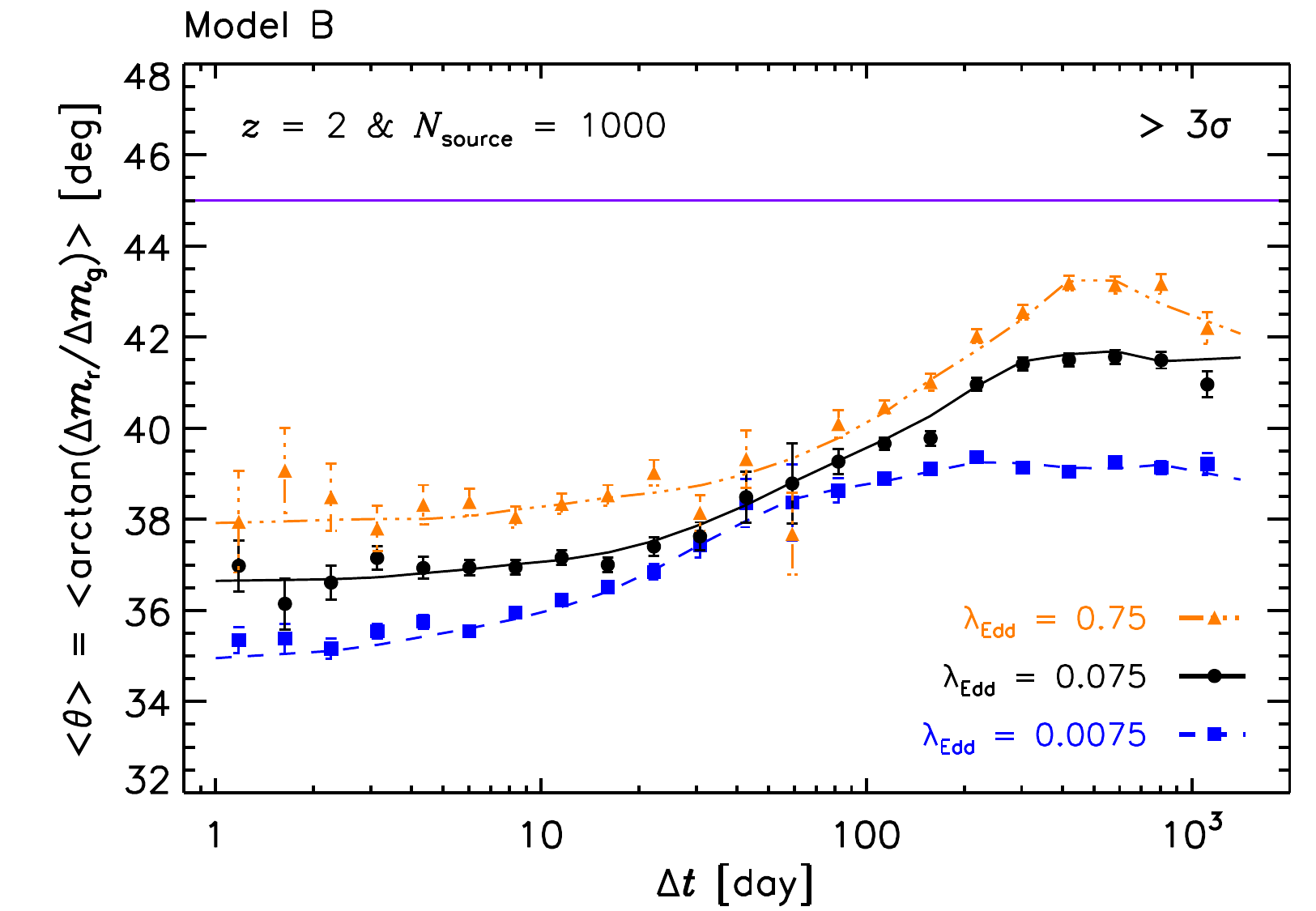}
\includegraphics[width=0.92\columnwidth]{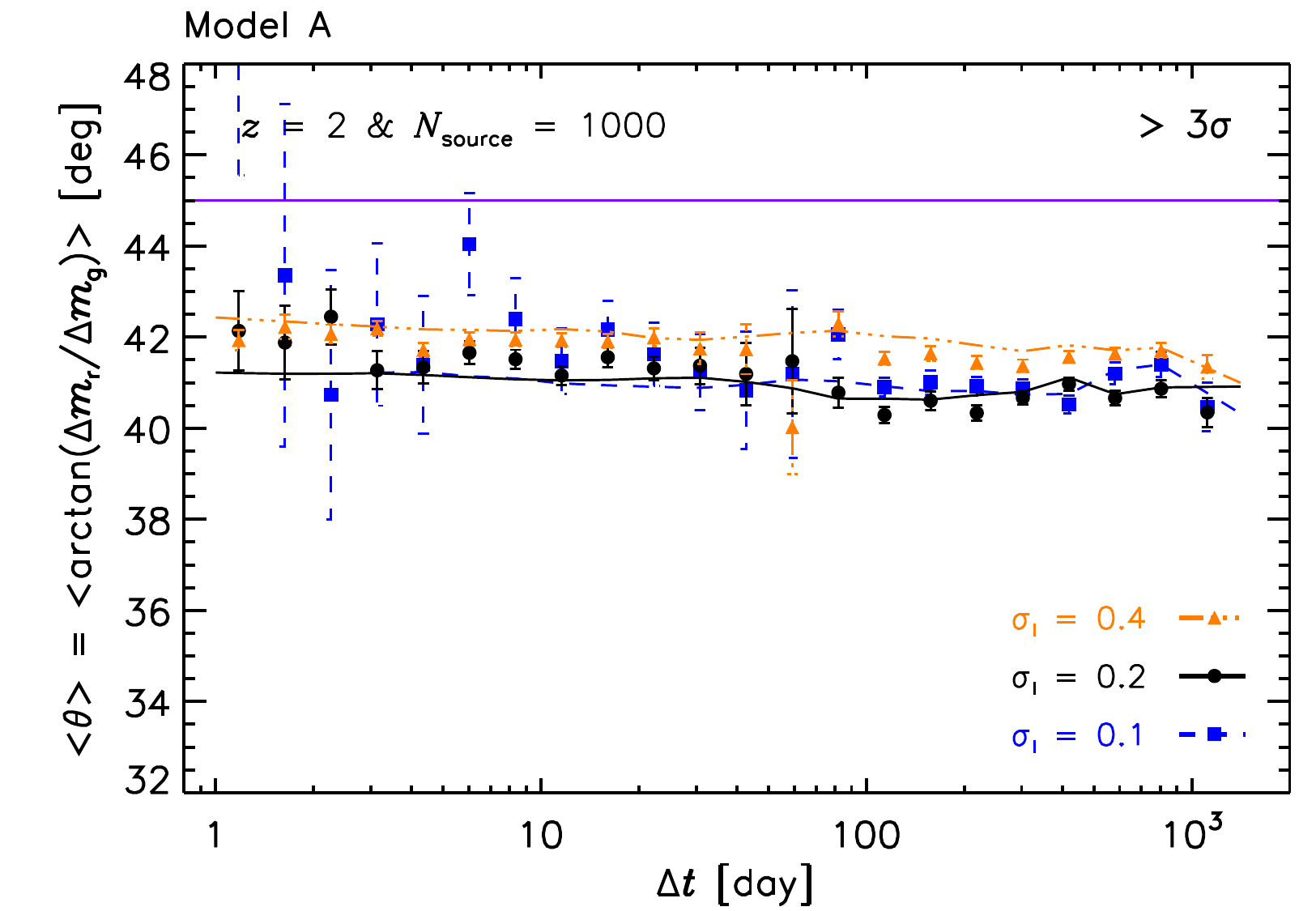}
\includegraphics[width=0.92\columnwidth]{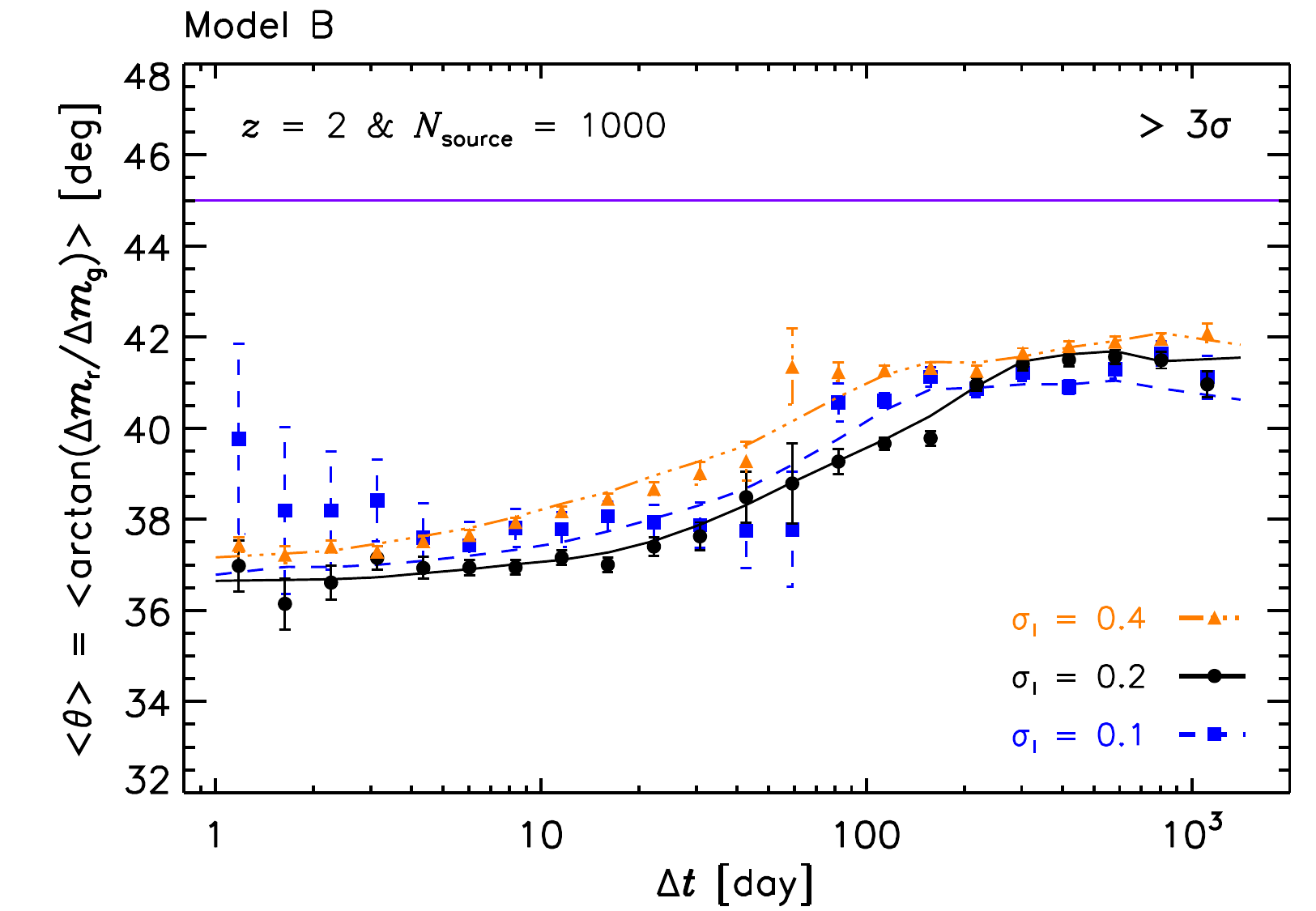}
\includegraphics[width=0.92\columnwidth]{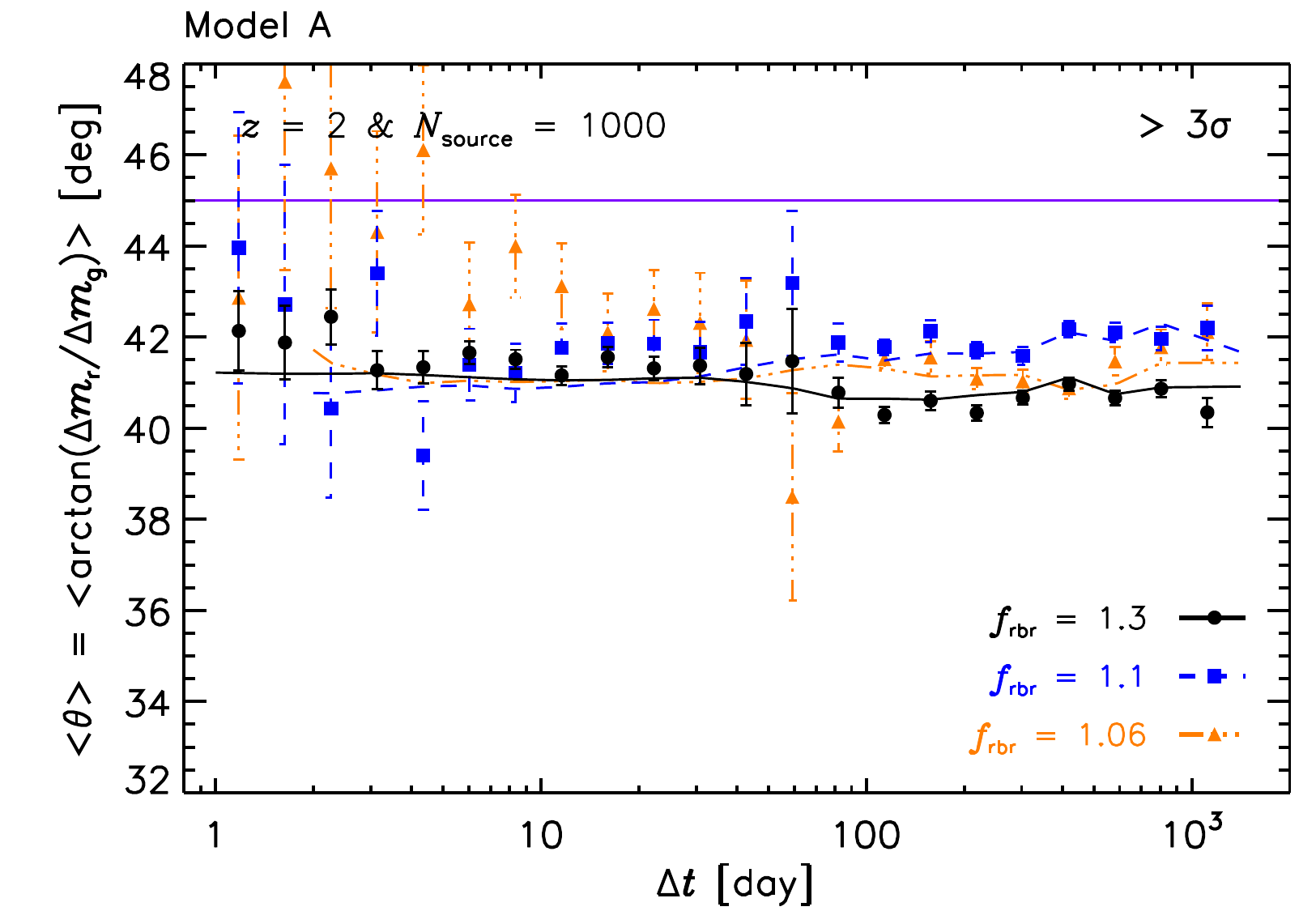}
\includegraphics[width=0.92\columnwidth]{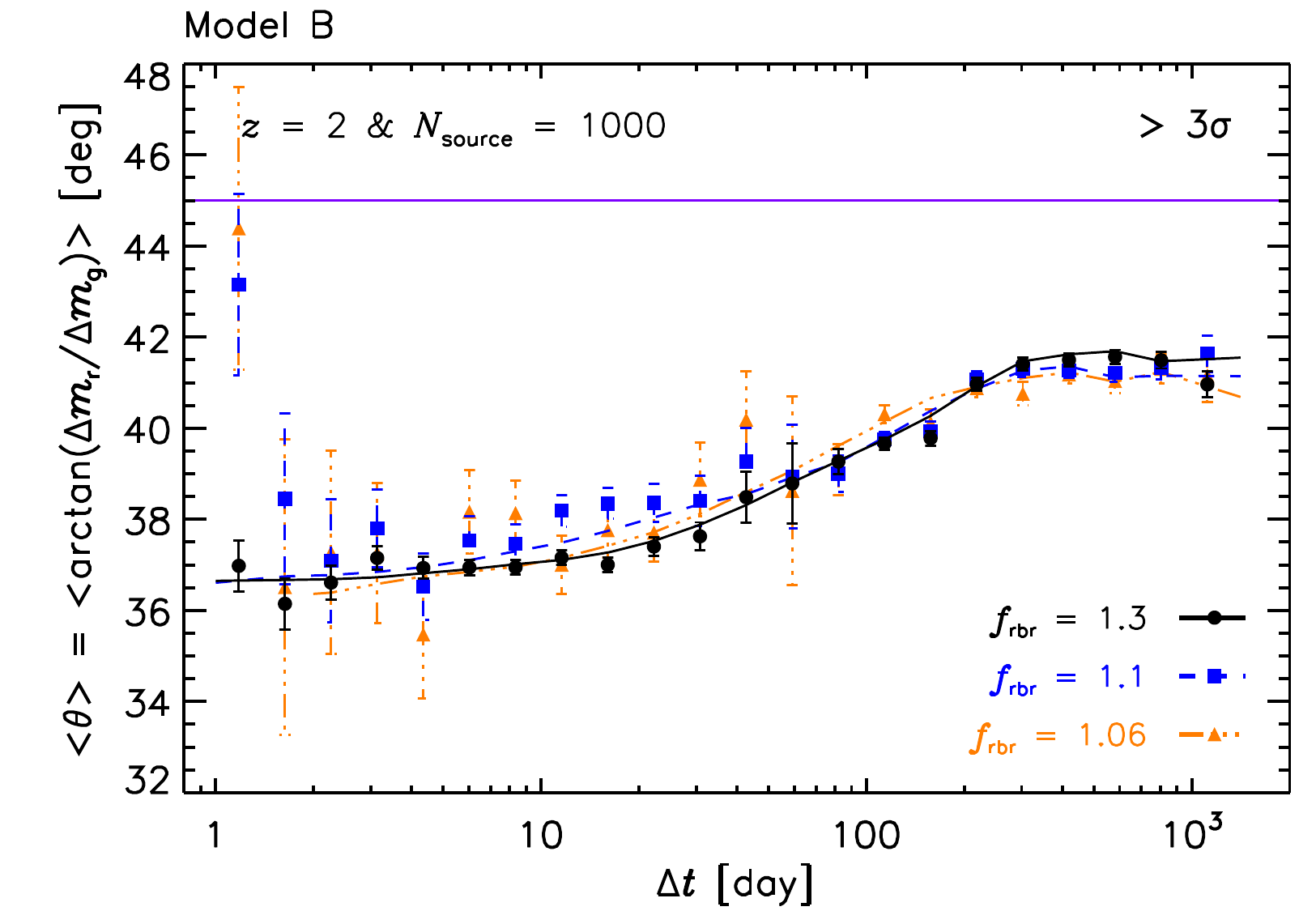}
\caption{Amplitudes of color variation, estimated at SDSS $g$- and $r$-bands for a sample of 1000 simulated ``{\it real}" quasars at $z = 2$ (symbols) with un-even sampling interval and photometric errors borrowed from the SDSS Stripe 82 data, as a function of the rest-frame timescale, $\Delta t$, implied by {\bf model A} (left panels) and {\bf model B} (right panels) with different parameters. 
The adjusted parameters for both models are the BH mass, $M_\bullet$, the Eddington ratio, $\lambda_{\rm Edd}$, the long-term temperature fluctuation amplitude, $\sigma_{\rm l}$, and the radial boundary ratio of split zones, $f_{\rm rbr}$, respectively, compared to the reference model with $M_\bullet = 5 \times 10^8\,M_\odot$, $\lambda_{\rm Edd} = 0.075$, $\sigma_{\rm l} = 0.2$, and $f_{\rm rbr} = 1.3$ (black filled circles). The lines, indicating the relations implied by the corresponding 1000 simulated ``{\it real}" quasars with photometric errors borrowed from the SDSS Stripe 82 data but with even sampling of one day in the observed frame, are shown in order to better demonstrate the dependence of the relation on these parameters.
}\label{fig:qv_BWB_tau_params}
\end{figure*}

\begin{figure*}[!t]
\centering
\includegraphics[width=\columnwidth]{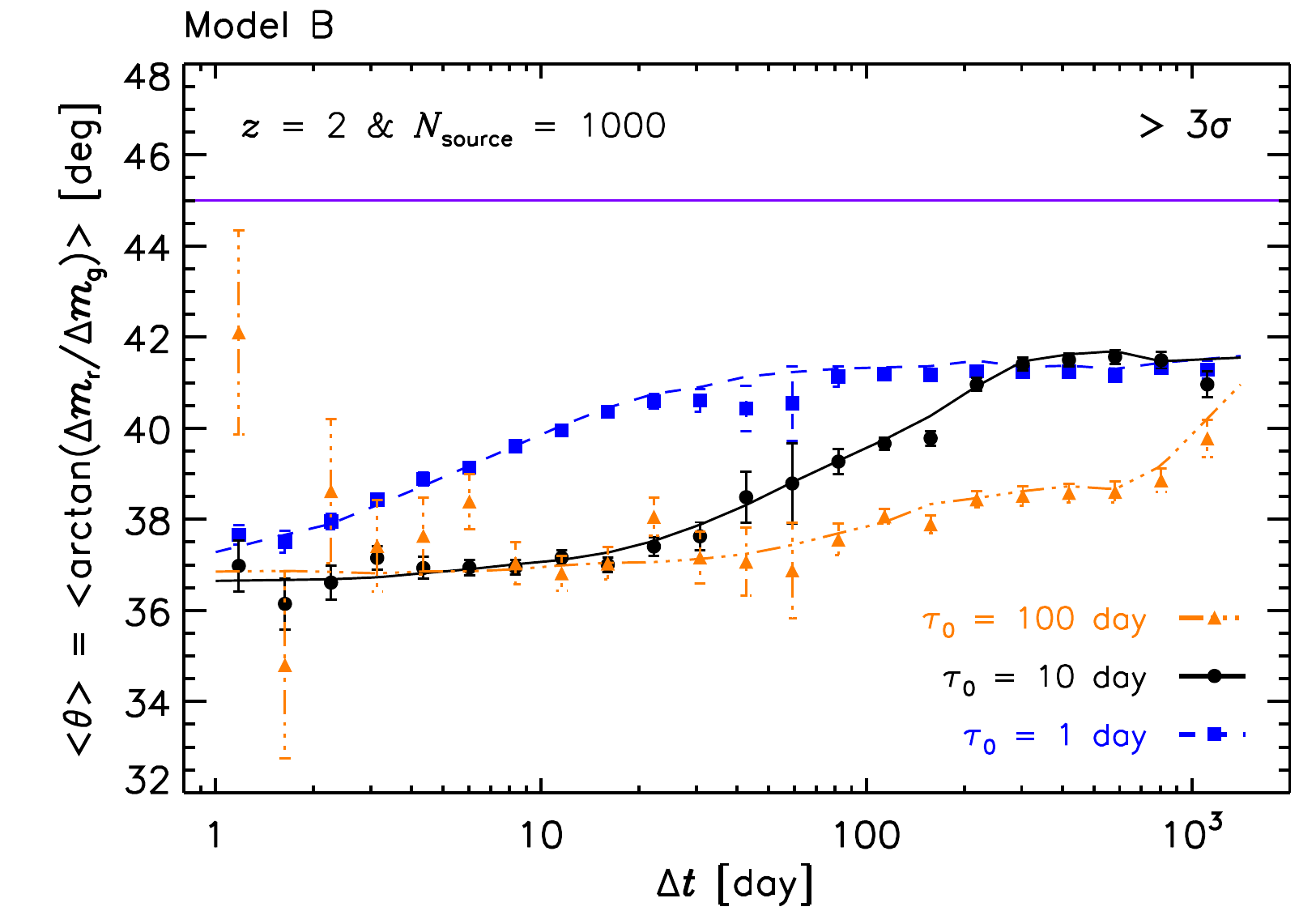}
\includegraphics[width=\columnwidth]{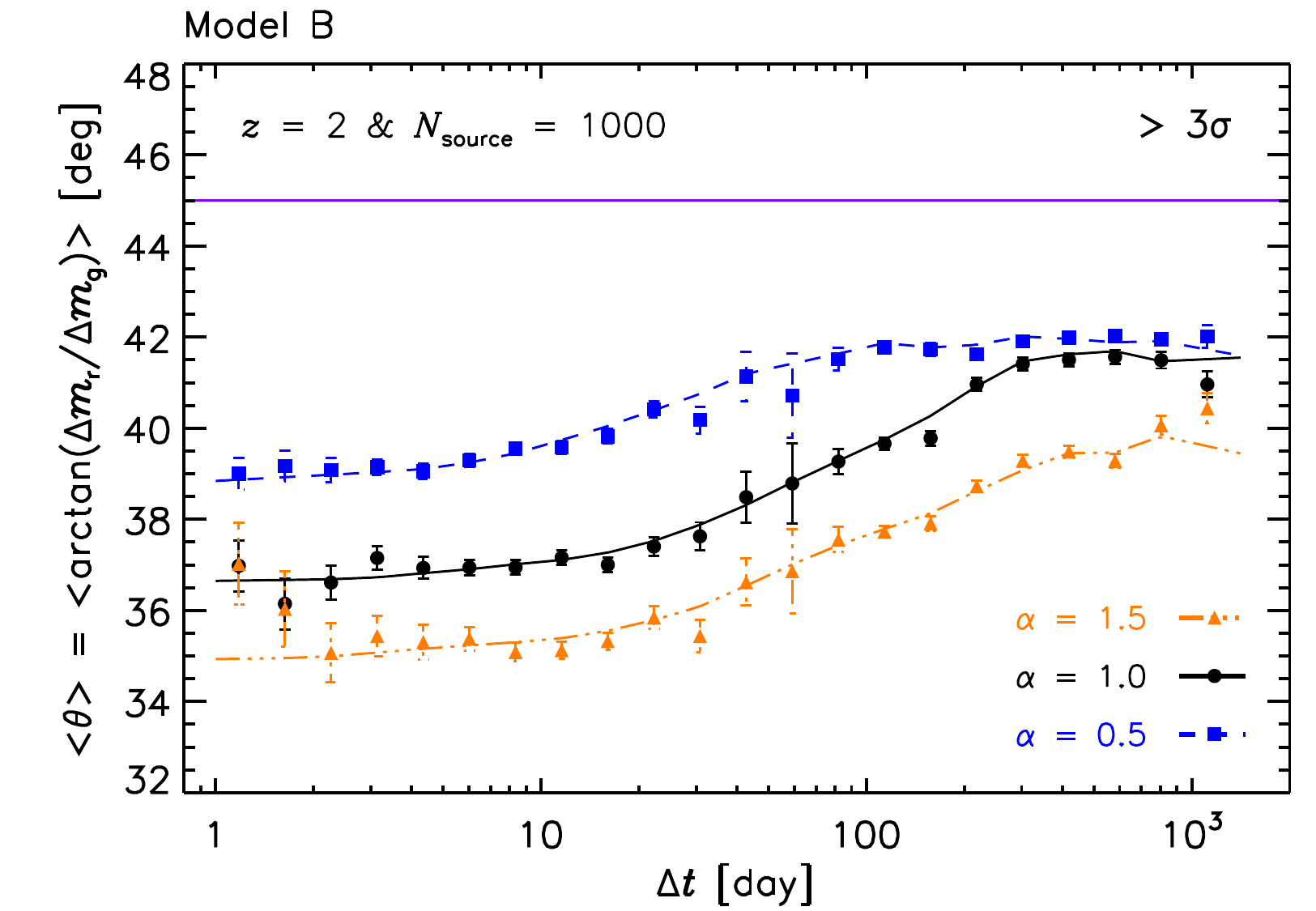}
\caption{Same as Figure~\ref{fig:qv_BWB_tau_params}, except that the adjusted parameters are two unique ones of {\bf model B}, i.e., $\tau_0$ (left panel) and $\alpha$ (right panel).
}\label{fig:qv_sed_comp_theta_dt_tau_alpha_B}
\end{figure*}

Up to now, we have mainly focused on the reference models with a single set of fixed parameters. Instead, \citet{DexterAgol2011} and \citet{Ruan2014} have explored a range of parameter space (i.e., the number of fluctuating zones per factor of two in radius, $n \simeq 300-1000$, the amplitude of long-term temperature fluctuation, $\sigma_{\rm T} \simeq 0.35-0.5$, the Eddington ratio, $\lambda_{\rm Edd} \simeq 0.07-0.16$, and the BH mass, $M_\bullet \simeq 3\times10^8-3\times10^9~M_\odot$) for the inhomogeneous disk model (analog to {\bf model A}) to explain observations, including the disk size measurements, the optical variability amplitudes, the UV spectral slope, and the composite different spectrum of quasars. 

In Figure~\ref{fig:qv_BWB_tau_params} we explore the impacts of changes of these parameters on the $\langle \theta \rangle$ -- $\Delta t$ relation (symbols), estimated at SDSS $g$- and $r$-bands for a sample of 1000 simulated ``{\it real}" quasars at $z = 2$ with un-even sampling interval and photometric errors borrowed from the SDSS Stripe 82 data, implied by both models. To better demonstrate the dependence of the relation on these parameters, the relations (lines) implied by the corresponding 1000 simulated ``{\it real}" quasars with photometric errors borrowed from the SDSS Stripe 82 data but with even sampling of one day in the observed frame are illustrated as well.

From upper to bottom panels of Figure~\ref{fig:qv_BWB_tau_params}, shown are the results by changing $M_\bullet = 5\times10^7-2\times10^9~M_\odot$, $\lambda_{\rm Edd} = 0.0075-0.75$, $\sigma_{\rm l} = 0.1-0.4$ (equivalent to $\sigma_{\rm T} = 0.14-0.56$), and $f_{\rm rbr} = 1.06-1.3$ (equivalent to $n = 1284-64$), respectively. 
For both models, the effects of changing these parameters on the $\langle \theta \rangle$ -- $\Delta t$ relation are qualitatively similar. Note that quasars at higher redshift generally possess smaller $\langle \theta \rangle$, estimated at SDSS $g$- and $r$-bands for all timescales, or at the shorter wavelength pair the smaller $\langle \theta \rangle$ (cf. Figure~\ref{fig:qv_BWB_tau}). 
Considering the $\langle \theta \rangle$ estimated at the same two bands for quasars with the same redshift but different other parameters as illustrated in Figure~\ref{fig:qv_BWB_tau_params}, it becomes larger with decreasing $M_\bullet$ or increasing $\dot M$ (or $\lambda_{\rm Edd}$ for a fixed BH mass). The dependence on $M_\bullet$ or $\dot M$ can be understood as follows. The lower $M_\bullet$ or higher $\dot M$ the hotter the disk at given radius, the emissions contributing to the $g$- and $r$-bands come from relatively outer radii, i.e., equivalent to longer wavelength pair, and therefore, larger $\langle \theta \rangle$. Meanwhile, for {\bf model B}, the outer radii correspond to longer characteristic timescales of temperature fluctuation and, consequently, the saturation of $\langle \theta \rangle$ happens at longer timescales.
Considering the dependence on $\sigma_{\rm l}$, since larger $\sigma_{\rm l}$ induces hotter mean SED (cf. the bottom panel of Figure~\ref{fig:qv_sed_comp_f_rbr_B}) and equivalent emissions come from outer radii, the $\langle \theta \rangle$ becomes larger with increasing $\sigma_{\rm l}$.
Finally, no dependence on $f_{\rm rbr}$ is found for both models. This suggests that the $\langle \theta \rangle$ -- $\Delta t$ relation can not provide any constraint on the number of fluctuating zones.

Comparing with {\bf model A}, {\bf model B} introduces two extra parameters, i.e., the minimal timescale at the inner most radius, $\tau_0$, and the index of radius dependence, $\alpha$.
The effect of $\tau_0$ on the relation is simply shifting the $\langle \theta \rangle$ -- $\Delta t$ plot horizontally (cf. the left panel of Figure~\ref{fig:qv_sed_comp_theta_dt_tau_alpha_B}).
The impact of changing $\alpha$ is however more complicated.
While it is expected that smaller $\alpha$ (e.g., 0.5) produces weaker correlation between $\langle \theta \rangle$ and $\Delta t$, a larger
$\alpha$ (e.g., 1.5) does not give a steeper slope comparing with $\alpha$ = 1.0 (cf. the right panel of Figure~\ref{fig:qv_sed_comp_theta_dt_tau_alpha_B}). 
Referring to explanations in last two paragraphs of Section~\ref{sect:stripe82}, larger $\alpha$, which produces relatively stronger inner disk fluctuations comparing with outer disk at short timescales, yields bluer variable emission thus smaller $\langle \theta \rangle$ at small $\Delta t$.
With increasing $\Delta t$, $\langle \theta \rangle$ starts to gradually increases to a saturation point when both inner and outer disk fluctuations do not grow anymore. While the value of saturated $\langle \theta \rangle$ is independent to $\alpha$, the saturation point moves to longer timescale for larger $\alpha$, thus further increasing $\alpha$ does not significantly alter the $\langle \theta \rangle$ -- $\Delta t$ slope.

\subsection{Comparison to the Stripe 82 data}\label{sect:stripe82}
\begin{figure}[!t]
\centering
\includegraphics[width=\columnwidth]{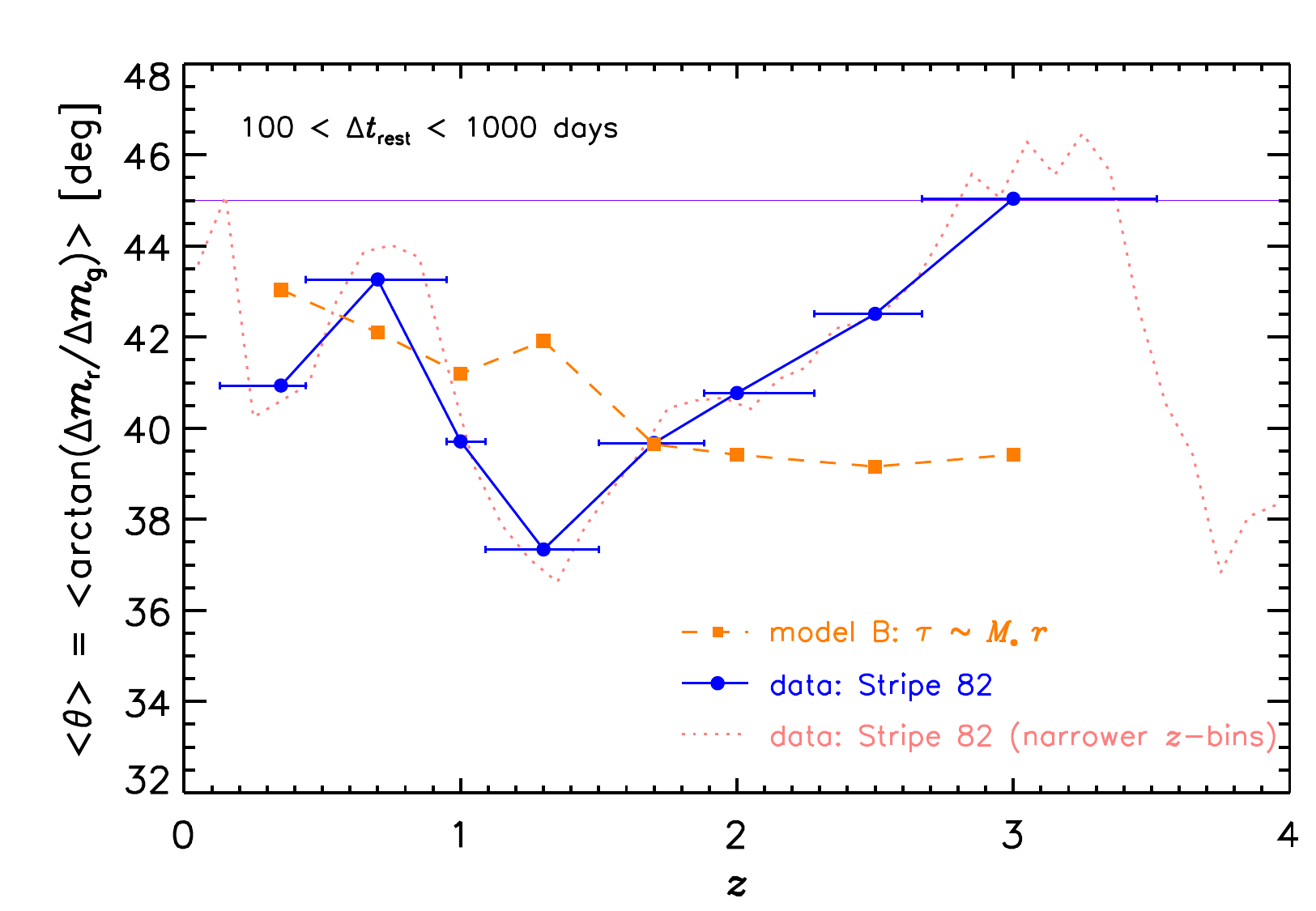}
\caption{Amplitudes of color variation of Stripe 82 quasars (blue solid line plus filled circles for the considered redshift bins; pink dotted line for narrower redshift bins) and of {\bf model B} (orange dashed line plus filled squares, inferred from the orange dashed lines in Figure~\ref{fig:qv_sed_comp_theta_dt_z_B}), averaged over rest-frame $10^2 < \Delta t < 10^3$ days.
The amplitude difference between blue filled circle and orange filled square is attributed to emission line contributions and adopted as the {\it empirical} factor for correction at each redshift.
}\label{fig:qv_sed_comp_theta_z_dt_B}
\end{figure}

\begin{figure*}[!t]
\centering
\includegraphics[width=0.8\textwidth]{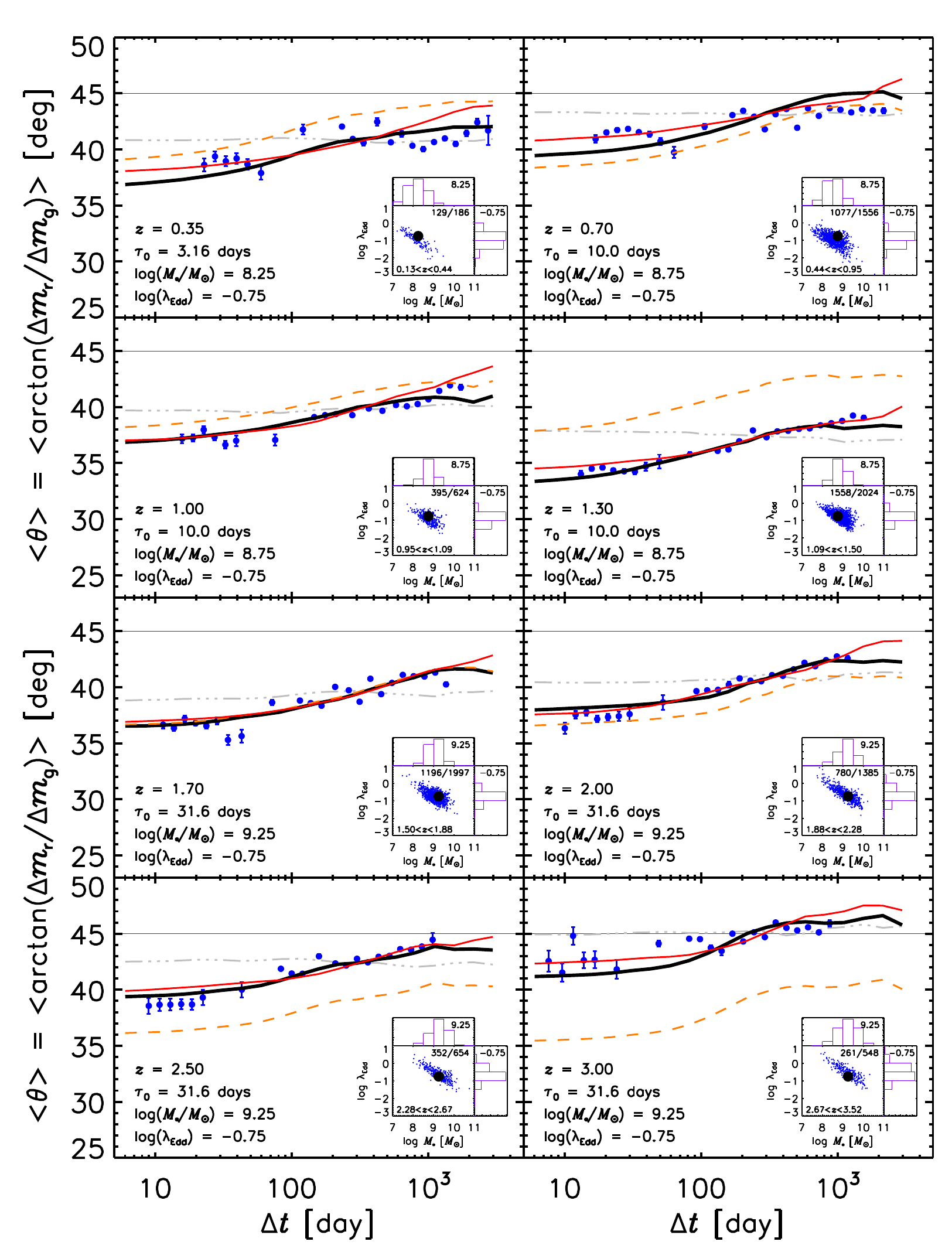}
\caption{Comparison between the color variations of Stripe 82 quasars in different redshift bins \citep[blue filled circles,][]{Sun2014} and those implied by the {\bf model B} with $\tau_0 = 10~(M_\bullet/10^{8.75}M_\odot)$ days and $\alpha = 1$, {\it empirically} corrected for emission line contributions (thick black lines; see also Figure~\ref{fig:qv_sed_comp_theta_z_dt_B}). The model predictions prior to corrections are also plotted (orange dashed lines).
The adopted model parameters are indicated at the lower-left corner of each panel.
In each redshift bin, the adopted typical redshift is chosen close to the mean/median of the subsample of Stripe 82 quasars, while the adopted typical BH mass and Eddington ratio are respectively the peaks of BH mass and Eddington ratio distributions for those quasars with both BH mass and Eddington ratio measurements \citep[small blue dots in the inserted plot, with the range of the redshift bin at the lower-left corner and the number ratio of quasars with both BH mass and Eddington ratio measurements to all sources in the bin at the upper-right corner;][]{Shen2011}.
For comparisons, the similarly corrected relations implied by {\bf model A} (light-gray thin triple-dot-dashed lines) and by {\bf model B} but with $\alpha = 1.5$ (red thin solid lines) are also shown.
}\label{fig:qv_sed_comp_theta_dt_z_B}
\end{figure*}

As demonstrated in the previous sections, the timescale independent color variations implied by {\bf model A}, similar to that of \citet{DexterAgol2011}, with radius-independent $\tau$ for temperature fluctuations conflict with the discovery of a timescale dependent color variations by \citet{Sun2014}. Neither the selection of model parameters nor the observational defects, e.g., un-even sampling and/or photometric error, can alleviate the tension. Therefore, later on, we will mainly compare the predictions implied by {\bf model B} with observations of \citet{Sun2014}.

Since only the continuum emission of the accretion disk is considered in the current model, one needs to correct for emission line contributions on the $\langle \theta \rangle$ at different redshift bins.
\citet{Schmidt2012} have shown the redshift dependence of the color variation amplitude of SDSS quasars (up to $z \sim 3$)
is more consistent with the assumption that the emission lines in quasar spectra simply do not follow the continuum variation (cf. their Figure~4).
Consequently, we estimate the $\langle \theta \rangle$ averaged over rest-frame $10^2 < \Delta t < 10^3$ days for both Stripe 82 quasars and simulated ``{\it real}" quasars in {\bf model B} with $\tau_0 = 10~(M_\bullet/10^{8.75} M_\odot)$ days and $\alpha = 1$, in several redshift bins as illustrated in Figure~\ref{fig:qv_sed_comp_theta_z_dt_B}, and attribute their differences to the contaminations of emission lines to certain bands. 
The model predicted $\langle \theta \rangle$ -- $\Delta t$ relation can then be {\it empirically} corrected for direct comparison with observations
(Figure~\ref{fig:qv_sed_comp_theta_dt_z_B}).
Note that this procedure is somewhat {\it arbitrary} (see Section~\ref{sect:caveats} for further discussion).
The same procedure has been applied for {\bf model A} to illustrate the completely different $\langle \theta \rangle$ -- $\Delta t$ relation. 
Within each redshift bin, the model predictions are inferred with the typical values of redshift, $\tau_0$, BH mass, and Eddington ratio, nominated at the lower-left corner of each panel (see the legend of Figure~\ref{fig:qv_sed_comp_theta_dt_z_B} for the selection of these typical values).

After correction, the general consistence between the shapes of the $\langle \theta \rangle$ -- $\Delta t$ relation implied by Stripe 82 quasars (blue filled circles) and by {\bf model B} with $\tau_0 = 10~(M_\bullet/10^{8.75} M_\odot)$ days and $\alpha = 1$ (black thick solid lines in Figure~\ref{fig:qv_sed_comp_theta_dt_z_B}) is encouraging, in consideration of the simplicity/caveats of the current model.
In the two lowest redshift bins, the observed $\langle \theta \rangle$ -- $\Delta t$ slopes however appear slightly flatter than the model prediction.
One possibility is that X-ray reprocessing \citep[e.g.,][]{Krolik1991} in quasars with lower luminosities, in which the X-ray to bolometric luminosity ratio is higher,
could contribute more to optical/UV variation, and its color variation dependence to timescale could be different from thermal fluctuation in the disk.
We will leave this issue to a future dedicated study.

Moreover, we also plot in Figure~\ref{fig:qv_sed_comp_theta_dt_z_B} the model prediction with $\alpha$ = 1.5, the agreement between which
and the observation is also remarkable, but not as good as $\alpha$ = 1.0. Considering the model predicted slopes in $\langle \theta \rangle$ -- $\Delta t$ rely on various parameters which we can not well constrain (see Figures~\ref{fig:qv_BWB_tau_params} and \ref{fig:qv_sed_comp_theta_dt_tau_alpha_B}), it is immature to conclude that which $\alpha$ is more favored by observations, and we are yet unable to pin down 
$\alpha$ and therefore possible physical mechanism responsible for the temperature fluctuations delineated by some characteristic timescales, e.g., the viscous timescale of $\tau \sim r^{5/4}$ \citep{Alloin1985} or the thermal (cooling) timescale of $\tau \sim r^{3/2}$ (\citealt{CollierPeterson2001}; see also \citealt{Lawrence2012}).

\begin{figure*}[!t]
\centering
\includegraphics[width=\columnwidth]{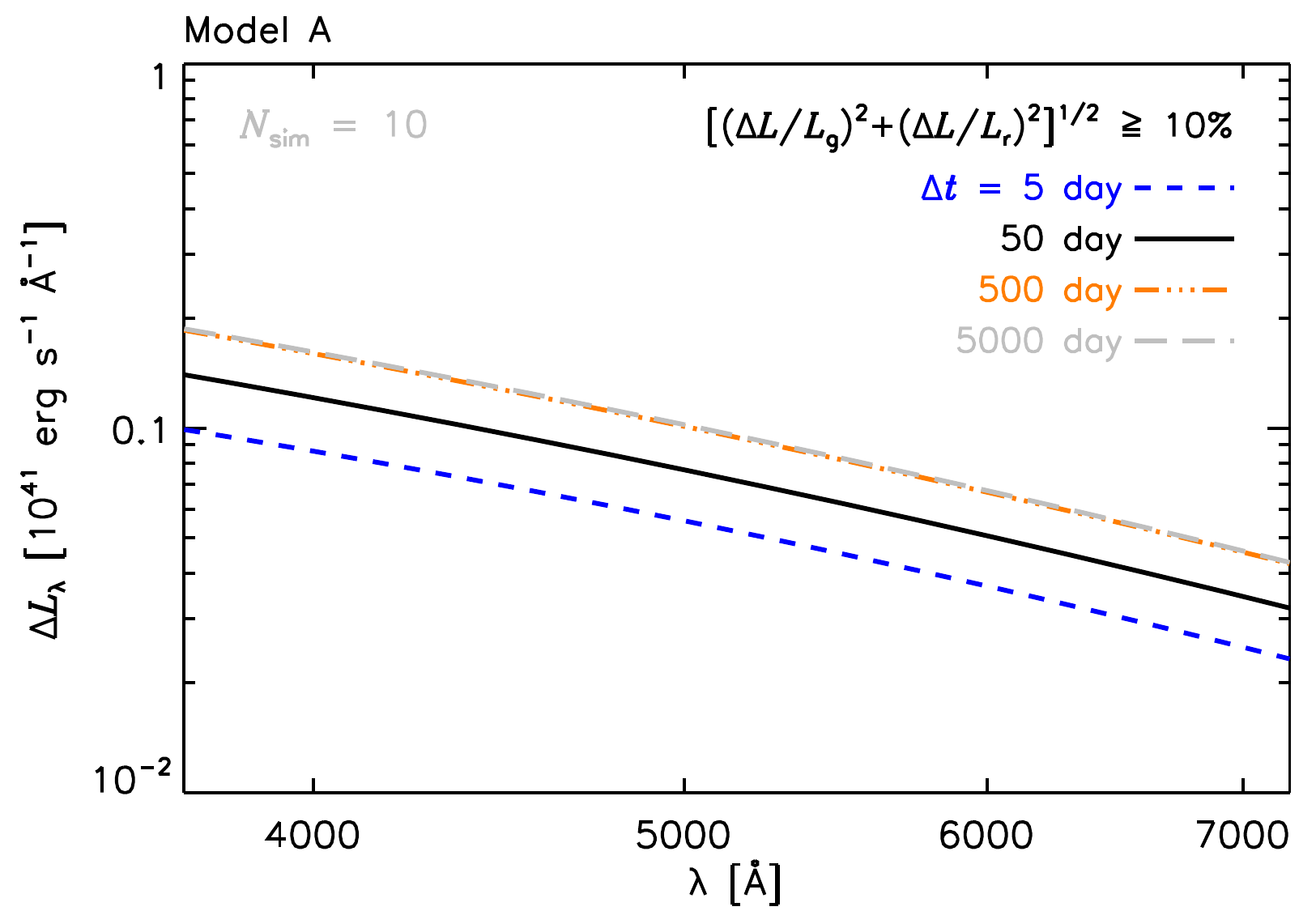}
\includegraphics[width=\columnwidth]{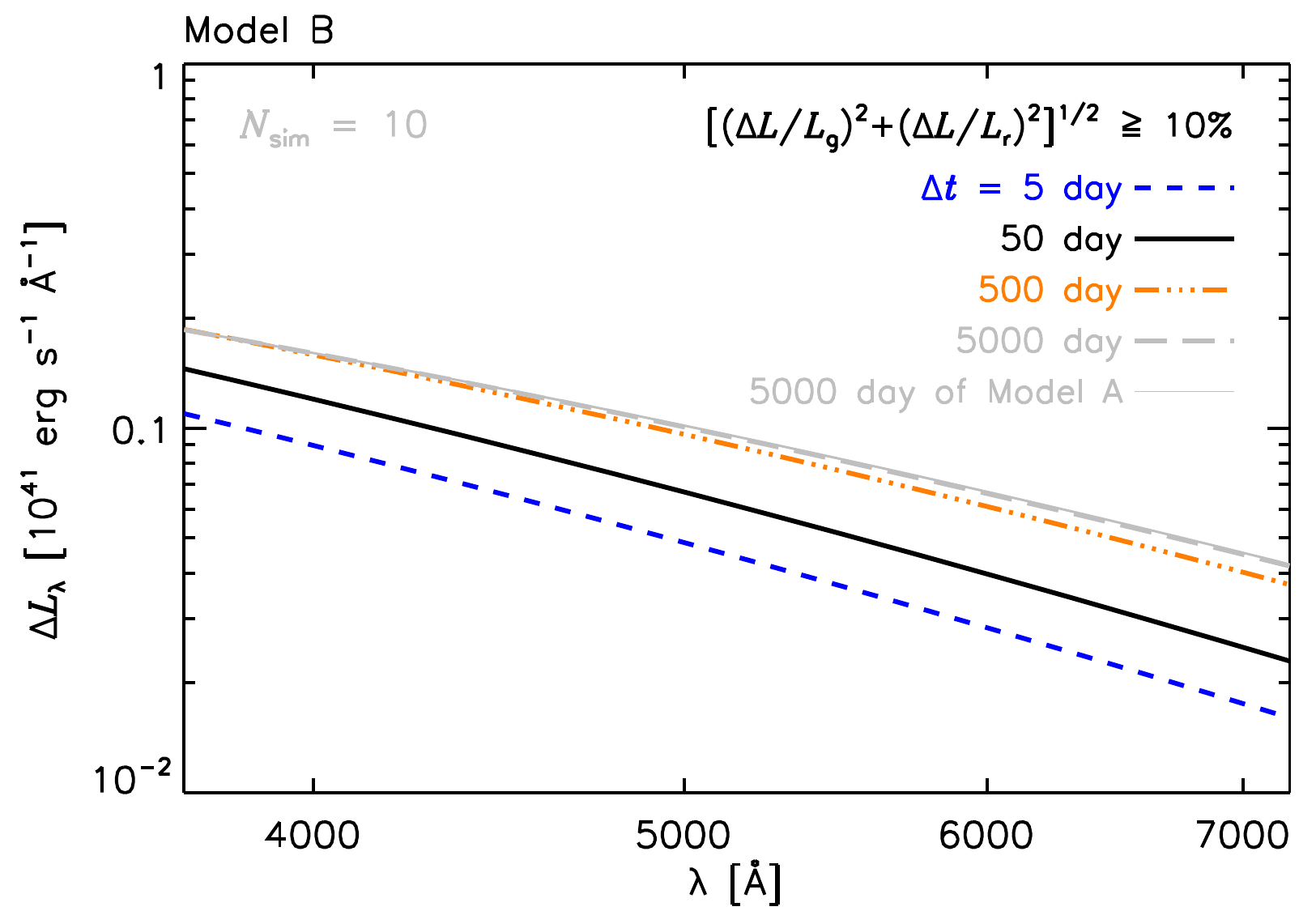}
\caption{The arithmetic mean composite difference spectra for {\bf model A} (left panel) and {\bf model B} (right panel), requiring $[(\Delta L/L_{\rm g})^2 + (\Delta L/L_{\rm r})^2]^{1/2} \geqslant 10\%$ at different $\Delta t = 5$ (blue dashed line), 50 (blue solid line), 500 (blue triple-dot-dashed line), and 5000 days (light-gray long-dashed line), and averaging over $N_{\rm sim} = 10$ times of {\it ideal} simulations. { The similar difference spectra at different timescales implies that the color variation is timescale independent, such as those given by {\bf model A}.} In {\bf model B} with radius-dependent $\tau$, the difference spectrum is prominent timescale dependent, i.e., bluer at shorter timescales. { At long timescales, both models present the same difference spectra as illustrated by the agreement between the light-gray thin solid line implied by {\bf model A} and that by {\bf model B} at $\Delta t = 5000$ days in the right panel.}
For simplicity the plotted wavelength range has been limited to be close to $g$- and $r$-bands, to alleviate suffering from the same selection effect as shown in Figure~\ref{fig:qv_BWB_Ruan14_wrange}.
}\label{fig:qv_BWB_Ruan14_dt}
\end{figure*}

The timescale dependence of the color variation in Figures~\ref{fig:qv_BWB_tau}, \ref{fig:qv_BWB_tau_params}-\ref{fig:qv_sed_comp_theta_dt_tau_alpha_B}, and \ref{fig:qv_sed_comp_theta_dt_z_B} implied by {\bf model B} can be interpreted as follows.
At given wavelength, the radiation mainly comes from a certain range of disk radius.
At very short timescales, the fluctuations at inner disk are stronger comparing with those at outer disk, thus the produced
variation emission is rather blue (with $\langle \theta \rangle$ significantly smaller than $45^\circ$). 
The fluctuations in disk zones at smaller radii (such as those contribute significantly 
to $g$-band emission) and at larger radii (such as those contribute to $r$-band emission) both gradually increase towards larger timescales.
This yields larger amplitude in flux variation in both $g$- and $r$-bands with increasing timescale, but the color of the variation remains constant.
Therefore $\langle \theta \rangle$ appears insensitive to $\Delta t$, e.g., in the top-right panel of Figure~\ref{fig:qv_BWB_tau} at small $\Delta t$.
At intermediate timescales, however, the fluctuations in disk zones at smaller radii begin to saturate, while the fluctuations at larger radii still steadily increase. 
Disk emission at larger radii which is redder thus makes relatively stronger contribution to the 
output flux variation with increasing timescales, and increases $\langle \theta \rangle$ accordingly.
At even larger $\Delta t$, the fluctuations in the outer disk zones also saturate, thus the color of the variation is again timescale independent. 
An interesting consequence is that for a single AGN one may not be able to detect the timescale dependency of the color variation at certain ranges of timescales. 

There is an intimate connection between the color variations and the mean difference spectra at different timescales. The similar shapes of difference spectra at different timescales would imply color variations are timescale independent, such as those given by {\bf model A} (the left panel of Figure~\ref{fig:qv_BWB_Ruan14_dt}). However, for {\bf model B} with radius-dependent $\tau$, the difference spectrum is prominent timescale dependent, that is, it is steeper and corresponds to more significant bluer-when-brighter at shorter timescale. Instead, at very long timescales, both models present the same difference spectra as well as the averaged $\theta$, as at long enough timescales, the fluctuations of the inner and outer disk zones all saturate at the same amplitude for both models.

\subsection{An artifact or not?}\label{sect:artifact}

Using the $B$- and $V$-band monitoring on 3C 120, \citet{Ramolla2015} did not find strong timescale dependence in its color variability, analyzed in flux-flux space on two ranges of timescales, i.e., 0-120 days and 1600-2000 days. They proposed that the reported timescale dependence of the color variability by \citet{Sun2014}  is likely an artifact caused by analyzing the data in magnitude space. Considering the host galaxy contamination (stronger in redder band) would reduce the variation amplitude in magnitude space, \citet{Ramolla2015} argued that, at longer timescales, as the flux variation amplitude is larger, the host contamination is reduced, and therefore, the color variation measured in mag-mag space appears weaker. 

However, the host galaxy contamination (in magnitude space) only depends on the brightness of the nucleus at given epoch, but not the time lags between epochs. In other words, photometric data points with longer time lags between epochs although tend to have
larger flux differences, these epochs do not correspond to brighter states on average in the nucleus emission. Therefore, the logic that the host contamination is weaker at longer timescales is incorrect, and analyzing the color variation in mag-mag space would not produce an artificial timescale dependence.

In fact, analyzing the Stripe 82 quasars in flux-flux space also produces clear timescale dependent color
variation pattern, in agreement with the results in mag-mag space (see also discussions in Section~\ref{sect:caveats}). Monte Carlo simulations also confirm that host galaxy contamination does not produce artificial timescale dependence in the color variation. Furthermore, using the same data of 3C 120 from \citet{Ramolla2015}, the timescale dependence in its color variation is found to be consistently weak in both mag-mag and flux-flux spaces. The absence of strong timescale dependence in the color variation in 3C 120 is likely due to 1) the light curve is not long enough to reveal the signal; 2) for AGNs with lower luminosities like 3C 120, X-ray reprocessing would be dominant and can yield timescale independent color variation (at timescales longer than the light travel time from corona to the corresponding disk radii); 3) 3C 120 contains a smaller SMBH with mass $\simeq 5 \times 10^7~M_\odot$ \citep{Nelson2000,Peterson2014} than quasars, thus the timescale dependence can only be detected at much shorter timescales (see the left panel of Figure~\ref{fig:qv_sed_comp_theta_dt_tau_alpha_B} for instance); or 4) radio loud AGNs behave differently. Details on these issues will be presented in an upcoming paper (Sun et al. in prep).

\subsection{Caveats}\label{sect:caveats}

\citet{Sun2014} discussed that neither the contamination from the host galaxy nor the changes of the global accretion rate can 
address the timescale dependent color variation. They suggested that shorter term variations could be attributed to thermal fluctuations in the inner region of the accretion disk, while longer term variations to larger scales. We examine this explanation by simply considering another revised reference {\bf model B} with radius-dependent $\tau$ for temperature fluctuations, which is in fact a natural assumption as the physical size of the disk cells is proportional to the radius. By doing so we achieve qualitatively similar timescale dependent color variations, as illustrated in the right panels of Figure~\ref{fig:qv_BWB_tau}, and further quantitative trends in shape, as illustrated in Figure~\ref{fig:qv_sed_comp_theta_dt_z_B}, after {\it empirically} correcting for emission line contributions.
We note that this revised reference model would also imply consistent microlensing disk sizes as in the original \citeauthor{DexterAgol2011} model since the time-averaged disk half-light radius is primarily determined by the long-term amplitude of temperature variance, $\sigma_{\rm l}$ or $\sigma_{\rm T}$ \citep[see Equation~4 of][]{DexterAgol2011}.

However, the contributions of emission lines to the broadband photometry at various redshifts have only been {\it empirically} taken in account in the present model (see Section~\ref{sect:stripe82}).
In principle, what is compared between observations and simulations in this work is the slope of $\langle \theta \rangle$ -- $\Delta t$ relation from simulations with observations, but not the absolute values of $\langle \theta \rangle$ (or the absolute amplitude of color variation). 
As broadband photometry is affected by not only the accretion disk continuum emission, but also the broad/narrow emission lines, the Balmer continuum \citep[e.g.,][]{Kokubo2014,Edelson2015}, and the host galaxy emission, recovering the absolute value of $\theta$ and its dependence to redshift through simulations is not straightforward and beyond the scope of this work.
\citet{Schmidt2012} have tried to construct a simple spectral variability model, including a single power-law (with variable slope and a pivot point in the IR) and an emission line template, to reproduce the observed color variability redshift dependence. However a single power-law, as assumed by \citet{Schmidt2012}, may be not a good enough description of real quasar spectrum when extending to UV wavelengths, and therefore, may bias the estimated color variation amplitudes (i.e., \citeauthor{Schmidt2012}'s $s_{\rm gr}$ or our $\theta$), especially at high redshifts. 
Furthermore, it is oversimplified to assume all quasars have the same emission line template spectrum, and these lines correspond or do not correspond to the continuum variation uniformly without lags.
These factors may in fact be responsible for the discrepancy between the observed redshift dependence of color variation and the one simulated by \citet[][see their Figure~4]{Schmidt2012}.

\begin{figure*}[!t]
\centering
\includegraphics[width=0.8\textwidth]{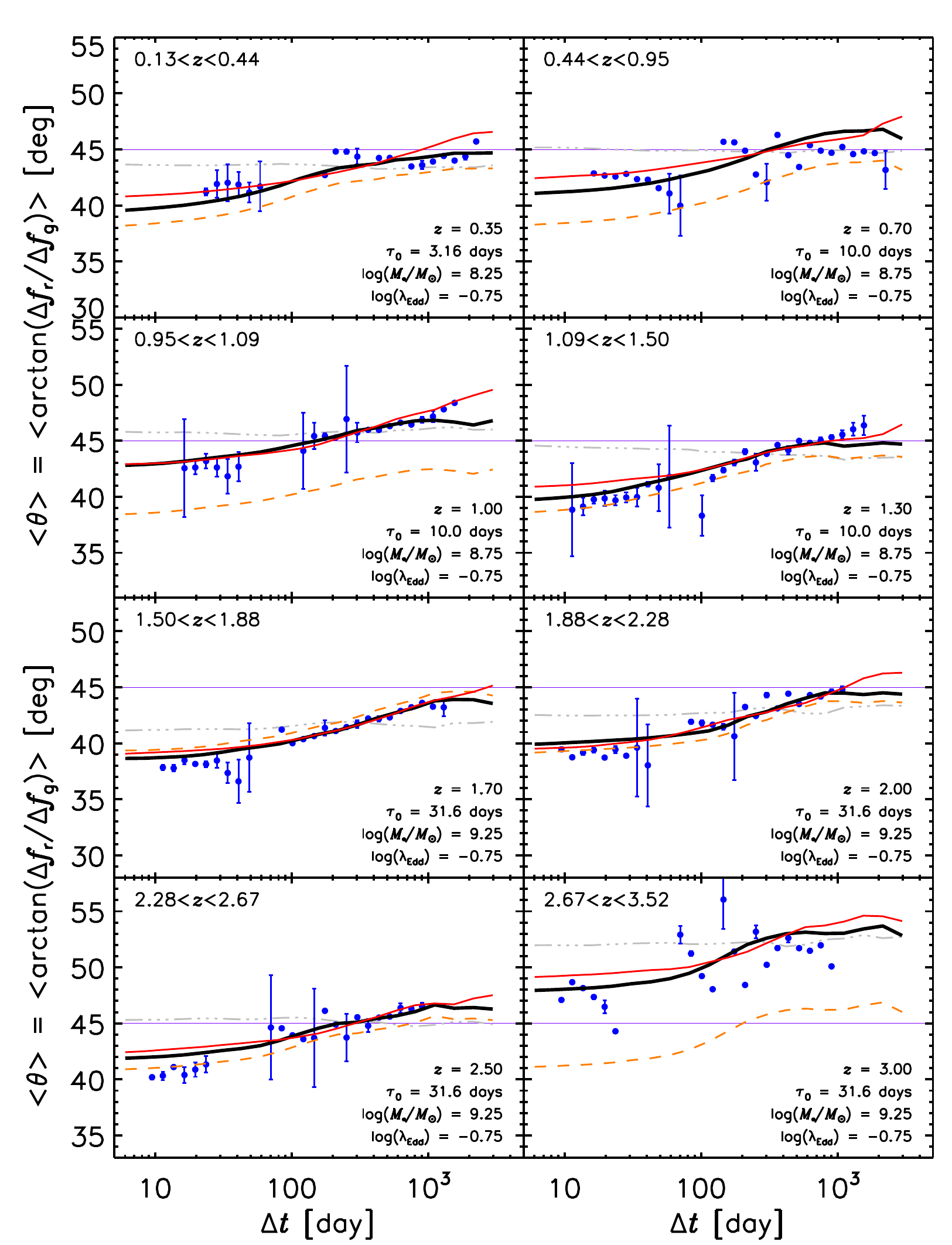}
\caption{Same as Figure~\ref{fig:qv_sed_comp_theta_dt_z_B}, except that the $\theta$ is estimated in flux-flux space, rather than mag-mag space. The data points (blue circles) are re-derived in flux-flux space as well using the same sources with bootstrapping errors. The results implied by the {\bf model B} with $\tau_0 = 10~(M_\bullet/10^{8.75}M_\odot)$ days and $\alpha = 1$, {\it empirically} corrected for emission line contributions, are illustrated as the thick black lines. The model predictions prior to corrections are also plotted (orange dashed lines). The adopted model parameters are indicated at the lower-right corner of each panel. For comparisons, the similarly corrected relations implied by {\bf model A} (light-gray thin triple-dot-dashed lines) and by {\bf model B} but with $\alpha = 1.5$ (red thin solid lines) are also shown.
}\label{fig:qv_sed_comp_theta_dt_z_ff_B}
\end{figure*}

Alternatively, the analyses can be performed in flux-flux space to directly measure the color of difference spectrum, which is free from contamination of non-variable component. 
However, as shown in \citet{Kokubo2014}, the color of difference spectrum of quasars shows also clear redshift dependence due to variable emission lines and Balmer continuum, indicating the effects of which can not be neglected either for analyses in flux-flux space.
In Figure~\ref{fig:qv_sed_comp_theta_dt_z_ff_B} we present plots similar to Figure~\ref{fig:qv_sed_comp_theta_dt_z_B}, but with $\theta$ measured in flux-flux space\footnote{The conversion from SDSS magnitudes, $b_{\rm SDSS}$, to flux densities is $f_{\nu, b} = 3631 \times 10^{-0.4(b_{\rm AB} - A_b)}$~Jy, where $b = \{g, r\}$, $g_{\rm AB} = g_{\rm SDSS}$, $r_{\rm AB} = r_{\rm SDSS}$, $A_g = 0.736 A_u$, $A_r = 0.534 A_u$, and $A_u$ the Galactic extinction \citep[e.g.,][or http://classic.sdss.org/dr7/algorithms/fluxcal.html]{MacLeod2012}.}, instead of in mag-mag space.
As expected, arbitrary shifts are still required to match the absolute values of $\theta$ from simulations to observations. Nevertheless, the slope in the $\theta$  -- $\tau$ relation is again nicely reproduced by our simulations.

The parameters of the inhomogeneous disk ($\tau$, $\sigma_{\rm l}$ and their possible radius dependence) could also depend on SMBH mass and accretion rate,  for which the observed sample span considerably large ranges, instead of the single values we adopted in this work.   
More fundamentally, the inhomogeneous disk model still encounters two problems. One of them is that the thin accretion model predicts bluer UV spectra than observed (see review by \citealt{Lawrence2012}).
Once introducing temperature fluctuations, the time-averaged SED is even bluer than the one without (see the top panels of Figures~\ref{fig:qv_SED_Lbol_Llambda} or \ref{fig:qv_BWB_Ruan14}). Therefore, the tension between the observed composite SED and the modeled one predicted by the thin disk would be further aggravated.
On the other hand, dust extinction and/or disk outflow could yield redder disk SEDs \citep[e.g.,][]{Capellupo2015}.

Nevertheless it is complicated by the extensive debate on whether the origin of AGN SED can be attributed to a geometrically thin, optically thick accretion disk or not \citep[see reviews by][]{KoratkarBlaes1999,DavisLaor2011}. 
To finally settle down the delicate dependence of $\tau$ and $\sigma_{\rm l}$ on radius for inhomogeneous disk model, we need to firstly work on a more realistic disk model.
It is possible as demonstrated by \citet{Capellupo2015} that the thin accretion disks are indeed, if not all, responsible for the origin of AGN SEDs once included various improvements, such as general relativistic corrections, radiation transfer in the disc atmosphere, dust extinction, and/or disk winds \citep[e.g.,][]{Hubeny2001,DavisLaor2011,SloneNetzer2012}.

\begin{figure}[!t]
\centering
\includegraphics[width=\columnwidth]{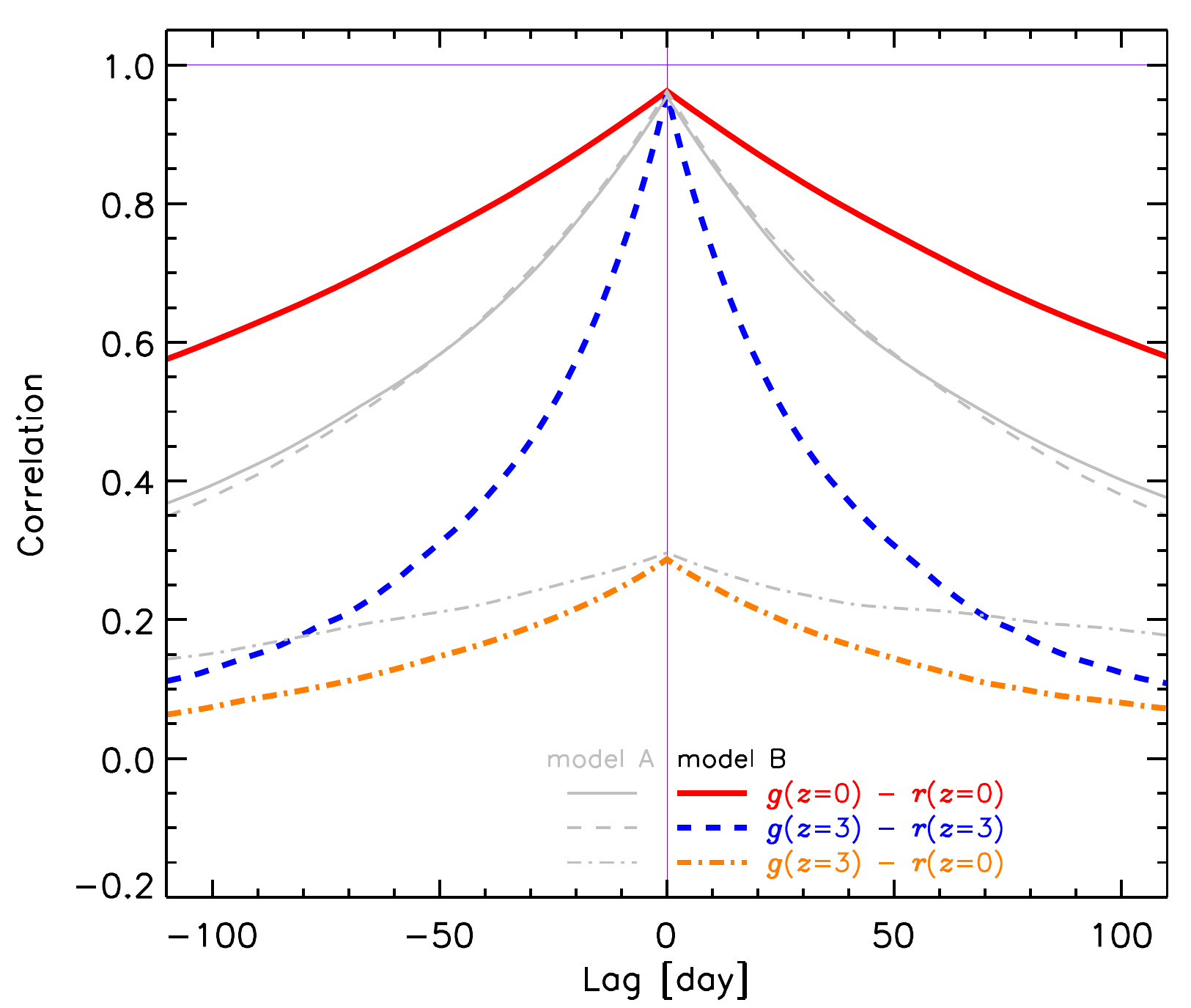}
\caption{Cross-correlation functions between simulated light curves at different wavelengths (labelled with SDSS $g$- and $r$-band at different redshifts) implied by {\bf model A} (light-gray thin lines) and {\bf model B} (colored thick lines). 
For both models, the correlations between $g$- and $r$-bands at all redshifts are strong and almost the same at zero lag, except that when the wavelength difference is larger, e.g., $g(z=3)$-$r(z=0)$.
The widths in the cross-correlation functions reflect the  characteristic timescales in the variation in corresponding light curves.
}\label{fig:qv_sed_comp_ccf}
\end{figure}

The second one is the so-called coordination problem.
Observationally, the variations of several Seyfert galaxies and few quasar at different UV/optical wavelengths are found to be quite coordinate (\citealt{Krolik1991}; see review by \citealt{Lawrence2012} and references therein). 
Delicately comparing the simulated inter-band correlations with observed ones are useful to testify and further improve the inhomogeneous disk models.
The bottom panel of Figure~\ref{fig:qv_SED_Lbol_Llambda} shows the simulated light curves at three UV/optical bands from the inhomogeneous disk models. Both (radius-independent and radius-dependent $\tau$) models produce well correlated light curves at the effective wavelengths of $\sim$ 4640\AA\ and $\sim$ 6122\AA, consistent with observations. 
More quantitative assessments are presented in the Figure~\ref{fig:qv_sed_comp_ccf} showing that the cross-correlation functions of the simulated $g$- and $r$-band light curves at two redshifts for both inhomogeneous models are found to be highly correlated ($\simeq 0.95$ for $g(z=0)-r(z=0)$ and $g(z=3)-r(z=3)$) at zero time lag. 
Instead, \citet{Kokubo2015} has shown the inhomogeneous disk model of \citet{DexterAgol2011} produces weaker inter-band correlations comparing with light curves of Stripe 82 quasars, suggesting further revisions to the models are required. 
It seems that our results somewhat contradicts that stated by \citet{Kokubo2015}, yet showing that the inter-band correlations with smaller wavelength differences, e.g., $g(z=0)-r(z=0)$, are indeed better. However, as \citet{Kokubo2015} adopted a different approach to assess the inter-band correlations, it is currently unclear that how good is the consistence both results even for the $g$-$r$ correlations and what is the reason for this potential difference, and we will leave this to a future work.

Notably, the inter-band correlations implied by both models indeed become less correlated with increasing the wavelength difference of the two bands, e.g., $g(z=3)-r(z=0)$ in Figure~\ref{fig:qv_sed_comp_ccf} or see Figure~\ref{fig:qv_SED_Lbol_Llambda} for $\sim$ 1160\AA\ and $\sim$ 4640\AA\//6122\AA\ light curves \citep[see also][]{Kokubo2015}.
In the current inhomogeneous disk models, the fluctuations at different disk regions were assumed to be totally independent to each other. 
The coordinate problem can be greatly alleviated if temperature fluctuations from different disk regions are somehow linked (such as through instability propagation, likely both inward and outward). 
With the propagation, the UV/optical variations could be more coordinated, at least at timescales longer than that for instability propagation.
In upcoming papers, we will delicately consider this problem as well as those aforementioned previously.

\section{Conclusion}\label{sect:conclusion}

We have demonstrated that the inhomogeneous accretion disk model, similar to that of \citet{DexterAgol2011} with radius-independent characteristic timescale for temperature fluctuations, conflicts with the discovery of a timescale dependent color variation of quasars by \citet{Sun2014}. The latter instead could be reproduced in an adjusted model whose characteristic timescales for temperature fluctuations are radius-dependent, unaffected by the un-even sampling and photometric error. This suggests that one may use quasar variations at different timescales to statistically probe accretion disk emission at different physical scales. We discuss further issues to be addressed in future simulations of inhomogeneous disks, including the discrepancy between the observed redder composite quasar SED and the modeled bluer one, and the coordination problem that variations at different UV/optical wavelengths occurs nicely in phase. 

Interesting by-products of this work include:

1. Assuming individual zones in an inhomogeneous accretion disk fluctuate in logarithm temperature following DRW process, the light curves of disk radiation (both bolometric and broadband) are not strict DRW mathematically, but can be approximately fitted with DRW. 

2. DRW fitting however yields smaller characteristic timescales than the input values for temperature fluctuations, thus the characteristic timescales measured from real quasar light curves can not be directly used as that of temperature fluctuations in the corresponding disk zones.

3. The inhomogeneous accretion disk models predict weak skewness in the magnitude (or $\log L$) distribution of quasar light curves, i.e., weak deviation from lognormal flux distribution. Such weak skewness, consistent with model prediction, is detected using quasar light curves in Stripe 82.

\section*{Acknowledgement}

We are grateful to Mou-Yuan Sun, John J. Ruan, Eric Agol, and Jason Dexter for valuable comments, and to the anonymous referee for his/her many constructive comments that led us to substantially improve our paper. 
This work is partly supported by National Basic Research Program of China (973 program, grant No. 2015CB857005), National Science Foundation of China (grants No. 11233002, 11421303, 11503024 $\&$ 11573023), and Specialized Research Fund for the Doctoral Program of Higher Education (20123402110030).
Z.Y.C. acknowledges support from the China Postdoctoral Science Foundation (grant No. 2014M560515) and the Fundamental Research Funds for the Central Universities.
J.X.W. thanks support from Chinese Top-notch Young Talents Program and the Strategic Priority Research Program ``The Emergence of Cosmological Structures" of the Chinese Academy of Sciences (grant No. XDB09000000).


\end{document}